\documentclass[11pt,a4paper]{article}
\usepackage[english]{babel}
\usepackage{a4wide}
\usepackage{ifthen}
\usepackage{amsmath}
\usepackage{amssymb}
\usepackage{array}
\usepackage{epsfig}
\usepackage{eepic}
\usepackage{color}

\begin{document}
\author{J.~Budczies and M.R.~Zirnbauer\\ {\footnotesize \it
    Universit\"at zu K\"oln, Institut f\"ur Theoretische Physik,
    Z\"ulpicher Str.~77, 50937 K\"oln, Germany}} \title{Howe Duality
  for an Induced Model of\\ Lattice ${\rm U}(N)$ Yang--Mills Theory}
\date{August 14, 2003} \maketitle

{\abstract We propose an approach that views ${\rm U}(N_{\rm c})$
  Yang--Mills theory as the critical point of an induced gauge model
  on the lattice.  Similar recent proposals based on the color--flavor
  transformation rely on taking the limit of an infinite number of
  infinitely heavy particles.  In contrast, we couple a {\it finite}
  number $N_{\rm b}$ of auxiliary boson flavors to the gauge field and
  argue that Yang--Mills theory is induced when $N_{\rm b}$ exceeds
  $N_{\rm c}$ and the boson mass is lowered to a critical point.
  Using the notion of Howe duality we transform the induced gauge
  model to a dual formulation in terms of local gauge invariant
  variables.  In the abelian case the Howe duality transform turns out
  to coincide with the standard one, taking weakly coupled ${\rm U}
  (1)_{d = 4}$ to strongly coupled ${\mathbb Z}_{d = 4}$ lattice gauge
  theory.}

\section{Introduction}

In what has come to be called ``induced QCD'', one starts from a
theory of auxiliary fields coupled minimally to a gauge field
background with gauge group $G$, the notable feature being that the
gauge field by itself has no dynamics.  A dynamical theory of gauge
fields, typically with nonlocal gauge interactions, is induced by
elimination of the auxiliary fields.  Such an approach
\cite{hamber,bander,hasenfratz,indQCD} has been suggested on the
lattice and in the continuum, using scalar fields as well as fermions
in the fundamental or adjoint representations of $G$.  Perhaps the
best known model in the induced QCD category is the Kazakov--Migdal
model \cite{indQCD}, which attracted a flurry of interest in 1992--93,
presumably because it admits a solution in the large--$N$ limit.
There exist two recent papers \cite{boristilo2,NS:02} motivated by the
color--flavor transformation that revive the old idea of inducing QCD
on the lattice. \smallskip

The big question looming over all these induced gauge models is
whether they do indeed lie in the universality class of Yang--Mills
theory with gauge group $G$ as intended.  In most of the past work,
the concern was with nonlocality of the effective gauge field action.
To suppress nonlocal contributions, one used the trick of sending both
the number of auxiliary fields and their mass to infinity in a
specific manner.  In the present paper, we propose a rather different
approach.  We choose the dynamics of the auxiliary fields to be local
to begin with, so that the induced gauge field action automatically
has that property.  Then, fixing the number $N_{\rm b}$ of auxiliary
fields, we tune the mass to a critical point so as to induce
Yang--Mills theory.  In order for this to work, it is crucial that
{\it bosonic} scalars (as opposed to fermions) be used. \smallskip

In more detail, the effective action of the induced gauge model we
propose is a sum over elementary plaquettes on a $d$--dimensional
lattice: $S = - 2 N_{\rm b} \sum_{\bf p} {\rm Re} \, {\rm Tr} \ln
\big( m - U(\partial {\bf p}) \big)$, with mass parameter $m > 1$.  At
unit mass the model has a critical point with a diverging correlation
length, allowing a continuum limit to be taken.  We claim that the $m
\to 1$ continuum model flows under renormalization to Yang--Mills
theory, provided that $N_{\rm b}$ exceeds a certain minimum value.
(What that minimum value is depends on the type of gauge group, its
rank, and the space--time dimension.)  The mathematical basis for this
claim is that the local weight function $U \mapsto |{\rm Det}(m - U)|
^{-2N_{\rm b}}$ on a classical compact Lie group $G$ approaches for $m
\to 1$ the $\delta$--function supported at $U = 1$, if $N_{\rm b}$ is
large enough.  \smallskip

The second major theme of the present paper is a duality
transformation for the induced gauge model.  While ``duality'' has
been very successful for gauge theories in the continuum and with
supersymmetries --- important examples are electric--magnetic (or
Montonen--Olive) duality, which constitutes an important aspect of the
Seiberg--Witten solution for the low--energy dynamics of ${\cal N} =
2$ supersymmetric gauge theory with gauge group ${\rm SU}(2)$; or the
Maldacena conjecture (or AdS/CFT--duality), by which ${\cal N} = 4$
super Yang--Mills theory with gauge group ${\rm SU}(N)$ is believed to
be dual to a type II-B superstring theory ---, our interest here will
be in pure gauge theories on the lattice, and without supersymmetry.
In this class of quantum field theories duality is a well developed
concept only for the {\it abelian} case
\cite{wegner,polyakov,banks,guth,juergtom,kogut83}, where passing to
the dual theory essentially amounts to taking a Fourier transform.
\smallskip

A direct transcription of the abelian duality transform to nonabelian
lattice gauge theories makes use of a character expansion of the
plaquette statistical weight function (see \cite{diakonov} for a
recent reference).  Integration over the gauge field then produces
sums over products of Racah coefficients of $G$ [these are
higher--rank generalizations of tensor invariants known as $3j$, $6j$,
$9j$, $12j$ symbols etc.~for the case of $G = {\rm SU} (2)$].  The
semiclassical asymptotics of such coefficients is not easy to handle,
and hence the continuum limit of the theory in this dual formulation
remains unclear except in some special situations. \smallskip

We consider it to be an interesting feature of the determinant--type
models we are going to introduce, that an alternative approach is
possible.  Our induced gauge models can be transformed to a dual
description --- for any one of the classical compact gauge groups $G =
{\rm U}(N)$, ${\rm Sp}(2N)$, and ${\rm O}(N)$ --- by viewing the
determinant weight functions as traces in an (auxiliary) Fock space.
The gauge group $G$ acts in this Fock space as one member of a dual
pair of Lie groups, where ``dual'' is meant in the sense of R.~Howe
\cite{Hw_pop,Howe}.  Integration over the gauge fields simply projects
on the $G$ invariant subspace of Fock space.  The other member of the
Howe dual pair is a noncompact Lie group acting irreducibly on that
subspace. This allows a fairly transparent description of what the
dual lattice theory is (although we do not yet understand its
continuum limit).  For $G = {\rm U}(1)$ our Howe duality transform
reproduces standard abelian duality upon elimination of some redundant
degrees of freedom. \smallskip

Howe pairs underlie the color--flavor transformation
\cite{circular,icmp97}, which was originally conceived in the context
of random matrix theory and disordered electron systems, and has
recently been applied to the strong coupling limit of QCD
\cite{Budczies:01,boristilo1,bnsz,thesis}.  Howe pairs have also been
used recently for the investigation of determinant correlations for
quantum maps \cite{detdet}. \smallskip

This paper is organized as follows.  In Section \ref{sec:induced} we
place (local) bosons and/or fermions on the sites of an arbitary
lattice and couple them to gauge fields in the standard manner to
induce a pure lattice gauge theory with gauge group $G$.  For the case
of $G = {\rm U}(N_{\rm c})$, Section \ref{sec:crit} establishes the
$\delta$--function property of the induced plaquette weight function
when the number of boson flavors $N_{\rm b} \ge N_{\rm c}$.  The
resulting boson induced gauge model is subjected to a careful
investigation in $d = 1 + 1$ dimensions in Section \ref{sec:2d}.  We
show that its partition function in the continuum limit (and on any
Riemann surface) agrees with the known partition function of some
${\rm U}(N_{\rm c})$ Yang--Mills theory if $N_{\rm b} \ge N_{\rm c} +
1$.  For $N_{\rm b} = N_{\rm c}$ we identify an exotic continuum gauge
theory, which is not Yang--Mills but of Cauchy type.  Then, after
introducing the notion of Howe pairs in the two--dimensional context,
we subject the boson induced gauge model to a duality transformation
in any space--time dimension in Section \ref{sec:dual}. We conclude
with a summary and listing of open questions in Section
\ref{sec:discuss}, commenting in particular on the issue of
convergence of the color--flavor transformation when the number of
boson flavors is large.

\section{Fermion and Boson Induced Gauge Model}\label{sec:induced}

It has become standard practice in lattice gauge theory to place the
fields on a simple hypercubic lattice.  While convenient for the
purpose of doing numerical calculations, the restriction to hypercubic
lattices is rather narrow and special from the perspective of
continuum field theory.  Since our interest ultimately is in defining
and taking a continuum limit, we will set up our formalism on lattices
more general than the hypercubic one.  This prevents us of from making
short cuts and helps us direct our attention to the proper structures.
\smallskip

We take discrete space--time to be some $d$--dimensional complex
$\Lambda$ built from oriented $k$--cells.  For the cases $k = 0, 1, 2$
these are referred to as {\it sites}, {\it links} (with a direction),
and {\it plaquettes} (with a sense of circulation).  It will often be
convenient to view the $k$--cells of $\Lambda$ as the generators of
abelian groups $C_k(\Lambda)$.  Their elements, called $k$--chains,
are linear combinations of $k$--cells with coefficients in ${\mathbb
  Z}$.  If $c$ is a $k$--chain, then so is $-c$; we get the latter
from the former by reversing the orientation for all its $k$--cells.
For any reasonable choice of $\Lambda$ there exists a boundary
operator $\partial$, which is a linear operator $\partial \, : \, C_k
(\Lambda) \to C_{k-1} (\Lambda)$ with the property $\partial \circ
\partial = 0$ (the boundary of a boundary always vanishes).  For
example, the boundary of an oriented link ${\bf l}$ that begins on
site $n_i$ and ends on site $n_f$ is $\partial {\bf l} = n_f - n_i$;
which is the chain consisting of the $0$--cells $n_f$ and $n_i$ with
coefficient $+1$ and $-1$ respectively.  The boundary of an oriented
plaquette ${\bf p}$ is the chain of $1$--cells ${\bf l}_i$ surrounding
it: $\partial {\bf p} = \sum_i \pm {\bf l}_i$, where the plus/minus
sign is chosen when the orientations agree/disagree.  \smallskip

The following elaboration on language might help prevent confusion
later on: each $k$--cell of $\Lambda$ comes with exactly {\it one} of
two possible orientations; i.e.~the choice of orientation for the
$k$--cells is {\it fixed}, albeit arbitrary.  This means in particular
that if the oriented link ${\bf l}$ is a $1$--cell, then $-{\bf l}$ is
{\it not} a $1$--cell (although it still is an oriented link).
\smallskip

On a $d$--dimensional lattice or cell complex of this general kind, we
are going to consider a gauge theory with partition function
\begin{equation}\label{partfunc}
  Z = \int [dU] \int [d \varphi] [d \psi] \, {\rm e}^{- S_{\rm f}
    [\psi, \bar{\psi},U] - S_{\rm b}[\varphi, \bar{\varphi}, U]} \;,
\end{equation}
where the action functionals $S_{\rm b}$ and $S_{\rm f}$ will be
specified shortly.  As usual, matrices $U$ taking values in a compact
gauge group $G$ are placed on the links of the lattice.  More
precisely speaking, a lattice gauge field configuration is a mapping
from the $1$--cells of $\Lambda$ into $G$.  The mapping extends to all
oriented links by the convention $U(-{\bf l}) \equiv U({\bf l})^{-1}$,
which is motivated by the interpretation of the $U$'s as discrete
approximations to the path--ordered line integrals of an underlying
gauge field $A$: $U({\bf l}) \approx P \exp(\int_{\bf l} A)$.  The a
priori statistical weight of the lattice theory is a product of Haar
measures $dU$ over the $1$--cells of $\Lambda$: $[dU] = \prod_{\bf l}
dU({\bf l})$.  Concrete calculations will be carried out for the
unitary groups $G = {\rm U}(N_{\rm c})$.  \smallskip

In addition to the gauge fields, the theory has a second ingredient:
complex bosonic and/or fermionic fields that are placed on the sites
of $\Lambda$ and transform according to the fundamental vector
representation of the gauge group.  The fermions are denoted by
$\psi$, the bosons by $\varphi$.  They are vectors not only in $N_{\rm
  c}$--dimensional color space, but also in a ``flavor'' space with
dimension $N_{\rm f}$ (fermions) and $N_{\rm b}$ (bosons). \smallskip

\begin{figure}
  \begin{center}
    \hspace{-1cm}
    \input{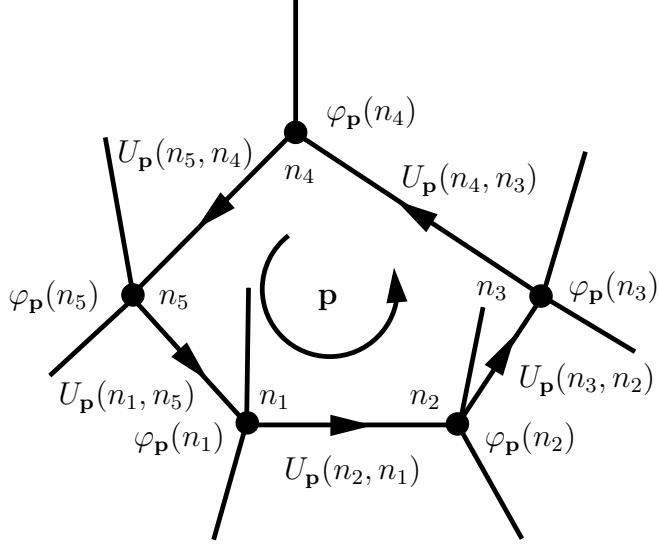}
  \end{center}
  \caption{Hopping of the auxiliary bosons and their coupling
    to the gauge field.}
  \label{fig:clock}
\end{figure}

These bosons and fermions are auxiliary (i.e.~unphysical) degrees of
freedom introduced solely for the purpose of inducing an effective
action for the gauge fields.  Unlike the conventional matter fields of
lattice gauge theory, they do not propagate all over the lattice.
Rather, each one of them is constrained to hop (in the presence of the
lattice gauge field $U$) along the boundary chain of some plaquette.
In other words, there is a one--to--one correspondence between the
$2$--cells ${\bf p}$ of $\Lambda$ and sets of complex boson and
fermion variables, $\{ \varphi_{\bf p} \}$ and $\{ \psi_{\bf p} \}$.
The orientation of ${\bf p}$ determines the sense of circulation of
the hopping of the auxiliary particles $\{ \varphi_{\bf p} \}$ and $\{
\psi_{\bf p} \}$; see Figure \ref{fig:clock}.  The integration measure
$[d\psi] [d\varphi]$ is taken to be the product of flat measures.
\smallskip

To fix the precise details, we must distinguish between gauge groups
of two types: those where the vector ($U$) and covector
(${U^{-1}}^{\rm T}$) representations are related by an inner
automorphism, and others where they are not.  The groups ${\rm SO}
(N_{\rm c})$, ${\rm O}(N_{\rm c})$ and ${\rm Sp}(2N_{\rm c})$ belong
to the former type, the unitary groups ${\rm U}(N_{\rm c})$ and ${\rm
  SU} (N_{\rm c})$ to the latter.  For gauge groups of the former
type, the setup we have described --- one set of variables per
$2$--cell --- would already suffice. For the latter type, however, and
especially for ${\rm U}(N_{\rm c})$, which the present paper focuses
on, we have to {\it double} the set of auxiliary variables: it will be
seen that, to induce a good effective action for the gauge fields,
hopping must take place in both the clockwise and the counterclockwise
sense for all plaquettes.  Thus for each $2$--cell ${\bf p}$ of
$\Lambda$, we introduce {\it two} sets of auxiliary variables, one
associated to the $2$--cell with its proper orientation ($+{\bf p}$),
and another one where the orientation is reversed ($-{\bf p}$).
\smallskip

These words are put in formulas as follows.  Let $p = \{ {\bf p},
-{\bf p} \}$, and let $L_p$ denote the ``length'' of $\pm \partial
{\bf p}$, i.e.~the number of links contained in the boundary chain of
$\pm {\bf p}$. Then, fixing some oriented plaquette ${\bf p}$, we
write its boundary as a formal sum of $L_p$ oriented links with
positive coefficients: $\partial{\bf p} = l_{n_1,n_2} + l_{n_2,n_3} +
\ldots$, where $\partial l_{n_j, n_{j+1}} = n_{j+1} - n_{j}$.  The
$0$--cells $n_1, \ldots, n_{L_p}$ are the sites visited by
$\partial{\bf p}$, arranged in ascending order as prescribed by the
sense of circulation of ${\bf p}$.  To take notational advantage of
the cyclic structure of the boundary chain $\partial{\bf p}$, we
identify $n_1 \equiv n_{{L_p}+1}$.  With these conventions, we put
$U_{\bf p} (n_{j+1}, n_j) \equiv U(l_{n_{j},n_{j+1}})$, and define the
actions $S_{\rm f}$ and $S_{\rm b}$ to be
\begin{eqnarray}
  S_{\rm f}[\psi, \bar{\psi},U] &=& \sum_{\pm{\bf p}} \sum_{j=1}^{L_p}
  \Big( m_{{\rm f},p} \, \bar\psi_{\bf p}(n_j) \psi_{\bf p}(n_j) -
  \bar\psi_{\bf p}(n_{j+1}) U_{\bf p}(n_{j+1},n_j) \psi_{\bf p}(n_j)
  \Big) \;, \\ S_{\rm b}[\varphi, \bar\varphi,U] &=& \sum_{\pm{\bf p}}
  \sum_{j=1}^{L_p} \Big( m_{{\rm b}, p} \, \bar \varphi_{\bf p}(n_j)
  \varphi_{\bf p}(n_j) - \bar\varphi_{\bf p}(n_{j+1}) U_{\bf p}
  (n_{j+1} ,n_j) \varphi_{\bf p}(n_j) \Big) \;.
  \label{bos_act}
\end{eqnarray}
To keep the expressions transparent, we have suppressed the color ($i
= 1, 2, \ldots, N_{\rm c}$) and flavor indices ($a = 1, 2, \dots,
N_{\rm f/b}$); it should be clear how to restore them.  For example,
$\bar\varphi U \varphi$ is short--hand for $\sum_{i_1,i_2 = 1}^{N_{\rm
    c}} \sum_{a=1}^{N_{\rm b}} \bar\varphi^{i_1,a} U^{i_1 i_2}
\varphi^{i_2 , a} $.  The notation $\sum_{\pm{\bf p}}$ means that each
plaquette occurs {\it twice} in the sum, once each for the two
possible orientations.  The parameters $m_{{\rm f},p}$ and $m_{{\rm
    b},p}$ are referred to as the (local) fermion and boson masses.
We allow for the possibility that they depend on the plaquette label
$p$ in general, but on a lattice with translational invariance and no
external gravitational field, we will take them to be constant.  The
field integral makes sense for any set of $m_{{\rm f},p} \in {\mathbb
  C}$, while the local boson masses must satisfy ${\rm Re} \, m_{{\rm
    b},p} > 1$ for convergence.  \smallskip

The lattice theory so defined is intended to be a discretization of
$d$--dimensional (Euclidean) Yang--Mills theory with gauge group $G =
{\rm U}(N_{\rm c})$ or ${\rm SU}(N_{\rm c})$.  In the second half of
the paper we will eliminate the gauge field so as to pass to a dual
description.  For the time being we address the question whether
(\ref{partfunc}) has a critical point with a continuum limit
equivalent to Yang--Mills theory. \smallskip

The field integrals for the auxiliary fermions $\psi, \bar\psi$ and
bosons $\varphi, \bar\varphi$ are Gaussian.  Carrying them out we
obtain a product of determinants (for fermions) and inverse
determinants (for bosons), one each for every oriented plaquette ${\bf
  p}$.  The determinants from the fermions associated with ${\bf p}$
are ${\rm Det}^{N_{\rm f}} \big( m_{{\rm f},p}^{L_p} - U(\partial{\bf
  p}) \big)$, with $U(\partial{\bf p})$ being the ordered product of
the $U$'s along the boundary chain of ${\bf p}$:
\begin{displaymath}
  U(\partial{\bf p}) \equiv U_{\bf p}(n_1,n_{L_p}) \, U_{\bf
    p}(n_{L_p}, n_{L_p - 1}) \cdots U_{\bf p}(n_3,n_2) \, U_{\bf
    p}(n_2,n_1) \;.
\end{displaymath}
The corresponding factor from the same plaquette with the opposite
orientation, $-{\bf p}$, is just the complex conjugate of this.  The
Gaussian boson integrals give a similar answer except that the
determinants in that case go in the denominator.  Thus, combining
factors we have
\begin{displaymath}
  Z = \int [dU] \, \prod_{\bf p} \frac{\big| {\rm Det} \big(m_{{\rm
        f},p}^{L_p} - U(\partial{\bf p}) \big) \big|^{2N_{\rm f}}}
  {\big| {\rm Det} \big(m_{{\rm b},p}^{L_p} - U(\partial{\bf p}) \big)
    \big|^{2N_{\rm b}}} \;.
\end{displaymath}
Note that this is a product over $2$--cells, i.e.~each oriented
plaquette ${\bf p}$ occurs only once (and the result does not depend
on the orientations chosen). \smallskip

By sending the result of the integration back to the exponent and
dropping an irrelevant constant, we obtain
\begin{equation}\label{ourmodel}
  Z = \int [dU] \, {\rm e}^{-S_{\text{ind}, {\rm f}}[U] - S_{\text{ind},
      {\rm b}}[U]}
\end{equation}
with fermion and boson induced actions
\begin{align}
  S_{\text{ind},{\rm f}}[U] &= - 2 N_{\rm f} \, {\rm Re} \sum_{\bf p}
  {\rm Tr} \ln \big(1 - \alpha_{{\rm f},p} \, U(\partial{\bf p}) \big)
  \;, \label{fAction} \\
  S_{\text{ind},{\rm b}}[U] &= + 2 N_{\rm b} \, {\rm Re} \sum_{\bf p}
  {\rm Tr} \ln \big(1 - \alpha_{{\rm b},p} \, U(\partial{\bf p}) \big)
  \;. \label{bAction}
\end{align}
In place of the fermion and boson masses, we have introduced the
coupling parameters $\alpha_{{\rm f},p} \equiv m_{{\rm f},p}^{- L_p}$
and $\alpha_{{\rm b},p} \equiv m_{{\rm b},p}^{- L_p}$.  We take the
range of these parameters to be $- \infty < \alpha_{{\rm f},p} <
\infty$ and $- 1 < \alpha_{{\rm b},p} < 1$.  Note that the partition
function (\ref{ourmodel}) becomes singular in the limit $\alpha_{{\rm
    b},p} \to 1$, since the function $U \mapsto | {\rm Det} (1 -
\alpha U)|^{-2N_{\rm b}}$ does so on the codimension one submanifold
of matrices $U$ with at least one eigenvalue equal to unity.
\smallskip

There is an obvious way \cite{NS:02} in which to approach Wilson's
lattice gauge theory \cite{wilson,kogut79,creutz} with action
\begin{equation}\label{wilson}
  S^{\text{W}}[U] = - \frac{\beta}{N_{\rm c}} \sum_{\bf p} {\rm Re} \,
  {\rm Tr} \, U(\partial{\bf p})
\end{equation}
from (\ref{fAction}) and (\ref{bAction}): specializing to the case of
a hypercubic lattice, we take the couplings $\alpha_{{\rm b},p}$ and
$\alpha_{{\rm f},p}$ to be independent of $p$.  We then send them to
zero, and the number of flavors $N_{\rm f/b}$ to infinity, while
keeping the products $\alpha_{\rm f/b} N_{\rm f/b}$ fixed.  Since
$\lim_{N \to \infty} N \ln (1 + x/N) = x$, this limit exactly
reproduces the Wilson action (\ref{wilson}), with $\beta / N_{\rm c} =
2(N_{\rm b} \alpha_{\rm b} - N_{\rm f} \alpha_{\rm f})$.  If only
fermions are used, the coupling $\alpha_{\rm f} = m_{\rm f}^{-L_p}$
must be negative, corresponding to a complex mass, say $m_{\rm f} =
|m_{\rm f}| \, {\rm e}^{{\rm i}\pi/L_p}$.  \smallskip

It is very unclear to us, however, whether this double limit of an
infinite number of infinitely heavy particles is going to lead to any
advance in our knowledge about Yang--Mills theory.  Certainly, the
case of a single boson flavor $(N_{\rm b}=1)$ treated in \cite{NS:02}
bears no relation with Yang--Mills theory, let alone QCD.  \smallskip

In the present paper we are going to propose and investigate a
more interesting possibility.  The partition function
(\ref{ourmodel}) becomes singular when all $\alpha_{{\rm b},p}$
are sent to unity (say, uniformly in $p$), which is in fact a
critical point with a diverging correlation length.  Our main
message will be that, on suitable lattices and in high enough
space--time dimension, the critical theory with $\alpha_{{\rm
b},p} \to 1$ (and $\alpha_{{\rm f},p} \not= 1$) is expected to be
in the universality class of ${\rm U} (N_{\rm c})$ Yang--Mills
theory, provided that $N_{\rm b} \ge N_{\rm c}$.  The universality
conjecture can readily be checked in dimension $d = 1 + 1$. There,
it will be demonstrated that universality holds for $N_{\rm b} >
N_{\rm c}$, while it fails for $N_{\rm b} = N_{\rm c}$.

\section{The Critical Point $\alpha_{\rm b} = 1$}\label{sec:crit}

\subsection{Lattice versus continuum}

To communicate our argument, we must first recall the general scenario
\cite{kogut79,creutz} by which lattice gauge and continuum gauge
theories are related.  Imagine some smooth $d$--dimensional
configuration of the gauge field $A$, and superimpose on it a fine
mesh in the form of some $d$--dimensional lattice or cell complex
$\Lambda$ with oriented elementary plaquettes ${\bf p}$.  The rule of
translation from the continuum gauge field $A$ to the lattice field
$U$ is given by
\begin{equation}\label{approx}
  U(\partial{\bf p}) = P \exp \oint_{\partial{\bf p}} A = 1 + F({\bf
    p}) + \ldots \;,
\end{equation}
where $P$ means path ordering, and $F({\bf p})$ is the field strength
(the curvature of the gauge connection $A$) evaluated on ${\bf p}$ (in
the standard setting of a hypercubic lattice, this means that if ${\bf
  p}$ is a plaquette based at the site $n$ and is parallel to the
$\mu\nu$--plane, then $F({\bf p}) = a^2 F_{\mu\nu}(n)$ with $a$ the
lattice constant). \smallskip

Clearly, the lattice approximation will be reasonable if the mesh
formed by the lattice is fine enough in order for field curvature
effects to be small on the lattice scale.  Under such conditions $P
\exp \oint_{\partial{\bf p}} A$, and hence $U(\partial{\bf p})$, will
be close to unity for every elementary plaquette ${\bf p}$.  Thus, if
the lattice approximates the continuum, the plaquette matrices $U(
\partial {\bf p})$ fluctuate only weakly around the zero--field
strength configuration $U(\partial{\bf p}) = 1$. \smallskip

Conversely, if the statistical weight of the lattice theory
sharply peaks at unity $U(\partial{\bf p}) = 1$ for all ${\bf p}$,
then the lattice theory has a very large correlation length and is
close to a continuum limit.  We may then pass to a continuum field
theory formulated in terms of the field strength $F$, by expanding
around $U(\partial{\bf p}) = 1$ and using the correspondence
(\ref{approx}). \smallskip

The main principle then is this.  Let the lattice gauge theory be
given by a product statistical measure $\prod_{\bf p} w_t \big( U(
\partial {\bf p}) \big) [dU]$, where $t$ is some coupling parameter
with critical value $t_c$.  We require the weight function $w_t(U)$
for $t$ close to $t_c$ to be very strongly localized at the unit
element so as to make excursions away from unity statistically very
rare.  More precisely, we want $w_{t_c}(U) \equiv \delta(U)$ to be the
Dirac $\delta$--function supported at the unit element of the gauge
group, and $w_{t}(U)$ for $t \not= t_c$ to be some smeared version
thereof.  When $t$ moves to $t_c$, the smearing is undone, the lattice
statistical weight approaches the product of $\delta$--functions
$\prod_{\bf p} \delta\big( U(\partial{\bf p}) \big)$, the correlation
length goes to infinity, and the lattice gauge theory converges to a
continuum limit.  Under favorable conditions, this continuum limit
will be Yang--Mills theory with action functional $- S_{\rm YM}
\propto \int {\rm Tr} \, F \wedge \star F$.  \smallskip

This scenario is realized for the Wilson weight function
\begin{displaymath}
  w^{\rm W}_\beta(U) = {\rm e}^{(\beta / N_{\rm c}) \, {\rm Re} \,
    {\rm Tr} U} \Big/ \int {\rm e}^{(\beta / N_{\rm c}) \, {\rm Re}\,
    {\rm Tr} U} dU
\end{displaymath}
when the parameter $\beta \equiv t$ is sent to $t_c = \infty$.  For
one thing, one easily shows
\begin{displaymath}
  \lim_{\beta \to \infty} \int\limits_{{\rm U}(N_{\rm c})} f(U)
  w_\beta^{\rm W}(U) dU = f(1)
\end{displaymath}
for any smooth function $f$ on ${\rm U}(N_{\rm c})$ (or another
compact gauge group, for that matter), which is equivalent to
saying $\lim_{\beta \to \infty} w_\beta^{\rm W} (U) = \delta(U)$.
Moreover, it is strongly suggested by numerical simulations that
the continuum limit approached by Wilson's lattice gauge theory
for $\beta \to \infty$ has all the properties expected of quantum
Yang--Mills theory in the continuum.  The ultraviolet stability of
Wilson's theory has been established by rigorous analysis
\cite{balaban}. \smallskip

There is of course nothing unique about the Wilson action and many
other one--parameter families of weight functions do the same or
an even better job.  Let us single out one example.  In the
representation theory of compact semisimple Lie groups $G$ there
exists a statement, called the Peter--Weyl theorem \cite{knapp},
which implies that the Dirac $\delta$--function on $G$ can be
built up from the complete set of irreducible representations
$D^\lambda$ of $G$ as follows:
\begin{equation}\label{peterweyl}
  \delta(U) = {\rm vol}(G)^{-1} \sum_{{\rm all} \, \lambda} d_\lambda
  \, {\rm Tr} D^\lambda (U) \;,
\end{equation}
where $d_\lambda = {\rm Tr} \, D^\lambda(1)$ is the dimension of
the representation.  Hence, if $c_2(\lambda) \ge 0$ is the
quadratic Casimir invariant evaluated in the representation
$D^\lambda$, the function
\begin{equation}\label{heatkernel}
  w_t^{\rm HK}(U) = {\rm vol}(G)^{-1} \sum_{{\rm all} \, \lambda} {\rm
    e}^{- c_2(\lambda) t} \, d_\lambda \, {\rm Tr} D^\lambda (U)
  \qquad (t > 0)
\end{equation}
approaches the $\delta$--function as $t \to t_c = 0$.  For this reason
one expects it to provide (in the limit of small $t$) a valid lattice
regularization of Yang--Mills theory with gauge group $G$.  \smallskip

This weight function is called the heat kernel: it solves the heat
equation $\partial_t w_t(U) = \triangle w_t(U)$ ($\triangle$ being the
Laplace--Beltrami operator) on $G$ with initial condition $\lim_{t \to
  0+} w_t(U) = \delta(U)$.  The ``time'' parameter $t$ is the analog
of $1 / \beta$ for the Wilson weight function.  In the abelian
case $G = {\rm U}(1)$, the model with weight function $w_t^{\rm
HK}$ is known in condensed matter theory as the Villain model.

\subsection{$\delta$--function limit}

After these preparations, we are ready to get to the point: we are
going to investigate the determinantal weight function
\begin{equation}\label{detw}
  w_{N_{\rm b},\alpha}(U) = \big| {\rm Det}(1 - \alpha \, U)
  \big|^{-2N_{\rm b}} \Big/ \int_{G} \big| {\rm Det}(1 - \alpha \, U)
  \big|^{ - 2 N_{\rm b}} dU
\end{equation}
of the boson induced gauge model (\ref{ourmodel}) with $\alpha_{\rm f}
= 0$ and $\alpha \equiv \alpha_{\rm b} < 1$ close to unity.  We will
be interested mostly in the case $G = {\rm U}(N_{\rm c})$, but note
that the definition of $w_{N_{\rm b},\alpha}$ still makes sense if $G
= {\rm U}(N_{\rm c})$ is replaced by any of its closed subgroups.
\smallskip

As a warm up, we will look at two special cases.  The simplest example
is $G = {\rm U}(1)$ with $N_{\rm b} = 1$, $U = {\rm e}^{{\rm i}
  \theta}$ and Haar measure $dU = d\theta$.  In that case, elementary
manipulations show
\begin{displaymath}
  w_{1,\alpha} ({\rm e}^{{\rm i}\theta}) = (2\pi)^{-1} \frac{1 -
  \alpha^2 } {|1 - \alpha \, {\rm e}^{{\rm i}\theta}|^2} = \frac{1}{2\pi}
  \sum_{n \in {\mathbb Z}} \alpha^{|n|} {\rm e}^{{\rm i}n \theta} \;,
\end{displaymath}
from which it is seen that $w_{1,\alpha}$ approaches the Dirac
$\delta$--function on ${\rm U}(1)$ in the limit $\alpha \to 1$.
The difference to the Villain model (\ref{heatkernel}) is that the
Gaussian cutoff ${\rm e}^{-c_2(n)t} = {\rm e}^{-n^2 t}$ has been
replaced by an exponential cutoff ${\rm e}^{-|n|t}$, with $t = \ln
(1/\alpha) > 0$. \smallskip

Another nice example is $G = {\rm SU}(2)$, still with $N_{\rm b} = 1$.
Again, by straightforward manipulations on the determinantal weight
function (\ref{detw}) evaluated on $U = {\rm e}^{{\rm i} \theta
  \sigma_3} \in {\rm SU}(2)$, one finds with $\int_{{\rm SU}(2)} dU
  = 2\pi^2$:
\begin{displaymath}
  w_{1,\alpha} ({\rm e}^{{\rm i}\theta\sigma_3}) = (2\pi^2)^{-1}
  \frac{1 - \alpha^2}{|1 - \alpha \, {\rm e}^{{\rm i} \theta}|^4} =
  \frac{1}{2\pi^2} \sum_{n=0}^\infty \alpha^{n} \, (n+1) \,
  \frac{ \sin \big( (n+1)\theta \big) }{\sin\theta} \;.
\end{displaymath}
On the right--hand side we recognize the character $\chi_S({\rm
  e}^{{\rm i}\sigma_3}) = \sin \left( (2S+1)\theta \right) / \sin
\theta$ of the spin $S = n/2$ representation of ${\rm SU}(2)$, and
its dimension $2S + 1 = n + 1$.  Both sides extend uniquely to
class functions of ${\rm SU}(2)$, i.e.~to functions on ${\rm SU}
(2)$ that are invariant under conjugation $U \mapsto g U g^{-1}$
by $g \in {\rm SU}(2)$.  As a result, we deduce $\lim_{\alpha\to
1} w_{1, \alpha}(U) = \delta^{{\rm SU}(2)} (U)$ from
(\ref{peterweyl}). \bigskip

\textsf{Fact}. {\it Let $f : {\rm U}(N_{\rm c}) \to {\mathbb C}$
be an analytic function.  Then, for any $N_{\rm b} \ge N_{\rm
c}$},
\begin{equation}\label{thm1}
  \lim_{\alpha \to 1} \int\limits_{{\rm U}(N_{\rm c})} f(U) w_{N_{\rm
      b},\alpha}(U) dU = f(1) \;.
\end{equation}

Thus we are claiming the desired $\delta$--function property for
$N_{\rm b} \ge N_{\rm c}$.  The proof will be given in the next
subsection.  For now we make two comments: (i) There is nothing
special about ${\rm U}(N_{\rm c})$ in this context, and we expect
a similar statement to be true for each of the compact matrix
groups $G = {\rm SU}(N_{\rm c})$, ${\rm SO}(N_{\rm c})$, ${\rm
O}(N_{\rm c})$, ${\rm Sp}(2N_{\rm c})$.  (ii) In view of the
Peter--Weyl theorem, the statement (\ref{thm1}) implies that the
complete set of irreducible representations of ${\rm U}(N_{\rm
c})$ occur in the character expansion of $w_{N_{\rm b},\alpha}(U)$
for $N_{\rm b} \ge N_{\rm c}$. Conversely, one can show that some
representations are missing for $N_{\rm b} < N_{\rm c}$, which
means that the inequality $N_{\rm b} \ge N_{\rm c}$ in the
statement cannot be relaxed but is optimal.

\subsection{Proof of fact}\label{sec:proof}

We will give an elementary proof, and start with a few considerations
that simplify it. \smallskip

First of all, if the function $U \mapsto |{\rm Det}(1 - \alpha \,
U)|^{-2N}$ is concentrated near unity $U = 1$ for some value of
the exponent $N$, then it will be even more so for exponents
larger than $N$.  Thus, if the statement is true for some value
$N$ of $N_{\rm b}$, it will certainly be true for all values of
$N_{\rm b}$ greater than $N$. It therefore suffices to show that
the statement holds for the border line case $N_{\rm b} = N_{\rm
c}$.  We abbreviate the notation by writing $w_\alpha (U) \equiv
w_{N_{\rm c},\alpha}(U)$. \smallskip

Second, since the weight function $w_\alpha$ is invariant under
conjugation $U \mapsto g U g^{-1}$, the operation of replacing the
analytic function $f$ by its average $f^{\rm av}$ over conjugacy
classes,
\begin{displaymath}
  f^{\rm av}(U) = {\rm vol}({\rm U}(N_{\rm c}))^{-1} \int\limits_{{\rm
      U} (N_{\rm c})} f(g U g^{-1}) dg \;,
\end{displaymath}
does not change the value of the integral (\ref{thm1}).  We may
therefore assume $f$ to be invariant. \smallskip

Third, given invariance under conjugation, we may view $w_\alpha$
and $f$ as functions on the maximal torus $T = {\rm U}(1)^{N_{\rm
c}}$ parameterized by the eigenvalues ${\rm e}^{{\rm i}\theta_1},
\ldots, {\rm e}^{{\rm i}\theta_{N_{\rm c}}}$ of $U$, and we may
reduce the integral over ${\rm U}(N_{\rm c})$ to an integral over
$T$.  Let $J({\rm e}^{{\rm i}\theta_1}, \ldots, {\rm e}^{{\rm i}
\theta_{N_{\rm c}}}) = \prod_{k < l} | {\rm e}^{{\rm i} \theta_k}
- {\rm e}^{{\rm i}\theta_l} |^2$ be the Jacobian of the polar
coordinate map $\left( {\rm U}(N_{\rm c}) / T \right) \times T \to
{\rm U}(N_{\rm c})$.  Then we have
\begin{displaymath}
  \int\limits_{{\rm U}(N_{\rm c})} f(U) w_\alpha (U) dU = \frac{\langle
    f \rangle_\alpha} {\langle 1 \rangle_\alpha} \;,
\end{displaymath}
where the angular brackets mean
\begin{displaymath}
  \langle f \rangle_\alpha = \int\limits_{[0,2\pi]^{N_{\rm c}}} f({\rm
    e}^{{\rm i}\theta_1}, \ldots, {\rm e}^{{\rm i} \theta_{ N_{\rm
        c}}}) \frac{ J({\rm e}^{{\rm i} \theta_1} , \ldots, {\rm
      e}^{{\rm i}\theta_{N_{\rm c}}}) } { \prod_{j=1}^{N_{\rm c}} | 1
    - \alpha \, {\rm e}^{{\rm i}\theta_j} |^{2N_{\rm c}} } \,
  d\theta_1 \cdots d\theta_{N_{\rm c}} \;.
\end{displaymath}
To prove the fact (\ref{thm1}) we must show
\begin{equation}\label{limit1}
  \lim_{\alpha \to 1} \frac{\langle f \rangle_\alpha} {\langle 1
    \rangle_\alpha} = f(1, \ldots, 1)
\end{equation}
for all analytic functions $f$ on the maximal torus $T = {\rm U}
(1)^{N_{\rm c}}$ which extend to functions on ${\rm U}(N_{\rm c})$.
(Such functions on $T$ are invariant under the Weyl group of ${\rm
  U}(N_{\rm c})$, i.e.~they do not change under permutations of their
arguments.) \smallskip

We will actually establish the limit (\ref{limit1}) for the larger
class of {\it all} analytic functions $F : T \to {\mathbb C}$.  By
definition, such functions are absolutely convergent sums of the basic
functions ${\rm e}^{{\rm i} \sum n_k \theta_k} $ with integer
exponents $n_k$.  It therefore suffices to prove (\ref{limit1}) for
the complete set of these basic functions.  So let
\begin{displaymath}
  F = {\rm e}^{{\rm i}(n_1 \theta_1 + \ldots + n_{N_{\rm c}}
    \theta_{N_{\rm c}})}
\end{displaymath}
with any $(n_1, \ldots, n_{N_{\rm c}}) \in {\mathbb Z}^{N_{\rm
c}}$. Without loss we may assume that the ordering of variables
has been adjusted so that the first $p$ integers $n_1, \ldots,
n_p$ are positive or zero, while the last $N_{\rm c} - p$ integers
$n_{p+1}, \ldots, n_{N_{\rm c}}$ are negative. \smallskip

Now we evaluate $\langle F \rangle_\alpha$ in the limit $\alpha \to
1$.  The first step is to switch to the variables $z_k = {\rm e}^{{\rm
    i} \theta_k}$, which yields the expression
\begin{displaymath}
  \langle F \rangle_\alpha = ({\rm i}/\alpha)^{N_{\rm c}^2}
  \oint\limits_{{\rm U}(1)^{N_{\rm c}}} \frac{z_1^{n_1} \cdots
    z_{N_{\rm c}}^{n_{N_{\rm c}}}} {\prod_{j=1}^{N_{\rm c}} (z_j -
    \alpha)^{N_{\rm c}} (z_j - \alpha^{-1})^{N_{\rm c}}} \prod_{k < l}
  (z_k - z_l)^2 dz_1 \cdots dz_{N_{\rm c}} \;.
\end{displaymath}
By the signs assumed for the integers $n_1, \ldots, n_{N_{\rm c}}$,
the integrand is obviously regular at zero in the complex plane for
each of the variables $z_1, \ldots, z_p$.  Simple power counting shows
that the same is true at infinity for $z_{p+1}, \ldots, z_{N_{\rm
    c}}$.  (This is to say that the integrand for $k = p+1, \ldots,
N_{\rm c}$ decays as $z_k^{-2}$ or faster at infinity.) Therefore, our
strategy now is to contract the contour of integration to zero for the
first $p$ variables, and expand it to infinity for the last $N_{\rm c}
- p$ variables.  In doing so, we pick up contributions from the poles
of order $N_{\rm c}$ at $z_1 = \ldots = z_p = \alpha$ inside the unit
circle ${\rm U}(1)\subset {\mathbb C}$, and at $z_{p+1} = \ldots =
z_{N_{\rm c}} = \alpha^{-1}$ outside the unit circle.  By the residue
theorem, we then arrive at the exact formula
\begin{eqnarray*}
  \langle F \rangle_\alpha &=& ({\rm i/\alpha})^{N_{\rm c}^2}
  (2\pi{\rm i})^p (-2\pi{\rm i})^{N_{\rm c}-p} (N_{\rm c} -
  1)!^{-N_{\rm c}} \\ &\times& \left( \frac{\partial}{\partial z_1}
    \cdots \frac{\partial}{\partial z_{N_{\rm c}}} \right)^{N_{\rm
      c}-1} \frac{ z_1^{n_1} \cdots z_{N_{\rm c}}^{n_{N_{\rm c}}}
    \prod_{k < l} (z_k - z_l)^2} {\prod_{j=1}^p (z_j - \alpha
    )^{N_{\rm c}} \prod_{j = p+1}^{N_{\rm c}} (z_j - \alpha^{-1}
    )^{N_{\rm c}} } \Bigg|_{z_1 = \ldots = z_p = z_{p+1}^{-1} = \ldots
    = z_{N_{\rm c}}^{-1} = \alpha} \;.
\end{eqnarray*}
This expression is divergent at $\alpha = 1$.  The highest--order
singularity, $(\alpha - \alpha^{-1})^{-N_{\rm c}^2}$, occurs when all
derivatives act on $\prod_{k < l} (z_k - z_l)^2$ or on the
denominator.  The order of the singularity is reduced if one or
several of the derivatives act on the monomial $F = z_1^{n_1} \cdots
z_{N_{\rm c}}^{n_{N_{\rm c}}}$.  To compute the limit $\alpha \to 1$,
it is enough to retain the leading--order singularity.  The
leading--order singularity is picked up by evaluating $F$ at $z_1 =
\ldots = z_p = \alpha$ and $z_{p+1} = \ldots = z_{N_{\rm c}} =
\alpha^{-1}$ and taking it outside of the expression.  What is left
behind is just the value of the integral $\langle 1 \rangle_\alpha$
obtained by replacing $F$ by unity.  Hence
\begin{displaymath}
  \langle F \rangle_\alpha = \alpha^{|n_1| + \ldots + |n_{N_{\rm c}}|}
  \, \langle 1 \rangle_\alpha + \mbox{less singular terms} \;.
\end{displaymath}
So we conclude $\lim_{\alpha \to 1} \langle F \rangle_\alpha / \langle
1 \rangle_\alpha = F \big|_{z_1 = \ldots = z_{N_{\rm c}} = 1}$, and
the proof of (\ref{thm1}) is complete.

\section{Continuum Limit in Two Dimensions}\label{sec:2d}

The result (\ref{thm1}) ensures that on sending all coupling
parameters $\alpha_{{\rm b},p} \to 1$, the induced gauge model
(\ref{bAction}) for $N_{\rm b} \ge N_{\rm c}$ becomes critical, which
allows a continuum limit to be taken on any reasonable direct system
of lattices.  Based on universality, we expect this continuum limit to
be quantum Yang--Mills theory, at least generically.  \smallskip

A precise investigation of the universality conjecture can be made,
and will now be made using harmonic analysis on the gauge group, in
the simple case of $d = 1 + 1$ dimensions.  There, the universality
mechanism at work is the central limit principle in its basic form:
computing the two--dimensional theory essentially amounts to taking
convolutions of the plaquette distribution $w_{N_{\rm b},\alpha}(U)
dU$ with itself and, by a central limit theorem for compact Lie
groups, multiple convolution sends a large class of distributions to
the universal heat kernel family $w_t^{\rm HK}(U) dU$ written down in
(\ref{heatkernel}).  \smallskip

In short, the central limit principle we will exploit is this.
(Although we are going to pursue the case of ${\rm U}(N_{\rm c})$,
we here state the principle for a semisimple compact Lie group
$G$. The extension to ${\rm U}(N_{\rm c})$ will cause minor
complications.) Let $w_t(U) dU$ $(t > 0)$ be a one--parameter
family of smooth ${\rm Ad} G$--invariant distributions on $G$ such
that $\lim_{t \to 0} w_t(U) = \delta(U)$.  Using the exponential
mapping $X \mapsto U = \exp X$ we can pull back the family to a
family of distributions $d\mu_t(X)$ on the Lie algebra of $G$ (or,
rather, to a domain of injectivity of $\exp$ in ${\rm Lie} G$).
With respect to $d\mu_t(X)$ we compute the expectation of the
Killing form $(X,X) = - {\rm Tr} \, {\rm ad}(X) {\rm ad}(X)$. If
there exists a ``diffusive scaling'' limit, i.e.~the expectation
of $(X,X) / t$ stays finite when $t$ is sent to zero, then a
central limit principle applies: denoting the $N^{\rm th}$
convolution of $w_t$ with itself by $w_t^{\star N}$, we have
$\lim_{N \to \infty} w_{t/N}^{\star N} = w_t^{\rm HK}$.
\smallskip

We will see that the plaquette distribution of the boson induced
model (\ref{bAction}) for $N_{\rm b} \ge N_{\rm c} + 1$ satisfies
the diffusive scaling criterion, with $t \sim (1 - \alpha_{{\rm
b}, p})^2$.  This will allow us to take a continuum limit which
can be considered as a rigorous definition of 2d quantum
Yang--Mills theory by the reasoning of Witten \cite{witten}.
\smallskip

On the other hand, for $N_{\rm b} = N_{\rm c}$ the diffusive scaling
criterion turns out to be violated!  A continuum limit can still be
defined, owing to (\ref{thm1}). This, however, is not Yang--Mills
theory but an unusual theory which, in the ``first--order formalism''
with an auxiliary ${\rm Lie} \, {\rm U}(N_{\rm c})$--valued scalar
field $\phi$, is given by an action functional
\begin{equation}\label{actfunc1}
  S = - {\rm i} \int_\Sigma {\rm Tr} \, \phi F + \mu \int_\Sigma
  \parallel \phi \parallel_1 \, d^2x \;,
\end{equation}
where $\parallel \phi \parallel_1 = \sum_{j=1}^{N_{\rm c}} |
\phi_j |$, the $\phi_j$ being the eigenvalues of $\phi$.  This
exists as a renormalizable theory because the Cauchy distribution
on $\mathfrak{u}(N_{\rm c}) \equiv {\rm Lie} \, {\rm U}(N_{\rm
c})$ approaches under subdivision a distribution which is stable,
and yet different from the heat kernel. \smallskip

We shall begin to substantiate these assertions in Section
\ref{sec:charexp}.

\subsection{One--Plaquette Model}\label{sec:0d}

Before we undertake the study of the two--dimensional models where
Yang--Mills universality rules, we dispose of those where it does not.
As a simple test, we look at a cell complex consisting of a single
plaquette ${\bf p}$ and consider the expectation value of ${\rm Tr} \,
U \equiv {\rm Tr} \, U(\partial {\bf p})$:
\begin{displaymath}
  W(\alpha_{\rm f},\alpha_{\rm b}) = \frac{1} {Z(\alpha_{\rm f},
    \alpha_{\rm b})} \int\limits_{{\rm U}(N_{\rm c})} {\rm Tr} \, U \,
  \frac{| {\rm Det}(1 - \alpha_{\rm f} U)|^{2N_{\rm f}}} {| {\rm Det}(1 -
    \alpha_{\rm b} U)|^{2N_{\rm b}}} \, dU \;.
\end{displaymath}
In ${\rm U}(N_{\rm c})$ Yang--Mills theory the expectation of ${\rm
  Tr} \, U(C)$ (the holonomy along any loop $C$) goes to ${\rm Tr} \,
1 = N_{\rm c}$ when the coupling is sent to zero. The same happens
with $W(\alpha_{\rm f},\alpha_{\rm b})$ in the limit $\alpha_{\rm
b} \to 1$, $N_{\rm b} \ge N_{\rm c}$, for in that case the statement
(\ref{thm1}) applies, and gives with $f(U) = {\rm Tr}\, U \, |
{\rm Det}(1 - \alpha_{\rm f} U) |^{2 N_{\rm f}}$:
\begin{displaymath}
  \lim_{\alpha_{\rm b} \to 1} W(\alpha_{\rm f},\alpha_{\rm b}) = {\rm
    Tr} \, U \big|_{U = 1} = N_{\rm c} \qquad (N_{\rm b} \ge N_{\rm
    c}, ~ \alpha_{\rm f} \not= 1) \;.
\end{displaymath}
While this is a statement made for the fundamental representation,
(\ref{thm1}) asserts that a similar result holds true for the trace
${\rm Tr} \, D^\lambda(U)$ in {\it any} representation $D^\lambda$.
\smallskip

\begin{figure}
  \begin{center}
  \epsfig{file=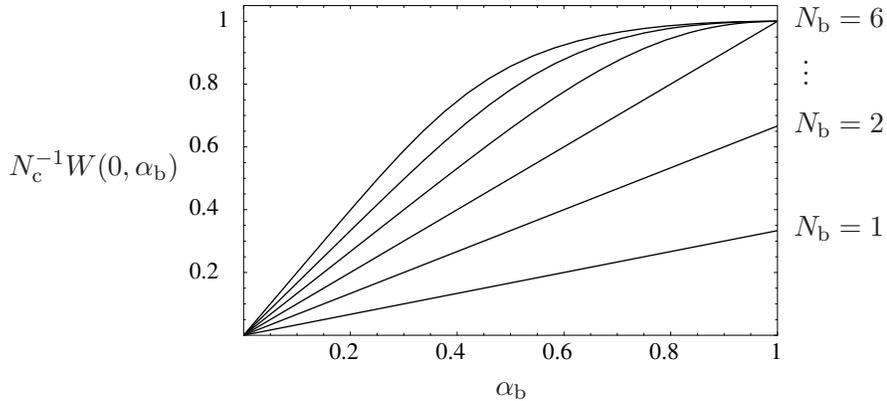,height=5cm}
  \begin{picture}(0,0)(300,0)
    \put(297,52){$N_{\rm b} = 1$} \put(297,91){$N_{\rm b} = 2$}
    \put(300,109){$\vdots$} \put(297,131){$N_{\rm b} = 6$}
    \put(0,75){$N_{\rm c}^{-1} W(0,\alpha_{\rm b})$}
    \put(184,-8){$\alpha_{\rm b}$}
  \end{picture}
  \end{center}
  \caption{Normalized one--plaquette expectation $\left\langle
      {\rm Tr}\, U \right\rangle$ for $N_{\rm c} = 3$ colors and
    $N_{\rm b} = 1, \ldots, 6$ boson flavors as a function of the
    parameter $\alpha_{\rm b}$ in the range $0 \le \alpha_{\rm b} <
    1$, and $\alpha_{\rm f} = 0$.} \label{fig:bosons}
\end{figure}

What happens in the other cases?  We separately look at the boson
induced models ($\alpha_{\rm f} = 0$) with $N_{\rm b} < N_{\rm
c}$, and at the fermion induced models ($\alpha_{\rm b} = 0$),
starting with the former.  Using complex contour integration and
residue calculus, we show in Appendix \ref{sec:appbos} that the
following holds true:
\begin{displaymath}
  \lim_{\alpha_{\rm b} \to 1} W(0,\alpha_{\rm b}) = N_{\rm b} \qquad
  (N_{\rm b} < N_{\rm c}) \;.
\end{displaymath}
Thus in that case $W(0,1)$ falls short of reaching the maximal value
$N_{\rm c}$ allowed by the bound $|{\rm Tr}\, U| \le N_{\rm c}$.  The
general dependence can be computed numerically (see Appendix
\ref{sec:appbos}), and is shown in Figure \ref{fig:bosons} for the case
$N_{\rm c} = 3$.  An interesting observation, proved in Appendix
\ref{sec:appbos}, is that for $N_{\rm b} \le N_{\rm c}$ the function
$W(0,\alpha_{\rm b})$ is exactly linear: $W(0,\alpha_{\rm b}) = N_{\rm
  b} \alpha_{\rm b}$. \smallskip

\begin{figure}
  \begin{center}
  \epsfig{file=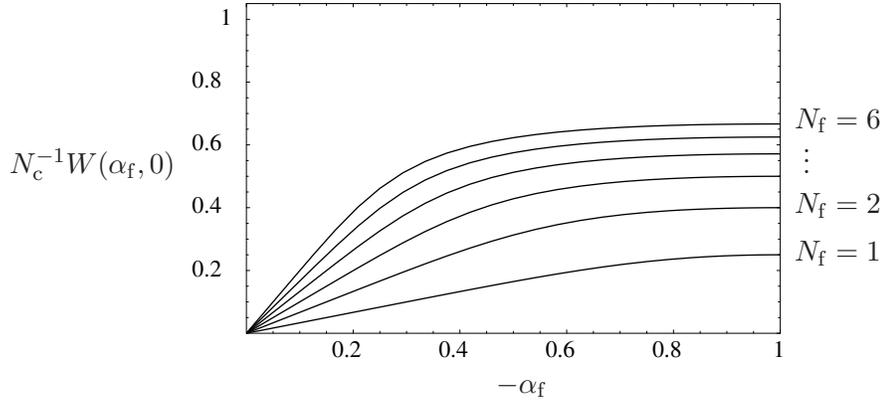,height=5cm}
    \begin{picture}(0,0)(300,0)
      \put(297,42){$N_{\rm f} = 1$} \put(297,60){$N_{\rm f} = 2$}
      \put(300,75){$\vdots$} \put(297,92){$N_{\rm f} = 6$}
      \put(0,75){$N_{\rm c}^{-1} W(\alpha_{\rm f},0)$}
      \put(184,-8){$- \alpha_{\rm f}$}
  \end{picture}
  \end{center}
  \caption{Same as Figure \ref{fig:bosons}, but now for the fermion
    induced model.}\label{fig:fermions}
\end{figure}

Turning to the fermion induced case, from the invariance of $dU$ under
$U \to -U$ and $U \to U^{-1}$, we deduce $W(\alpha_{\rm f}, 0) = -
W(-\alpha_{\rm f},0) = W(1/ \alpha_{\rm f},0)$, so it suffices to
restrict attention to the interval $-1 \le \alpha_{\rm f} \le 0$.
Figure \ref{fig:fermions} shows numerical results (Appendix
\ref{sec:appferms}) in that range.  We observe that $W(\alpha_{\rm
  f},0)$ is a monotonically increasing function of $-\alpha_{\rm f}$
for $|\alpha_{\rm f}| \le 1$.  The global maximum, attained at
$\alpha_{\rm f} = - 1$, can be computed analytically by a fermionic
variant of the method of Howe pairs described in Section
\ref{sec:howe}:
\begin{displaymath}
  W(-1,0) = \frac{N_{\rm f} N_{\rm c}}{N_{\rm f} + N_{\rm c}} \;.
\end{displaymath}
Thus, the maximum again falls short of reaching the limit posed by the
bound $| {\rm Tr} \, U | \le N_{\rm c}$. \smallskip

These results show that in the fermion induced model with any finite
number $N_{\rm f}$ of species, and in the boson induced model with
$N_{\rm b} < N_{\rm c}$, the single--plaquette action cannot ever
enforce complete suppression of the fluctuations of $U$ away from the
unit element.  Consequently, we do not expect a continuum limit of
Yang--Mills type in these models.  \smallskip

\subsection{Character expansion}\label{sec:charexp}

We now embark on a detailed study of the boson induced gauge model
(\ref{bAction}) with at least as many flavors as colors ($N_{\rm c}
\ge N_{\rm b}$).  As a preparatory step, we do some harmonic analysis
for the one--plaquette distribution
\begin{equation}\label{plaqdist}
  d\mu_{N_{\rm b},\alpha}(U) = \big| {\rm Det}\left( 1 - \alpha U
  \right) \big|^{ -2N_{\rm b}} dU \;.
\end{equation}
Our motivation is that we wish to compute multiple convolution
integrals of this distribution with itself, and transforming to the
appropriate Fourier (or harmonic) space turns convolutions into simple
multiplications.  \smallskip

Recall that by the Peter--Weyl theorem any $L^2$--function on a
compact Lie group can be expanded in the matrix entries of its
complete set of irreducible representations $D^\lambda$.  Since the
distribution function in (\ref{plaqdist}) is a class function of ${\rm
  U}(N_{\rm c})$, the expansion proceeds by ${\rm U}(N_{\rm c})$
characters $\chi_\lambda(U) \equiv {\rm Tr} \, D^\lambda(U)$:
\begin{equation}\label{charexp}
  \big| {\rm Det}\left( 1 - \alpha U \right) \big|^{ -2N_{\rm b}} =
  \sum_{{\rm all}~\lambda} c_{\lambda} (\alpha) \, \chi_\lambda(U) \;,
\end{equation}
and, by the orthonormality of characters, the expansion coefficients
are
\begin{equation}\label{coeffs}
  c_{\lambda}(\alpha) = \int\limits_{{\rm U}(N_{\rm c})} \big| {\rm
    Det}\left( 1 - \alpha U \right) \big|^{ -2N_{\rm b}} \chi_\lambda
  (U^{-1}) dU \;.
\end{equation}
Here, and throughout this section, we normalize Haar measures by
$\int_{{\rm U}(N_{\rm c})} dU = 1$.  From statement (\ref{thm1})
we already know the limit
\begin{displaymath}
  \lim_{\alpha \to 1} \frac{c_{\lambda}(\alpha)}{c_{0} (\alpha)} =
  \chi_\lambda(1) = d_\lambda \;,
\end{displaymath}
which asserts convergence of the ratio of coefficients to the
dimension of the representation $\lambda$.  The goal of the current
subsection is to gain a more precise understanding of the expansion
coefficients $c_{\lambda}(\alpha)$ close to $\alpha = 1$. \smallskip

To that end we will use the Cayley map
\begin{displaymath}
  X \mapsto \gamma(X) = \frac{1 + X}{1 - X}
\end{displaymath}
to express the integral (\ref{coeffs}) over ${\rm U}(N_{\rm c})$ as an
integral over the anti--Hermitian $N_{\rm c} \times N_{\rm c}$
matrices, $X \in \mathfrak{u}(N_{\rm c})$.  The Jacobian of the Cayley
map is ${\rm Det}^{- N_{\rm c}}(1 - X^2)$, which is to say that there
exists a flat positive density $dX$ on $\mathfrak{u} (N_{\rm c})$ such
that
\begin{displaymath}
  \gamma^*(dU) = {\rm Det}^{- N_{\rm c}}(1 - X^2) \, dX \;.
\end{displaymath}
Note $-X^2 \ge 0$ for $X \in \mathfrak{u}(N_{\rm c})$.  We then
immediately have the statements
\begin{eqnarray}
  &&\int\limits_{\mathfrak{u}(N_{\rm c})} {\rm Det}^{-N_{\rm c}} (1 -
  X^2) \, dX = \int\limits_{{\rm U}(N_{\rm c})} dU = 1 \;, \nonumber
  \\ &&\int\limits_{\mathfrak{u}(N_{\rm c})} {\rm Det}^{ -N_{\rm b}}
  (1 - X^2) \, dX < 1 \quad \mbox{for} \quad N_{\rm b} > N_{\rm c} \;,
  \quad \mbox{and} \label{ineq1} \\ && \int\limits_{\mathfrak{u}
    (N_{\rm c})} {\rm Tr}(-X^2) \, {\rm Det}^{-N_{\rm b}} (1 - X^2) \,
  dX \le 1 \quad \mbox{for} \quad N_{\rm b} \ge N_{\rm c} + 1 \;.
  \label{ineq2}
\end{eqnarray}
The last of these follows from the elementary inequality ${\rm Tr}
(-X^2) \le {\rm Det}(1 - X^2)$. \smallskip

Making two variable transformations in sequence,
\begin{displaymath}
  U = \frac{1+x}{1-x} \;, \qquad x = \frac{1-\alpha}{1+\alpha} X
  \, \in \mathfrak{u}(N_{\rm c}) \;,
\end{displaymath}
we bring the integral formula for the expansion coefficients into the
form
\begin{displaymath}
  c_{\lambda}(\alpha) = (1 - \alpha)^{-2N_{\rm c}N_{\rm b}} \left(
    \frac{1 - \alpha}{1 + \alpha} \right)^{N_{\rm c}^2}
  \int\limits_{\mathfrak{u}(N_{\rm c})} \frac{\chi_\lambda \big( (1 -
    x)/(1 + x) \big)} {{\rm Det}^{N_{\rm c} - N_{\rm b}} \left( 1 -
      x^2 \right)} \Bigg|_{x = \frac{1 - \alpha}{1+\alpha} X} {\rm
    Det}^{-N_{\rm b}}(1 - X^2) \, dX \;.
\end{displaymath}
In view of the finiteness statement (\ref{ineq1}) it is clear that
the integral for $c_{\lambda} (\alpha)$ will localize at unity $U
= 1$ in the limit $\alpha \to 1$, for $N_{\rm b} \ge N_{\rm c}$.
Hence, to get an accurate approximation to these $c_{\lambda}
(\alpha)$ near $\alpha = 1$, the only further input we need is an
understanding of the characters $\chi_\lambda (U)$ close to unity.
\smallskip

Let $\lambda$ denote a set of $N_{\rm c}$ integers, $\lambda = (
\lambda_1 , \lambda_2 , \ldots, \lambda_{N_{\rm c}} )$, let ${\rm
  e}^{{\rm i}\theta} = ({\rm e}^{{\rm i}\theta_1} , \ldots, {\rm
  e}^{{\rm i}\theta_{N_{\rm c}}}) \in {\rm U}(1)^{N_{\rm c}}$, and
define the elementary antisymmetric torus function $\xi_\lambda$ by
\begin{displaymath}
  \xi_\lambda({\rm e}^{{\rm i}\theta}) = \sum_{\pi \in S_{N_{\rm c}}}
  {\rm sgn}(\pi) \, {\rm e}^{{\rm i} \lambda_1 \theta_{\pi(1)}} \,
  {\rm e}^{{\rm i} \lambda_2 \theta_{\pi(2)}} \, \cdots \, {\rm
    e}^{{\rm i} \lambda_{N_{\rm c}} \theta_{\pi(N_{\rm c})}} \;,
\end{displaymath}
where the sum is over all permutations of $\{ 1, 2, \ldots, N_{\rm
c} \}$.  By a classic result of Weyl \cite{weyl}, the irreducible
representations of ${\rm U}(N_{\rm c})$ are in one--to--one
correspondence with ordered sets $\lambda$, $\lambda_1 \ge
\lambda_2 \ge \ldots \ge \lambda_{N_{\rm c}}$, and the character
associated with $\lambda$ is the class function $\chi_\lambda(U)$
determined by
\begin{equation}\label{schur}
  \chi_\lambda({\rm e}^{{\rm i}\theta}) = \xi_{\lambda + \rho} ({\rm
    e}^{{\rm i}\theta}) / \xi_{\rho} ({\rm e}^{{\rm i}\theta})
\end{equation}
where $\rho = (N_{\rm c} - 1, N_{\rm c} - 2, \ldots, 1, 0)$.  The
character at unity computes the dimension of the representation, which
by the Weyl dimension formula is
\begin{equation}\label{weyldim}
  \chi_\lambda({\rm e}^0) = d_\lambda = \triangle(\lambda + \rho) /
  \triangle(\rho) \;,
\end{equation}
$\triangle(\lambda) = \prod_{k < l}(\lambda_k - \lambda_l)$ being the
Vandermonde determinant. \smallskip

Because characters are joint eigenfunctions of the full ring of
invariant differential operators, the characters $\chi_\lambda (U)$ of
${\rm U}(N_{\rm c}) = {\rm U}(1) \times {\rm SU}(N_{\rm c})$ with
central factor ${\rm U}(1)$ separate.  If we put $U = {\rm e}^X$, a
${\rm U}(1)$ factor ${\rm e}^{(q/N_{\rm c}) {\rm Tr} X}$ with charge
$q(\lambda) = \sum_j \lambda_j$ splits off.  Moreover, since the ${\rm
  SU}(N_{\rm c})$ part only depends on the traceless part of $X \in
\mathfrak{u}(N_{\rm c})$, the characters expand around unity as
\begin{eqnarray*}
  \frac{\chi_\lambda({\rm e}^X)}{\chi_\lambda({\rm e}^0)} &=& {\rm
    e}^{(q/N_{\rm c}) {\rm Tr} X} \times f^{{\rm SU}(N_{\rm c})}({\rm
    e}^X \cdot {\rm e}^{-{\rm Tr} X / N_{\rm c}} ) \\ &=& 1 +
  \frac{q}{N_{\rm c}} {\rm Tr} X + \frac{1}{2} \left( \frac{q^2}{
      N_{\rm c}^2 } - \frac{A}{N_{\rm c}} \right) ({\rm Tr} X)^2 +
  \frac{A}{2} {\rm Tr} ( X^2 ) + \ldots \;,
\end{eqnarray*}
with some coefficient $A = A(\lambda)$.  In parallel, one can directly
expand Weyl's formula (\ref{schur}) for $\chi_\lambda$ to second order
in the angles $\theta_j$.  By doing so and comparing coefficients (or
by more sophisticated techniques not discussed here), one finds
\begin{displaymath}
  A(\lambda) = (N_{\rm c}^2 - 1)^{-1} \left( {\rm Cas}_2(\lambda) -
    q(\lambda)^2 / N_{\rm c} \right) \;,
\end{displaymath}
where ${\rm Cas}_2(\lambda)$ is a quadratic Casimir element of
${\rm
  U}(N_{\rm c})$,
\begin{equation}\label{casimir}
  {\rm Cas}_2(\lambda) = \sum_{j = 1}^{N_{\rm c}} \lambda_j (
  \lambda_j + N_{\rm c} + 1 - 2j ) \;,
\end{equation}
which is associated with the invariant quadratic form $- {\rm
  Tr}(X^2)$ in the standard way \cite{knapp}. \smallskip

Having gathered enough information about the characters $\chi_\lambda$
near unity, we now return to the task of computing the behavior of the
expansion coefficients $c_{\lambda}(\alpha)$ close to $\alpha = 1$.
If we insert the small--$X$ expansion of $\chi_\lambda({\rm e}^X)$
into the integral formula for $c_{\lambda}(\alpha)$ and take the ratio
$c_\lambda(\alpha) / c_0(\alpha)$, we encounter the second moments of
the distribution ${\rm Det}^{-N_{\rm b}} (1 - X^2)\, dX$:
\begin{eqnarray}
  {\mathbb E}_{N_{\rm b}} \, {\rm Tr}(X^2) &\equiv& Z^{-1}
  \int\limits_{\mathfrak{u} (N_{\rm c})} {\rm Tr}(X^2) \, {\rm
    Det}^{-N_{\rm b}} (1 - X^2) \, dX \;, \\ {\mathbb E}_{N_{\rm b}}
  \, ({\rm Tr}X)^2 &\equiv& Z^{-1} \int\limits_{\mathfrak{u} (N_{\rm
      c})} ({\rm Tr}X)^2 \, {\rm Det}^{-N_{\rm b}} (1 - X^2) \; dX \;,
\end{eqnarray}
with $Z = \int_{\mathfrak{u}(N_{\rm c})} {\rm Det}^{-N_{\rm b}} (1 -
X^2) \, dX$.  These are finite for $N_{\rm b} \ge N_{\rm c} + 1$, by
the inequality (\ref{ineq2}) and $0 \le - ({\rm Tr}X)^2 \le - N_{\rm
  c} {\rm Tr}(X^2)$.  The first moment ${\mathbb E}_{N_{\rm b}} \,
{\rm Tr}X$ vanishes by parity. It is therefore easy to prove the
following statement:

\bigskip\textsf{Fact}. {\it The coefficients $c_{\lambda}
  (\alpha)$ for $N_{\rm b} \ge N_{\rm c} + 1$ have a Taylor expansion}
\begin{equation}\label{thm2}
  \frac{c_\lambda (\alpha)}{c_{0}(\alpha)} = d_\lambda \left( 1 -
    \frac{1}{2} (1 - \alpha)^2 \big( B_1 \, q(\lambda)^2 + B_2 \, {\rm
      Cas}_2(\lambda) \big) + {\cal R} (\alpha) \right) \;,
\end{equation}
{\it with a remainder term ${\cal R}(\alpha)$ that vanishes faster
  than $(1-\alpha)^2$ in the limit $\alpha \to 1$.  The coefficients
  $B_1$ and $B_2$ are determined by the linear system of equations}
\begin{equation}\label{B1B2}
\begin{pmatrix}
  - {\mathbb E}_{N_{\rm b}} ({\rm Tr} X)^2 / N_{\rm c} \\ - {\mathbb
    E}_{N_{\rm b}} \, {\rm Tr} (X^2) / N_{\rm c} \end{pmatrix} =
\begin{pmatrix} N_{\rm c} &1 \\ 1 &N_{\rm c} \end{pmatrix}
\begin{pmatrix} B_1 \\ B_2 \end{pmatrix} \;.
\end{equation}
{\it The leading singularity in the normalization $c_0(\alpha)$ is}
\begin{equation}\label{thm2c}
  c_0(\alpha) \sim (1-\alpha)^{-2N_{\rm b}N_{\rm c} + N_{\rm c}^2} \;.
\end{equation}

\subsection{Continuum limit}\label{contlimit}

We now place the boson induced gauge model (\ref{bAction}) with
$N_{\rm b} \ge N_{\rm c} + 1$ on a two--dimensional cell complex
$\Lambda$ approximating an orientable compact Riemann surface
$\Sigma$.  Such a complex consists of plaquettes, links and sites,
with every link joining at most two plaquettes, and the plaquettes
receiving an orientation from $\Sigma$.  \smallskip

To each plaquette $p$ of $\Lambda$, we assign an area $A_p$
determined by some choice of Riemannian structure of $\Sigma$.
Measuring area in units of a fundamental area $a^2$, we put
$\alpha_p = 1 - \sqrt{A_p} / a$.  This specifies the set $\{
\alpha \}$ of coupling parameters of our boson induced gauge model
on $\Lambda$. \smallskip

We now focus on the partition function
\begin{displaymath}
  Z_\Lambda \left( \{ \alpha \} \right) = \int [dU] \prod_{\bf p}
  \left| {\rm Det} \big( 1 - \alpha_p \, U(\partial{\bf p}) \big)
  \right|^{-2 N_{\rm b}} \;.
\end{displaymath}
The goal is to demonstrate that, when the cell complex $\Lambda$ is
refined so as to approximate $\Sigma$ ever more closely, $Z_\Lambda
(\{ \alpha \})$ converges to the known partition function of ${\rm
  U}(N_{\rm c})$ Yang--Mills theory on $\Sigma$, with a particular
choice for the ${\rm U}(1)$ coupling.  \smallskip

Since any $\Sigma$ can be made by gluing of surfaces with disk
topology, the first step we take is to compute the (boundary--value)
partition function, $\Gamma_\Lambda({\cal U}, \{ \alpha \})$, for a
two--dimensional cell complex $\Lambda$ approximating a disk, where
\begin{displaymath}
  {\cal U} = U(\partial \Lambda) = \prod_{{\bf l} \in \partial
    \Lambda} U({\bf l})
\end{displaymath}
is the product of $U$'s along the boundary $\partial \Lambda$, and
these boundary $U$'s are held fixed to prescribed values.  To
compute $\Gamma_\Lambda({\cal U}, \{ \alpha \})$, we need to
integrate over all the matrices $U$ associated with the links in
the interior of $\Lambda$.  These link integrations can be carried
out in any order, and because $\Lambda$ has disk topology
(i.e.~every interior link joins exactly two plaquettes), it
suffices to show how to do one of them.  \smallskip

Hence, let ${\bf l}$ be any interior link of $\Lambda$, and let ${\bf
  p}$ and ${\bf p}^\prime$ be the two oriented plaquettes joined by
${\bf l}$.  We are going to do the integral over the link matrix
$U({\bf l})$.  For that purpose, let the holonomy along the boundary
chain of ${\bf p}$ be
\begin{displaymath}
  U(\partial{\bf p}) = U_{\bf p}(n_1,n_{L_p}) \cdots U_{\bf p} (n_3,
  n_2) \, U_{\bf p}(n_2,n_1) \;,
\end{displaymath}
and similar for ${\bf p}^\prime$, where our notational conventions are
as specified in Section \ref{sec:induced}.  Without loss we may assume
that ${\bf l}$ is the link joining the sites $n_1 \equiv n_2^\prime$
and $n_2 \equiv n_1^\prime$:
\begin{displaymath}
  U \equiv U({\bf l}) = U_{\bf p}(n_2,n_1) = \left( U_{{\bf p}^\prime}
    (n_2^\prime , n_1^\prime) \right)^{-1} \;.
\end{displaymath}
Abbreviating the notation by putting
\begin{displaymath}
  U(\partial{\bf p}) = W U \;, \qquad U(\partial{\bf p}^\prime) =
  V U^{-1} \;,
\end{displaymath}
we are then faced with the convolution integral
\begin{displaymath}
  I(V,W) = \int\limits_{{\rm U}(N_{\rm c})} \big| {\rm Det} (1 -
  \alpha_{p^\prime} V U^{-1}) \big|^{-2N_{\rm b}} \big| {\rm Det} (1 -
  \alpha_{p} \, U W) \big|^{-2N_{\rm b}} \, dU \;.
\end{displaymath}
To carry it out, we use the character expansion $| {\rm Det}(1 -
\alpha \, U) |^{-2N_{\rm b}} = \sum_\lambda c_\lambda(\alpha)
\chi_\lambda(U)$ and the fact that characters reproduce under
convolution,
\begin{equation}\label{convchar}
  \int\limits_{{\rm U}(N_{\rm c})} \, \chi_\lambda(V U^{-1}) \,
  \chi_{\lambda^\prime} (UW) \, dU = \delta_{\lambda, \lambda^\prime}
  \, \frac{\chi_{\lambda}(VW)}{d_\lambda} \;,
\end{equation}
to obtain
\begin{displaymath}
  I(V,W) = \sum_\lambda \frac{c_\lambda (\alpha_p) \, c_\lambda
    (\alpha_{p^\prime})} {d_\lambda} \, \chi_\lambda(VW) \;.
\end{displaymath}
We now iterate the procedure, and successively do all inner link
integrations using the convolution law (\ref{convchar}) for the
characters. The resulting expression for the boundary--value partition
function is
\begin{displaymath}
  \Gamma_\Lambda \big( {\cal U} ,\{ \alpha \} \big) = \sum_\lambda
  d_\lambda \left( \prod_p \frac{c_\lambda(\alpha_p)} {d_\lambda}
  \right) \chi_\lambda( {\cal U} ) \;,
\end{displaymath}
where ${\cal U} \equiv U(\partial\Lambda) = U_{\Lambda}
(n_{L_\Lambda}, n_{L_\Lambda - 1}) \cdots U_{\Lambda}(n_2,n_1)$ is
the holonomy along $\partial\Lambda$, as before. \smallskip

Here is the point where we take the continuum limit.  If we refine the
lattice discretization, and keep on refining it so that $\Lambda$
becomes a closer and ever closer approximation to a disk $D$ (or some
other surface $D$ diffeomorphic to a disk), the elementary plaquette
areas $A_p$ go to zero, the parameters $\alpha_p = 1 - \sqrt{A_p} / a$
approach unity, and we may eventually use the asymptotic law
(\ref{thm2}), giving
\begin{displaymath}
  \prod_p \frac{c_\lambda(\alpha_p)}{d_\lambda} = \prod_p \left( 1 -
    \frac{A_p}{2a^2} \big( B_1 \, q(\lambda)^2 + B_2 \, {\rm Cas}_2
    (\lambda) \big) \right) c_0(\alpha_p) + \ldots \;,
\end{displaymath}
with corrections that vanish in the limit.  Since $\lim_{N \to
\infty} (1 - x/N)^N = {\rm e}^{-x}$, the plaquette areas $A_p$
exponentiate and piece together to yield the total area $\sum_p
A_p$ of the surface $D$.  To eliminate one irrelevant constant
from our expressions, we adopt the convention of measuring area in
suitable units: $\mu \equiv (B_2/a^2) \times \sum_p A_p$. Taking
the continuum limit then leads to
\begin{equation}\label{localdata}
  \Gamma_D({\cal U},\mu) = \sum_\lambda d_\lambda \, \exp \left( -
    \frac{\mu} {2} \big( {\rm Cas}_2 (\lambda) + (B_1/B_2)
    q(\lambda)^2 \big) \right) \, \chi_\lambda({\cal U}) \;.
\end{equation}
We have dropped a (diverging) multiplicative constant $\prod_p c_0
(\alpha_p)$ that arose from our using an unnormalized statistical
weight function. (Physically speaking we made use of our freedom to
set the vacuum energy of flat space to zero.) \smallskip

An expression for $\Gamma_D({\cal U},\mu)$ of the given form
constituted the starting point of Witten's combinatorial treatment
\cite{witten}.  From it, he computed (among many other things) the
Yang--Mills partition function $Z_g(\mu)$ for any orientable
compact Riemann surface of genus $g$ and dimensionless area $\mu$.
Instead of repeating that computation here, we just quote the
answer it entails in the present case:
\begin{equation}\label{Zg_rho}
  Z_g(\mu) = \sum_\lambda d_\lambda^{-2g + 2} \, {\rm e}^{ - \mu
    \big( {\rm Cas}_2(\lambda) + (B_1/B_2) q(\lambda)^2 \big) /2 } \;,
\end{equation}
and verify it for the two simplest examples, the sphere and the torus.
\smallskip

A sphere $(g = 0)$ with area $\mu$ can be made by gluing together
two disks $D_+$ and $D_-$, say with areas $\mu_+$ and $\mu_- = \mu
- \mu_+$.  The boundary--value partition functions $\Gamma_{D_+}$
and $\Gamma_{D_-}$ depend only on the total holonomies along
$\partial D_+$ and $\partial D_-$ respectively. If the two disks
are to fit together to give an oriented sphere, we must have
$\partial D_+ = -
\partial D_-$, which means that the holonomies are inverse to each
other: $U(\partial D_+) = U(\partial D_-)^{-1}$.  Thus, introducing $U
\equiv U(\partial D_+)$ as the variable of final integration, we
obtain
\begin{displaymath}
  Z_{\rm sphere}(\mu) = \int\limits_{{\rm U}(N_{\rm c})} \Gamma_{D_+}
  (U,\mu_+) \, \Gamma_{D_-}(U^{-1},\mu_-) \, dU \;.
\end{displaymath}
Inserting (\ref{localdata}) and doing the resulting integral over
a product of two characters with the help of (\ref{convchar}), we
indeed get the answer (\ref{Zg_rho}) for genus $g = 0$.
\smallskip

To manufacture a torus $(g = 1)$ with area $\mu$, we take a rectangle
$Q$ with the same area, and identify opposite edges.  Writing the
holonomy along $\partial Q$ as a product over its four edges, $U(
\partial Q) = U V U^{-1} V^{-1}$, we have
\begin{displaymath}
  Z_{\rm torus}(\mu) = \int\limits_{{\rm U}(N_{\rm c})}
  \int\limits_{{\rm U}(N_{\rm c})} \Gamma_Q(UVU^{-1}V^{-1}, \mu) \,
  dU \, dV \;.
\end{displaymath}
To do the first of these two integrals, we use
\begin{displaymath}
  \int_G \chi_\lambda(UAU^{-1}B) \, dU = \chi_\lambda(A)
  \chi_\lambda(B) / d_\lambda \;,
\end{displaymath}
which is a consequence of the orthogonality relations obeyed by
the matrix entries of an irreducible representation $D^\lambda$.
Doing the second integral with (\ref{convchar}), we again
reproduce the formula (\ref{Zg_rho}), now with genus $g = 1$.
\smallskip

Let us summarize.  From the $\delta$--function property
(\ref{thm1}) we always expect the boson induced gauge model
(\ref{bAction}) with $N_{\rm b} \ge N_{\rm c}$ to admit a
continuum limit.  In two dimensions --- and on a direct system of
two--dimensional cell complexes converging to a compact Riemann
surface ---, making the natural assignment $(1 - \alpha_p)^2 \sim
A_p$ and taking the continuum limit $A_p \to 0$ sends the disk
boundary--value partition function $\Gamma_D$ of the boson induced
gauge model to the expression (\ref{localdata}), if $N_{\rm b} \ge
N_{\rm c} + 1$.  By the reasoning of \cite{witten}, this
expression is the proper local data to use for ${\rm U} (N_{\rm
c})$ Yang--Mills theory, and gives the partition function
(\ref{Zg_rho}).  \smallskip

More precisely speaking, the situation is this.  The gauge group ${\rm
  U} (N_{\rm c})$ is not semisimple; the number of its quadratic
Casimir invariants is two (as opposed to one for a semisimple Lie
group), and therefore ${\rm U}(N_{\rm c})$ Yang--Mills theory is not
unique but comes as a one--parameter family.  The procedure of
canonical quantization associates the Casimir invariants ${\rm Cas}_2
(\lambda)$ and $q(\lambda)^2$ with the action densities $\sum |F_{\mu
  \nu}^{ij}|^2 d^2x = - {\rm Tr}( F \wedge \star F)$ and $-\sum F_{\mu
  \nu}^{ii} F_{\mu\nu}^{jj} d^2x = - ({\rm Tr} F) \wedge \star ({\rm
  Tr} F)$, respectively.  Thus the combination
\begin{displaymath}
  {\rm Cas}_2(\lambda) + (B_1/B_2) \, q(\lambda)^2
\end{displaymath}
in (\ref{Zg_rho}) is the Hamiltonian arising from the action
functional
\begin{displaymath}
  - S = \frac{1}{2} \int_\Sigma {\rm Tr} (F \wedge \star F) + \frac{
    B_1} {2B_2} \int_\Sigma ({\rm Tr} F) \wedge \star ({\rm Tr} F)
\end{displaymath}
by canonical quantization.  The Lagrangian depends on the parameters
$B_1$ and $B_2$, which in turn are given by the second moments of the
distribution ${\rm Det}^{-N_{\rm b}}(1 - X^2)\, dX$; see (\ref{B1B2}).
\smallskip

Note that this distribution becomes Gaussian for $N_{\rm b} \to
\infty$.  In that limit, $- {\mathbb E}_{N_{\rm b}} \, ({\rm Tr} X)^2
\sim N_{\rm c} / 2 N_{\rm b}$ and $-{\mathbb E}_{N_{\rm b}} \, {\rm
  Tr} (X^2) \sim N_{\rm c}^2 / 2 N_{\rm b}$, and we get $B_1 / B_2 =
0$.  Thus, the coupling $B_1 / B_2$ gives some measure of how much the
distribution ${\rm Det}^{- N_{\rm b}}(1 - X^2)\, dX$ differs from a
Gaussian distribution.  For $N_{\rm c} = 2$ an easy computation gives
\begin{displaymath}
  - {\mathbb E}_{N_{\rm b}} \, {\rm Tr}(X^2) = \frac{3}{2N_{\rm b} -
    5} + \frac{1}{2N_{\rm b} - 3} \;, \quad -{\mathbb E}_{N_{\rm b}}
  \, ({\rm Tr} X)^2 = \frac{3}{2N_{\rm b} - 5} - \frac{1}{2N_{\rm b} -
    3} \;,
\end{displaymath}
which yields $B_1 / B_2 = \frac{1}{2} (N_{\rm b} - 2)^{-1}$.  Thus in
this case the ratio $B_1 / B_2$ for $N_{\rm b} \ge N_{\rm c} + 1 = 3$
is always positive.

\subsection{The Cauchy case $N_{\rm b} = N_{\rm c}$}
\label{sec:cauchy}

The continuum limit taken in the previous subsection does not go
through for $N_{\rm b} = N_{\rm c}$.  The main obstacle is that the
second moments of the Cauchy distribution ${\rm Det}^{-N_{\rm c}}(1 -
X^2) \, dX$ do not exist.  The nonexistence can be verified by power
counting, and is compatible with the failure of the inequality
(\ref{ineq2}) in that case.  It is also signalled by the expectation
values of the positive quantities $-{\rm Tr}(X^2)$ and $-({\rm
  Tr}X)^2$ we have just given for $N_{\rm b} > N_{\rm c} = 2$, which
are seen to become negative at $N_{\rm b} = N_{\rm c} = 2$.  Thus,
although we still have a good integral formula,
\begin{displaymath}
  c_{\lambda}(\alpha) = (1 - \alpha^2)^{-N_{\rm c}^2}
  \int\limits_{\mathfrak{u}(N_{\rm c})} \chi_\lambda \left( \frac{1 -
      \frac{1 - \alpha}{1+\alpha} X} {1 + \frac{1 - \alpha}{1 +
        \alpha} X} \right) \, {\rm Det}^{- N_{\rm c}} \left( 1 - X^2
  \right) \, dX \;,
\end{displaymath}
we can no longer extract the behavior of $c_\lambda(\alpha)$ near
$\alpha = 1$ by expanding the argument of the character $\chi_\lambda$
under the integral sign, as this would lead to a divergent integral.
Hence a different approach is needed. \smallskip

To get some inspiration, we turn to the simple case $N_{\rm c} =
N_{\rm b} = 1$ with distribution
\begin{displaymath}
  (2\pi)^{-1} \frac{d\theta}{| 1-\alpha \, {\rm e}^{{\rm i}\theta}
    |^{-2}} \Big|_{{\rm e}^{{\rm i}\theta} = \frac{1 + \alpha + (1 -
      \alpha) x}{1 + \alpha - (1-\alpha)x}} = ({\rm i}\pi)^{-1} (1 -
  \alpha^2)^{-1} \frac{dx}{1 - x^2} \qquad (x \in {\rm i}{\mathbb R})
  \;.
\end{displaymath}
The primitive characters of ${\rm U}(1)$ are $\chi_n({\rm e}^{{\rm i}
  \theta}) = {\rm e}^{{\rm i}n\theta}$ $(n \in {\mathbb Z})$.  The
coefficients $c_n(\alpha)$ can easily be written down in closed form:
\begin{displaymath}
  c_n(\alpha) = (1 - \alpha^2)^{-1} \alpha^{|n|} \;.
\end{displaymath}
Obviously, the effect of having a divergent second moment here is to
cause a nonanalyticity in the Fourier variable $n$ at $n = 0$.  In the
previous case of $N_{\rm b} \ge N_{\rm c} + 1 = 2$, the ratio $c_n
(\alpha) / c_0(\alpha)$ had zero derivative at $\alpha = 1$ by parity
symmetry $(x \mapsto -x)$, but now
\begin{displaymath}
  \lim_{\alpha \to 1-} \frac{d}{d\alpha} \frac{c_n(\alpha)} {c_0
    (\alpha)} = |n| \;,
\end{displaymath}
which forces a different scaling of the parameters $\alpha_p$, and
thus a different continuum limit.  We are going to show that the same
situation occurs for all $N_{\rm b} = N_{\rm c} \ge 1$.  \smallskip

Consider the formula
\begin{displaymath}
  r_\lambda(\alpha) \equiv \frac{c_{\lambda}(\alpha)}{c_0(\alpha)} =
  \int\limits_{\mathfrak{u} (N_{\rm c})} \chi_\lambda \left( \frac{1 -
      \frac{1 - \alpha} {1+\alpha} X} {1 + \frac{1 - \alpha}{1 +
        \alpha} X} \right) \, {\rm Det}^{- N_{\rm c}} \left( 1 - X^2
  \right) \, dX \;,
\end{displaymath}
which expresses the ratio $r_\lambda(\alpha)$ as an expectation of the
character $\chi_\lambda$ with respect to the normalized Cauchy
distribution ${\rm Det}^{-N_{\rm c}}(1 - X^2) \, dX$.  Because
$|\chi_\lambda(U)|$ is bounded from above by the dimension
$d_\lambda$, the integral makes sense and defines $r_\lambda(\alpha)$
as a continuous function of $\alpha$ in the range $-1 < \alpha <
\infty$, and we have
\begin{displaymath}
  \left| r_\lambda (\alpha) \right| \le d_\lambda \qquad (-1 < \alpha
  < \infty) \;.
\end{displaymath}
$r_\lambda(\alpha)$ is maximal at $\alpha = 1$, where the upper
bound $r_\lambda(1) = \chi_\lambda(1) = d_\lambda$ is attained.
Going away from that maximum, $r_\lambda(\alpha)$ has to decrease
(if $\lambda \not= 0$), since nontrivial characters oscillate.
From the $N_{\rm b} = N_{\rm c} = 1$ example (and for reasons
which, if not clear already, will soon become evident), we expect
a finite first derivative and hence a cusp singularity in
$r_\lambda(\alpha)$ at $\alpha = 1$.  The task is to compute the
slope of $r_\lambda(\alpha)$ on the $\alpha < 1$ side of that
cusp.  \smallskip

To that end, we start from the formula
\begin{displaymath}
  2 \lim_{\alpha \to 1-} \frac{d}{d\alpha} r_\lambda(\alpha) = -
  \lim_{\epsilon \to 0+} \frac{d}{d\epsilon} \int\limits_{\mathfrak{u}
    (N_{\rm c})} \chi_\lambda \left( \frac{1 - \epsilon X} {1 +
      \epsilon X} \right) \, {\rm Det}^{- N_{\rm c}} \left( 1 - X^2
  \right) \, dX \;.
\end{displaymath}
We would like to take the derivative inside the integral but, as it
stands, are not allowed to do so.  Since the obstacle arises from the
noncompactness of the integration domain, the trick will be to
compactify it. \smallskip

In the first step, we introduce the eigenvalues of $X$ and the
diagonalizing matrix as the new variables of integration.  Thus, let
$x = (x_1, \ldots, x_{N_{\rm c}}) \in ({\rm i}{\mathbb R})^{N_{\rm
    c}}$ be the set of eigenvalues of the anti--Hermitian matrix $X$,
and let $\mathfrak{t}^+ \subset \mathfrak{t} \equiv ({\rm i}{\mathbb
  R})^{N_{\rm c}}$ be the positive Weyl chamber given by ${\rm i}x_1 <
{\rm i}x_2 < \ldots < {\rm i}x_{N_{\rm c}}$.  If $T = {\rm U}(1)^{
  N_{\rm c}}$ is a maximal torus, say the diagonal matrices in ${\rm
  U} (N_{\rm c})$, some dense open set in $\mathfrak{u}(N_{\rm c})$ is
diffeomorphic to $\mathfrak{t}^+ \times {\rm U}(N_{\rm c}) / T$ by the
polar coordinate map $\psi \, : \, (x,gT) \mapsto g x g^{-1} = X$.
The Jacobian of this map is $\triangle({\rm i}x)^2 = {\rm i}^{N_{\rm
    c}(N_{\rm c} - 1)} \prod_{k < l} (x_k - x_l)^2$, which means there
exists some positive flat density $dx$ on $\mathfrak{t}^+$ such that
\begin{displaymath}
  \psi^\ast (dX) = dg_T \cdot \triangle({\rm i}x)^2 \, dx \;,
\end{displaymath}
with $dg_T$ an invariant volume form on ${\rm U}(N_{\rm c}) / T$.
\smallskip

The next step is to write the character $\chi_\lambda$ as a sum over
the integer weight lattice $L_\lambda$ of the representation
$\lambda$:
\begin{displaymath}
  \chi_\lambda ({\rm e}^{{\rm i} \theta}) = \sum_{ \{n\} \in
    L_\lambda} {\rm e}^{{\rm i}\sum n_k \theta_k} \;.
\end{displaymath}
$L_\lambda$ is determined in principle by division of the elementary
antisymmetric torus functions in Weyl's formula $\chi_\lambda ({\rm
  e}^{{\rm i} \theta}) = \xi_{\lambda+\rho} ({\rm e}^{{\rm i} \theta})
/ \xi_{\rho} ({\rm e}^{{\rm i} \theta})$.  Inserting the sum into the
integral we obtain
\begin{displaymath}
  - 2 \lim_{\alpha \to 1-} \frac{d}{d\alpha} r_\lambda(\alpha) = {\rm
    vol} \left( {\rm U}(N_{\rm c}) / T \right) \lim_{\epsilon \to 0+}
  \frac{d}{d\epsilon} \int\limits_{ \mathfrak{t}^+} \sum_{ \{n\} \in
    L_\lambda} \prod_{k = 1}^{N_{\rm c}} \left( \frac{1 - \epsilon
      x_k} {1 + \epsilon x_k} \right)^{n_k} \, (1 - x_k^2)^{- N_{\rm
      c}} \, \triangle({\rm i}x)^2 \, dx \;.
\end{displaymath}
For further treatment, the restriction of the domain of integration to
the positive Weyl chamber $\mathfrak{t}^+ \subset \mathfrak{t}$ is
inconvenient.  Because the integrand is invariant w.r.t.~the symmetric
group (the Weyl group of ${\rm U}(N_{\rm c})$), we may actually lift
the restriction and integrate over all of $\mathfrak{t} = ({\rm i}
{\mathbb R})^{N_{\rm c}}$.  The new integral is $N_{\rm c}!$ times the
old one, so we divide by that factor. \smallskip

Now is the point where we compactify.  Observe that
\begin{displaymath}
  f(z) = \frac{1 - z}{1 + z}
\end{displaymath}
as a function of $z \in {\mathbb C}$ satisfies $|f(z)| < 1$ for ${\rm
  Re} z > 0$, and $|f(z)|^{-1} < 1$ for ${\rm Re} z < 0$.  Therefore,
given some fixed term $\{ n \}$ in the sum over weights, we may modify
the integration domain at infinity without changing the value of the
integral, as follows: for all $k$ with $n_k > 0$ we close the
integration contour for the variable $x_k \in {\rm i} {\mathbb R}$
around the right half--plane ${\rm Re} z > 0$, whereas for all $k$
with $n_k < 0$ we close around the left half--plane (for $n_k = 0$ we
may close either way).  Having closed the contours for all the
variables, we pull them in from infinity, by holomorphicity.  The
integration domain, say $C_{ \{n\} }$, is now compact. \smallskip

After compactification, we are permitted to differentiate under the
integral sign and set $\epsilon$ to zero, which gives
\begin{displaymath}
  \lim_{\alpha \to 1-} \frac{d}{d\alpha} r_\lambda(\alpha) = N_{\rm
    c}!^{-1} {\rm vol} \left( {\rm U}(N_{\rm c}) / T \right) \sum_{
    \{n\} \in L_\lambda} \sum_{k = 1}^{N_{\rm c}} n_k \int\limits_{
    C_{ \{ n \} }} x_k \prod_{l} (1 - x_l^2)^{- N_{\rm c}} \,
  \triangle({\rm i}x)^2 \, dx \;.
\end{displaymath}
To compute the remaining integral, we first make the closed
integration contours identical for all the variables $x_1, \ldots,
x_{N_{\rm c}}$.  This is possible to arrange because after setting
$\epsilon$ to zero, the point at infinity is no longer a singularity
for the variables $x_l$ with $l \not= k$ (although for $l = k$ it
still is, because of the presence of the factor $x_k$ in the
integrand).  Thus we deform from $C_{ \{n\} }$ to some
$C_{\sigma(n_k)} \times \ldots \times C_{\sigma(n_k)} =
C_{\sigma(n_k)}^{N_{\rm c}}$.  The subscript $\sigma(n_k)$ reminds us
that $C_{\sigma (n_k)}$ lies in the right or left half--plane --- and
encircles the pole of $(1 - x_k^2)^{-N_{\rm c}}$ at $x_k = \pm 1$ ---
depending on whether $n_k$ is positive or negative, respectively.
\smallskip

Next we symmetrize the integrand, replacing $x_k$ by $N_{\rm
c}^{-1} \sum_l x_l$, and we revert to integrating over a
restricted domain (and drop the factor $1/N_{\rm c}!$), by
requiring the variables $x_1, \ldots, x_{N_{\rm c}}$ to be
arranged in ascending order on $C_{\sigma(n_k)}$ according to the
orientation of $C_{\sigma(n_k)}$. \smallskip

Now let $M_{\sigma(n_k)}$ be the adjoint orbit of ${\rm U}(N_{\rm c})$
on $C_{\sigma(n_k) }^{N_{\rm c}}$.  Then, using the polar coordinate
map in reverse we arrive at
\begin{displaymath}
  \lim_{\alpha \to 1-} \frac{d}{d\alpha} r_\lambda(\alpha) = \sum_{
    \{n\} \in L_\lambda} \sum_{k = 1}^{N_{\rm c}} n_k \int\limits_{
    M_{ \sigma(n_k) }} \frac{ N_{\rm c}^{-1} {\rm Tr}X \, dX} {{\rm
      Det}^{N_{\rm c}}(1 - X^2)} \;.
\end{displaymath}
This integral is easy to compute.  The integration domain $M_{\sigma
  (n_k)}$ is a closed orientable $N_{\rm c}^2$--manifold in the
complex space $\mathfrak{gl} (N_{\rm c},{\mathbb C})$, and the
integrand, a holomorphic density, can be viewed as a differential
form (of degree $N_{\rm c}^2$) which is closed.  Therefore, the
value of the integral does not change if we contract the
integration manifold $M_{\sigma(n_k)}$ to another one in the same
homology class enclosing the singular point $X = {\rm sgn}(n_k)
\times 1$.  At that singularity, the scalar factor ${N_{\rm c}}^{
-1} {\rm Tr} X$ takes the value ${\rm sgn}(n_k)$.  We extract the
factor with this value from the integral.  Having done so, we
expand the integration manifold to its original form, and then see
that the remaining integral is unity by normalization of the
Cauchy distribution ${\rm Det}^{-N_{\rm c}}(1 - X^2) \, dX$.  Thus
we get the result
\begin{displaymath}
  \lim_{\alpha \to 1-} \frac{d}{d\alpha} r_\lambda(\alpha) = d_\lambda
  \, {\rm Cas}_1(\lambda) \;,
\end{displaymath}
where
\begin{equation}\label{Cas1}
  {\rm Cas}_1(\lambda) = d_\lambda^{-1} \sum_{ \{n\} \in L_\lambda}
  \sum_{k = 1}^{N_{\rm c}} | n_k |
\end{equation}
is some sort of Casimir invariant, as will be explained shortly.
\smallskip

An equivalent way of stating the result is
\begin{displaymath}
  \frac{c_\lambda(\alpha)}{c_0(\alpha)} = d_\lambda \big( 1 - (1 -
  \alpha) {\rm Cas}_1(\lambda) + {\cal R}(\alpha) \big) \qquad (\alpha
  < 1) \;,
\end{displaymath}
with a remainder term ${\cal R}(\alpha)$ that vanishes faster than $1
- \alpha$ as $\alpha \to 1$, from which we deduce
\begin{displaymath}
  \lim_{N \to \infty} \left( \frac{1}{d_\lambda} \, \frac{c_\lambda(1
      - \mu / N)} {c_0(1 - \mu / N)} \right)^N = {\rm e}^{ - \mu \,
    {\rm Cas}_1 (\lambda)} \;.
\end{displaymath}

Given this formula, the rest of the calculation goes the same way as
in Section \ref{contlimit}, and we just write down the answer.  The
continuum limit is defined by putting $\alpha_p = 1 - A_p / a^2$ and
taking the plaquette areas $A_p$ to zero. Then, setting $\mu = \sum_p
A_p / a^2$ we have the local data
\begin{equation}\label{localdat1}
  \Gamma({\cal U},\mu) = \sum_\lambda d_\lambda \, \exp \left( -
  \mu \, {\rm Cas}_1(\lambda) \right) \, \chi_\lambda({\cal U}) \;,
\end{equation}
and hence the genus $g$ partition function
\begin{displaymath}
  Z_g(\mu) = \sum_\lambda d_\lambda^{-2g + 2} \, {\rm e}^{ - \mu \,
    {\rm Cas}_1(\lambda) } \;,
\end{displaymath}
as follows again by the reasoning of \cite{witten}. \smallskip

What is the corresponding continuum field theory, i.e.~what's the
Lagrangian that gives rise to the Hamiltonian ${\rm Cas}_1$ upon
canonical quantization?  To answer that question, we first need to
sharpen our understanding of ${\rm Cas}_1$.  Recall from Lie
theory that Casimir invariants are elements in the center of the
universal enveloping algebra (of the Lie algebra at hand), which
consists of polynomials in the generators.  The invariant ${\rm
Cas}_1$ certainly does not come from a finite--degree polynomial,
so we are not entitled to call it a Casimir invariant in the
strict sense.  However, it is something quite similar.  Fix some
basis $\{ \tau_A \}$ of $\mathfrak{u}(N_{\rm c})$ and write $X \in
\mathfrak{u} (N_{\rm c})$ as $X = \sum_A X_A \tau_A$. Reinterpret
$\tau_A$ as an abstract operator $\hat\tau_A$ that acts in the
linear space $V_\lambda$ of any representation $D^\lambda$ by
$D_\ast^\lambda (\tau_A)$.  Then, viewing the coefficients $X_A$
as real--valued functions on $\mathfrak{u}(N_{\rm c})$, consider
the formal expression
\begin{displaymath}
  K_\mu = \int\limits_{\mathfrak{u}(N_{\rm c})} \frac{{\rm e}^{\sum_A
      X_A \hat\tau_A} \, \mu^{N_{\rm c}^2} dX} {{\rm Det}^{N_{\rm
        c}} \left( \mu^2 - X^2 \right)} \;.
\end{displaymath}
By the replacement $\hat\tau_A \to D_\ast^\lambda(\tau_A)$, this
makes sense as a compact operator in every unitary representation
space $V_\lambda$.  Moreover, by the invariance of the Cauchy
distribution w.r.t.~conjugation $X \to U X U^{-1}$, this operator
commutes with all the ${\rm U}(N_{\rm c})$ generators, so it is a
multiple of unity in every irrep by Schur's lemma.  We claim
\begin{displaymath}
  {\rm Cas}_1 = \lim_{\mu \to 0+} \frac{1 - K_\mu}{\mu} \;.
\end{displaymath}
Indeed, since $K_\mu$ acts as a multiple of the identity on every
irreducible $V_\lambda$, we might as well take the trace over
$V_\lambda$ and divide by the dimension $d_\lambda$.  What we then
encounter is an integral of the character $\chi_\lambda$ against
the Cauchy distribution, and by a slight variant of the
calculation done earlier for $r_\lambda(\alpha)$, we find that
$K_\mu \big|_{V_\lambda}$ has a right--hand derivative at $\mu =
0$ and the value of this derivative is ${\rm Cas}_1(\lambda)$.
\smallskip

When acting in the Hilbert space of square--integrable class functions
$f(U)$ on ${\rm U}(N_{\rm c})$, the operator ${\rm e}^{\sum X_A \hat
  \tau_A}$ causes translations $f(U) \mapsto f(U {\rm e}^X) = f({\rm
  e}^{-X} U)$.  It follows that the invariant operator $K_\mu$ acting
in the same Hilbert space has the integral kernel
\begin{displaymath}
  \left\langle U^\prime \big| K_\mu \big| U \right\rangle =
  \int\limits_{\mathfrak{u}(N_{\rm c})} \delta \left( U^{-1} U^\prime
    {\rm e}^X \right) \, {\rm Det}^{-N_{\rm c}} \left( \mu^2 - X^2
  \right) \, \mu^{N_{\rm c}^2} \, dX \;.
\end{displaymath}
By definition, $\Gamma(U^{-1} U^\prime, \mu)$ in (\ref{localdat1}) is
the kernel of the one--parameter semigroup generated by ${\rm Cas}_1$.
From what we have just shown, the same holds true of $\left \langle
  U^\prime \big| K_\mu \big| U \right \rangle$ asymptotically for
small $\mu$. Hence we infer the limit relation
\begin{displaymath}
  \Gamma(U^{-1} U^\prime, \mu) = \lim_{N \to \infty} \left\langle
    U^\prime \big| \left( K_{\mu / N} \right)^N \big| U \right\rangle
  \;,
\end{displaymath}
which allows us to construct a functional integral representation for
$\Gamma(U,\mu)$ in the standard fashion.  The resulting field theory
is given by the action functional anticipated in the introductory part
of the current section.  Indeed, expressing the tangent--space
$\delta$--function $\delta(U^{-1}U^\prime {\rm e}^X)$ by Fourier
integration over a conjugate variable $\phi$ with values in
$\mathfrak{u}(N_{\rm c})$, we encounter the Fourier transform of the
Cauchy distribution:
\begin{displaymath}
  \int\limits_{\mathfrak{u}(N_{\rm c})} {\rm e}^{{\rm i} {\rm Tr} \phi
    X} {\rm Det}^{-N_{\rm c}} \left( \mu^2 - X^2 \right) \, \mu^{N_{
      \rm c}^2} \, dX \;.
\end{displaymath}
On substituting $X$ by $\mu X$ this becomes precisely the type of
integral we computed in the current subsection, albeit with the
character $\chi_\lambda \left( (1 - \epsilon X) / (1 + \epsilon X)
\right)$ replaced by the bounded function ${\rm e}^{{\rm i} {\rm
    Tr}\phi X}$.  Proceeding in the same manner as before we find that
the result in the small--$\mu$ limit is approximated by ${\rm e}^{-
  \mu \parallel \phi \parallel_1}$ with $\parallel \phi \parallel_1$
the linear potential in (\ref{actfunc1}).

\subsection{A different perspective: Howe duality}\label{sec:howe}

In treating the two--dimensional theory, crucial use was made of the
expansion of the plaquette distribution function in terms of ${\rm U}
(N_{\rm c})$ characters $\chi_\lambda(U)$:
\begin{displaymath}
  \big| {\rm Det}\left( 1 - \alpha \, U \right) \big|^{ -2N_{\rm b}} =
  \sum_{\lambda} c_{\lambda} (\alpha) \, \chi_\lambda(U) \;.
\end{displaymath}
We now wish to communicate the intriguing fact that the expansion
coefficients $c_\lambda (\alpha)$ {\it themselves are characters}, of
a noncompact Lie group dual to ${\rm U}(N_{\rm c})$ in the sense of
R.~Howe \cite{Howe,Hw_pop}.  Explaining this duality will go a long
way toward preparing and setting up the duality transformation carried
out in the next section. \smallskip

The general framework for the kind of duality we are about to describe
is what is called the bosonic Fock space in physics (and the
Shale--Weil, or metaplectic, or oscillator representation in
mathematics).  To get started, let there be a single oscillator or
boson mode with operators $b, b^\dagger$ satisfying the canonical
commutation relations $[ b , b^\dagger ] = 1$.  There is a vacuum $| 0
\rangle$, which is annihilated by $b$, and the bosonic Fock space
${\cal V}$ is the linear span of the state vectors $(b^\dagger)^n | 0
\rangle$ with $n \in {\mathbb N} \cup \{ 0 \}$.  The space ${\cal V}$
comes with a natural Hermitian scalar product, with respect to which
$b^\dagger$ is the adjoint of $b$.  In the presence of a chemical
potential $\mu > 0$, the partition sum of a single boson mode is
\begin{displaymath}
  {\rm Tr}_{\cal V} \, {\rm e}^{-\mu b^\dagger b} = \sum_{n =
    0}^\infty {\rm e}^{- \mu n} = \left( 1 - {\rm e}^{-\mu}
  \right)^{-1} \;.
\end{displaymath}
Now we enlarge the formalism by adding a color degree of freedom $i =
1, 2, \ldots, N_{\rm c}$ to the boson operators $b^i, {b^i }^\dagger$.
(Of course the canonical commutation relations still hold, the
Hermitian scalar product extends in the obvious manner, and the vacuum
$| 0 \rangle$ is annihilated by each of the $b^i$.)  Let $E^{kl}$ be
the $N_{\rm c} \times N_{\rm c}$ matrix with entry $1$ at the
intersection of the $k^{\rm th}$ row with the $l^{\rm th}$ column, and
zeroes everywhere else.  To a ``Hamiltonian'' $X = \sum_{kl} X^{kl}
E^{kl} \in \mathfrak{u} (N_{\rm c})$, we associate the partition sum
\begin{displaymath}
  {\rm Tr}_{\cal V} \, {\rm e}^{-\mu \sum_j {b^j}^\dagger b^j} {\rm
    e}^{ \sum_{kl} {b^k}^\dagger X^{kl} b^l} = {\rm Det} \left( 1 -
    {\rm e}^{-\mu + X} \right)^{-1} \;.
\end{displaymath}
Now we extend the formalism further by adding flavor $a = 1, 2,
\ldots, N_{\rm b}$, and distinguishing between bosons of opposite
${\rm U}(1)$ charges, $\pm$.  Then, if $\hat N_{\rm bos} = \sum_{j,a}
\big( {b_+^{j,a}}^\dagger b_+^{j,a} + {b_-^{j,a}} ^\dagger b_-^{j,a}
\big)$ is the total boson number, we get
\begin{displaymath}
  {\rm Tr}_{\cal V} \, {\rm e}^{-\mu \hat N_{\rm bos}} \, {\rm
    e}^{\sum_{kl} X^{kl} \sum_a \left( {b_+^{k,a}}^\dagger b_+^{l,a} -
      {b_-^{l,a}}^\dagger b_-^{k,a} \right)} = \left| {\rm Det} \left(
      1 - {\rm e}^{-\mu} {\rm e}^X \right) \right|^{- 2N_{\rm b}} \;,
\end{displaymath}
which becomes the distribution function we have been working with on
identifying ${\rm e}^{-\mu} = \alpha$ and ${\rm e}^X = U \in {\rm
  U}(N_{\rm c})$. \smallskip

The linear mapping
\begin{displaymath}
  X \mapsto \sum_{a = 1}^{N_{\rm b}} \sum_{k,l=1}^{N_{\rm c}} X^{kl}
  \left( {b_+^{k,a}}^\dagger b_+^{l,a} - {b_-^{l,a}}^\dagger b_-^{k,a}
  \right)
\end{displaymath}
is an isomorphism of Lie algebras (i.e.~it preserves the commutator),
and it is easily seen to exponentiate to a homomorphism of Lie groups,
$U \mapsto T_U$. Thus, $U \in {\rm U}(N_{\rm c})$ acts on the bosonic
Fock space ${\cal V}$ by $T_U$.  Using $T_U$ we can write the previous
formula in the concise form
\begin{displaymath}
  \left| {\rm Det} \left( 1 - \alpha \, U \right) \right|^{- 2 N_{\rm
      b}} = {\rm Tr}_{\cal V} \, \alpha^{\hat N_{\rm bos}} \, T_U \;.
\end{displaymath}
This identifies the plaquette statistical weight function of the boson
induced gauge model as a trace, or character, in Fock space.
\smallskip

In addition to ${\rm U}(N_{\rm c})$ there is a second group that
naturally acts on the bosonic Fock space ${\cal V}$.  This is the
noncompact Lie group ${\rm U}(N_{\rm b},N_{\rm b})$.  To describe its
action, we decompose Lie algebra elements $Y \in \mathfrak{u} (N_{\rm
  b}, N_{\rm b})$ into $N_{\rm b} \times N_{\rm b}$ blocks as $Y =
\begin{pmatrix} A &B \\ C &D \end{pmatrix}$, with anti--Hermitian $A =
\sum A^{ab} E^{ab}$, $D = \sum D^{ab} E^{ab}$, and with $B = \sum
B^{ab} E^{ab}$, $C = \sum C^{ab} E^{ab}$ being adjoints of each other;
and we assign to $Y$ an operator $\hat Y = - Y^\dagger$ on ${\cal V}$
by
\begin{displaymath}
  \hat Y = \sum_{a,b=1}^{N_{\rm b}} \sum_{j = 1}^{N_{\rm c}} \left(
    A^{ab} {b_+^{j,a}}^\dagger b_+^{j,b} + B^{ab} {b_+^{j,a}}^\dagger
    {b_-^{j,b}}^\dagger - C^{ab} b_-^{j,a} b_+^{j,b} - D^{ab}
    b_-^{j,a} {b_-^{j,b}}^\dagger \right) \;.
\end{displaymath}
The mapping $Y \mapsto \hat Y$ is an isomorphism of $\mathfrak{u}
(N_{\rm b},N_{\rm b})$ Lie algebras.  By exponentiating it, we get a
unitary representation of ${\rm U}(N_{\rm b},N_{\rm b})$ on ${\cal
  V}$. \smallskip

Thus we have two groups, ${\rm U}(N_{\rm c})$ and ${\rm U}(N_{\rm
  b},N_{\rm b})$, acting on the bosonic Fock space.  Because ${\rm U}
(N_{\rm c})$ acts on color and ${\rm U} (N_{\rm b},N_{\rm b})$ on
flavor, the two group actions commute. This turns out to be a maximal
property: you cannot enlarge either one of the two groups without
compromising it. Put differently, ${\rm U} (N_{\rm b} , N_{\rm b})$ is
the centralizer of ${\rm U}(N_{\rm c})$ inside the big group of
symplectic (actually, metaplectic) transformations of ${\cal V}$.
R.~Howe calls such a pair of Lie groups a {\it dual pair}; see
\cite{Hw_pop} for a pedagogical introduction to the subject.
\smallskip

Because ${\rm U}(N_{\rm c})$ is compact, general theory guarantees
that we can decompose ${\cal V}$ into irreducible representation
spaces $V_\lambda$ for this group.  All ${\rm U}(N_{\rm c})$
representations of a given type $\lambda$ are collected into what is
called an isotypic component for ${\rm U}(N_{\rm c})$ in ${\cal V}$.
Here comes the main point of the present discussion: as a consequence
of the dual pair property, the product ${\rm U}(N_{\rm b} , N_{\rm b})
\times {\rm U}(N_{\rm c})$ acts irreducibly \cite{Hw_pop} on every
such isotypic component. Thus the decomposition of ${\cal V}$ takes
the form of a multiplicity--free sum,
\begin{displaymath}
  {\cal V} = \sum_\lambda \tilde V_\lambda \otimes V_{\lambda} \;,
\end{displaymath}
where $\tilde V_\lambda$ and $V_\lambda$ are irreducible
representation spaces for ${\rm U}(N_{\rm b},N_{\rm b})$ and ${\rm
  U}(N_{\rm c})$, respectively, and the correspondence $V_\lambda
\leftrightarrow \tilde V_\lambda$ is one--to--one. \smallskip

We can now use this to decompose the trace of the product
$\alpha^{\hat N_{\rm bos}} T_U$ over ${\cal V}$. The operator
$\alpha^{\hat N_{\rm bos}}$ is trivial on the second factor of any
isotypic component $\tilde V_\lambda \otimes V_\lambda$ in ${\cal V}$,
while $T_U$ is trivial on the first factor.  As a result, the trace
over every isotypic component separates into two factors, one
depending on $\alpha$ and the other on $U$:
\begin{displaymath}
  {\rm Tr}_{\tilde V_\lambda \otimes V_\lambda} \, \alpha^{\hat N_{\rm
      bos}} T_U = {\rm Tr}_{\tilde V_\lambda} \alpha^{\hat N_{\rm
      bos}} \times {\rm Tr}_{V_\lambda} T_U = c_\lambda(\alpha)
  \chi_\lambda(U) \;.
\end{displaymath}
The factors are of a similar kind: both are primitive characters;
$\chi_\lambda(U)$ for ${\rm U}(N_{\rm c})$, and $c_\lambda (\alpha)$
for ${\rm U}(N_{\rm b}, N_{\rm b})$.  What we have learned then is
this: the coefficient $c_\lambda(\alpha)$ in the character expansion
of the induced statistical weight function,
\begin{displaymath}
  \left| {\rm Det} \left( 1 - \alpha \, U \right) \right|^{- 2 N_{\rm
      b}} = {\rm Tr}_{\cal V} \, \alpha^{\hat N_{\rm bos}} \, T_U =
  \sum_\lambda c_\lambda(\alpha) \chi_\lambda(U) \;,
\end{displaymath}
is itself a character; it is the value on $\alpha^{\hat N_{\rm bos}}$
of the ${\rm U}(N_{\rm b},N_{\rm b})$ character associated to the
${\rm U}(N_{\rm c})$ representation $\lambda$ by the dual pair
correspondence. \smallskip

Precisely speaking, $\hat N_{\rm bos}$ does not represent a generator
of the real form $\mathfrak{u}(N_{\rm b},N_{\rm b})$, but a generator
of the complexified Lie algebra $\mathfrak{gl} (2N_{\rm b}, {\mathbb
  C})$. (The generator lying in $\mathfrak{u} (N_{\rm b},N_{\rm b})$
is ${\rm i} \hat N_{\rm bos}$.)  Thus $\alpha^{\hat N_{\rm bos}}$ is
to be viewed as representing an element of ${\rm GL}(2N_{\rm
  b},{\mathbb C})$, and the $c_\lambda(\alpha)$ are obtained from
${\rm U}(N_{\rm b}, N_{\rm b})$ characters by analytically continuing
to that element.  \smallskip

Because the representation spaces $\tilde V_\lambda$ are
infinite--dimensional, the characters $c_\lambda(\alpha) = {\rm
  Tr}_{\tilde V_\lambda} \, \alpha^{\hat N_{\rm bos}}$ all diverge at
the unit element, $\alpha = 1$. One of the results we found earlier is
$c_0(\alpha) \sim (1 - \alpha)^{-2 N_{\rm b} N_{\rm c} + N_{\rm c}^2}$
near that point. This has a transparent interpretation from the
present perspective: taking the trace of $\alpha^{\hat N_{\rm bos}}$
over all states of ${\cal V}$ would give $(1 - \alpha)^{-2 N_{\rm b}
  N_{\rm c}}$, since there are $2 N_{\rm b} N_{\rm c}$ boson species.
The projection onto ${\rm U}(N_{\rm c})$ singlets (the trivial
representation, $\lambda = 0$) amounts to imposing $N_{\rm c}^2$
constraints, which reduces the degree of the singularity to $2 N_{\rm
  b} N_{\rm c} - N_{\rm c}^2$. \smallskip

Every irreducible representation $\lambda$ of ${\rm U}(N_{\rm c})$
need not occur in the sum ${\cal V} = \sum_\lambda \tilde V_\lambda
\otimes V_\lambda$.  Howe's statement is that the sum is
multiplicity--free, i.e.~the multiplicity is at most one, but it can
also be zero.  For example, for $N_{\rm c} = 2$ and $N_{\rm b} = 1$
all the representations $\lambda = (\lambda_1, \lambda_2)$ with
$\lambda_1 \ge \lambda_2 > 0$ or $0 > \lambda_1 \ge \lambda_2$ are
missing.  For $N_{\rm b} \ge N_{\rm c}$, however, all irreducible
representations of ${\rm U}(N_{\rm c})$ do occur, as is implied by the
$\delta$--function property (\ref{thm1}).

%

\section{Duality Transformation}\label{sec:dual}

Having established the continuum limit of the 2d induced gauge model
(\ref{ourmodel}) (with $N_{\rm b} \ge N_{\rm c} + 1$ boson species) to
lie in the universality class of 2d Yang--Mills theory, we certainly
expect a similar scenario to be true in higher dimension. The simple
reason is that going up in dimension enhances the collectivity of the
fields, and thereby works in favor of universality. \smallskip

The 2d Cauchy--type theory we found for $N_{\rm b} = N_{\rm c}$
(Section \ref{sec:cauchy}) is a low--dimensional gimmick, and is
highly unlikely to persist above two space--time dimensions.  Thus in
three dimensions and higher, we expect universal Yang--Mills physics
in the continuum limit of the boson induced gauge model for all
$N_{\rm b} \ge N_{\rm c}$. \smallskip

Motivated by this expectation, we will show in the current section how
to pass to a dual version of the induced gauge model.  The main tool
we are going to use is a variant of the color--flavor transformation
\cite{circular,icmp97}, which is based on Howe duality and the notion
of dual pairs we have just sketched.  \smallskip

As an aside, we mention that the same goal of constructing a dual
theory was pursued in \cite{NS:02}, by similar techniques.  However,
the transformation carried out there does not extend to the good range
$N_{\rm b} \ge N_{\rm c}$ (in fact, only the case $N_{\rm b} = 1$ was
addressed in that reference) and therefore fails to be of relevance
for Yang--Mills theory.

\subsection{Abelian duality (review)}\label{sec:abelian}

By way of preparation and for future reference, we first review the
standard construction of a dual theory in the abelian case.  The
nonabelian duality transform described later on will be seen to reduce
to the standard one in the abelian limit. \smallskip

Consider the partition function of the induced ${\rm U}(1)$ gauge
model (\ref{ourmodel}) with $N_{\rm b} = 1$ (and no fermions, $N_{\rm
  f} = 0$):
\begin{displaymath}
  Z(\alpha) = \int [dU] \prod_{\bf p} \left| 1 - \alpha_p \,
    U(\partial{\bf p}) \right|^{-2} \;.
\end{displaymath}
As before, the model is placed on a $d$--dimensional cell complex
$\Lambda$ with boundary operator $\partial$, and all of the real
parameters $\alpha_p$ are chosen close to (but less than) unity.  To
dualize such a model in a concise manner, we need a few basic facts
\cite{sternberg} from discrete ex\-ter\-ior calculus, which are
rapidly summarized in the next paragraph. \smallskip

Recall that the boundary operator $\partial$ is a linear mapping from
the vector space of $k$--chains into the vector space of
$(k-1)$--chains, and the boundary of a boundary is always zero:
$\partial \circ \partial = 0$.  A $k$--chain $b$ is called closed if
$\partial b = 0$.  The Poincar\'e lemma (when applicable) states that
every closed $k$--chain $b$ is a boundary: $b = \partial c$.  Objects
dual to chains are called {\it cochains}: a $k$--cochain $\omega$ (the
discrete version of a $k$--form) is a linear function that assigns to
every $k$--chain $c$ a real number, say $\langle c , \omega \rangle$.
The coboundary operator ${\rm d}$ : $k$--cochains $\to$
$(k+1)$--cochains (the discrete version of the exterior derivative on
forms) is defined by demanding that the discrete analog of Stokes'
theorem be valid: $\langle c , {\rm d} \omega \rangle = \langle
\partial c , \omega \rangle$.  \smallskip

Turning to the case at hand, we write the ${\rm U}(1)$ elements on
links as $U({\bf l}) = {\rm e}^{{\rm i} \theta ({\bf l}) }$, where the
gauge field $\theta({\bf l})$ takes values in the additive group
${\mathbb R} / 2\pi{\mathbb Z}$.  We regard the $\theta({\bf l})$'s as
constituting a 1--cochain $\theta$, and may then write $U({\bf l}) =
{\rm e}^{{\rm i} \theta ({\bf l}) } = {\rm e}^{{\rm i} \langle {\bf l}
  , \theta \rangle}$.  In terms of $\theta$, the ${\rm U}(1)$ field
strengths on plaquettes are $U(\partial{\bf p}) = {\rm e}^{{\rm i}
  \langle \partial {\bf p} , \theta \rangle} = {\rm e}^{{\rm i}
  \langle {\bf p} , {\rm d}\theta \rangle }$. \smallskip

Now, inserting the Fourier expansion of the induced statistical weight
function,
\begin{displaymath}
  | 1 - \alpha {\rm e}^{{\rm i}\vartheta} |^{-2} = (1 - \alpha^2)^{-1}
  \sum_{\nu \in {\mathbb Z}} {\rm e}^{- |\nu| \ln(1/\alpha)} {\rm
    e}^{{\rm i}\nu \vartheta} \;,
\end{displaymath}
into the formula for the partition function, we get
\begin{displaymath}
  Z(\alpha) = \prod_p (1 - \alpha_p^2)^{-1} \int [d\theta] \sum_n {\rm
    e}^{- \parallel n \parallel_\alpha + {\rm i} \langle n , {\rm
      d}\theta \rangle } \;,
\end{displaymath}
where the sum is over all $2$--chains $n = \sum n_{\bf p}\, {\bf p}$
($n_{\bf p} \in {\mathbb Z}$) on $\Lambda$.  The notation $\parallel n
\parallel_\alpha$ means a weighted sum of absolute values: $\parallel
n \parallel_\alpha = \sum_{\bf p} \ln(1/\alpha_p) \, | n_{\bf p} |$.
\smallskip

The information contained in the partition function is somewhat
limited, and to expose the full physics of the model we need to couple
it to an external source, say by inserting a Wegner--Wilson electric
current loop \cite{wegner,wilson}.  Let $C$ be any closed loop on
$\Lambda$.  Viewed as a 1--chain, $C$ defines a linear function
$\langle C, \theta \rangle$ on $\theta$; and this function is
invariant under gauge transformations $\theta \mapsto \theta + {\rm
  d}f$, by Stokes' theorem and the fact that $C$ is closed.  The
Wegner--Wilson loop is $W(C) = {\mathbb E} \, {\rm e}^{{\rm i} \langle
  C , \theta \rangle}$, where the expectation value is taken with
respect to the statistical measure given by the partition function.
\smallskip

Next we use Stokes' theorem to go from $\langle n , {\rm d}\theta
\rangle + \langle C , \theta \rangle$ to $\langle \partial n + C ,
\theta \rangle$.  We then change the order of summation and
integration and carry out the $\theta$--integral, which results in the
$1$--chain $\partial n + C$ being constrained to vanish:
\begin{displaymath}
  \sum_n {\rm e}^{- \parallel n \parallel_\alpha} \int [d\theta /
  2\pi] \, {\rm e}^{{\rm i} \langle n , {\rm d}\theta \rangle + {\rm
      i}\langle C , \theta \rangle} = \sum_{n : \partial n = -C} {\rm
    e}^{-\parallel n \parallel_\alpha} \;.
\end{displaymath}
It is in this dual form --- as a sum over $2$--chains $n$ with
boundary $-C$ --- that $W(C)$ will be seen to emerge as a special case
from the nonabelian duality transform described in the sequel;
cf.~Section \ref{sec:U1_limit}. \smallskip

Let us quickly review how further analysis of the dual theory is
carried out, and what is the physics to be expected.  For brevity, we
do this only for the special case of $d = 3 + 1$ dimensions and
lattices $\Lambda$ with trivial homology groups. \smallskip

Choosing some surface $S$ with boundary $\partial S = C$, one shifts
$n \to n - S$ in order for the sum to be over $n$ with $\partial n =
0$. Since we are assuming trivial homology, the Poincar\'e lemma
applies and guarantees for every closed $2$--chain $n$ the existence
of a $3$--chain $a$ such that $n = \partial a$. Of course, $a$ is not
uniquely determined: if $a$ has boundary $n$, then so does the gauge
transform $a + \partial \varphi$ by any $4$--chain $\varphi$. Thus,
solving the constraint $\partial n = 0$ by $n = \partial a$, one
arrives at a sum over gauge equivalence classes $[a] = [a + \partial
\varphi]$.  For the Wegner--Wilson loop one obtains
\begin{equation}\label{dualWW}
  W(C) = \sum_{[a]} {\rm e}^{- \parallel \partial a - S
    \parallel_\alpha} \Big/ \sum_{[a]} {\rm e}^{- \parallel \partial a
    \parallel_\alpha } \;.
\end{equation}

At this point one usually transcribes the theory from $\Lambda$ to the
dual lattice $\Lambda^*$.  The transcription proceeds by a canonical
isomorphism (known in the cohomological setting as Poincar\'e duality)
which turns $k$--chains on $\Lambda$ into $(4-k)$--cochains on
$\Lambda^*$.  In particular, the $3$--chains $a$ on $\Lambda$ become
${\mathbb Z}$--valued $1$--cochains $a^*$ on $\Lambda^*$, and the
boundary $\partial a$ becomes the coboundary ${\rm d}a^*$.  Passing to
the dual lattice has the virtue of revealing the true meaning of the
dual theory as a lattice gauge theory with gauge group ${\mathbb Z}$.
Other than that, the passage is really quite unnecessary, and one may
as well continue to work on $\Lambda$ as we do here. \smallskip

Finally, using Poisson summation one relaxes the values of $a$ from
${\mathbb Z}$ to ${\mathbb R}$, at the expense of inserting a factor
${\rm e}^{{\rm i}\langle a , m \rangle}$ and summing over all closed
${\mathbb Z}$--valued 3--cochains $m$ on $\Lambda$ (they must be
closed in order to be well--defined functions on gauge equivalence
classes $[a]$).  The idea behind this last step is that the ${\mathbb
  R}$--valued gauge field $a$ should become the dual of the
``photon'', and the closed $3$--cochain $m$ (which turns into a closed
1--chain on passing to the dual lattice $\Lambda^*$) is to be
interpreted as the world lines of magnetic monopoles. \smallskip

On a hypercubic lattice, and in the Villain approximation
\begin{displaymath}
  {\rm e}^{- \parallel \partial a \parallel_\alpha} \to {\rm e}^{- t
    \sum_{\bf p} (\partial a)_{{\bf p}}^2 } \;,
\end{displaymath}
the photon field $a$ is now readily integrated out to produce an
effective action for the magnetic monopole current $m$.  The
physics of the resulting model is readily understood \cite{banks}
when $t$ is sufficiently small: in that case the magnetic
monopoles are bound in neutral clusters, leaving the system in a
Coulomb phase with a massless photon.  Rigorous mathematical
control on this scenario has been achieved in
\cite{guth,juergtom}. \smallskip

In the present model, that conclusion is less immediate.  At very
short scales of a few lattice units the photon cannot be free, as the
action $\parallel \partial a \parallel_\alpha$ is not quadratic but
linear (and in fact nonanalytic, by the use of the absolute value)!
However, we expect that a sequence of suitable real space
renormalization group transformations
will cause flow toward the quadratic action.  (The renormalization
process of thinning the degrees of freedom by summing over the
short--distance fluctuations of $n$, should have a similar effect
as taking convolutions of the nonanalytic distribution $n \mapsto
{\rm e}^{- \parallel n \parallel_\alpha}$, and a generalized
version of the central limit theorem should take effect.)  We note
that the method of Fr\"ohlich and Spencer \cite{juergtom}, which
does \emph{not} rely on integrating out the ``photon'' to produce
an effective action for the magnetic monopoles, looks promising
for a check on this picture. \smallskip

In summary, our induced ${\rm U}(1)$ gauge model is definitely
interacting at short distances, but Coulomb behavior and a free photon
are expected to emerge at large scales in four dimensions (provided
that the $\alpha_p$'s are close enough to unity).

\subsection{Nonabelian transform: one--link integral}

Our starting point is the partition function (\ref{partfunc}) on any
$d$--dimensional cell complex $\Lambda$. Although our approach is
general and can handle fermions as well as bosons, we will restrict
our attention to the case of bosons only. Switching the order of
integrations, we will first do the gauge field integral and afterwards
the auxiliary boson field integral. \smallskip

For a fixed configuration of the complex boson fields $\varphi$, the
Boltzmann weight of the field theory partition function
(\ref{partfunc}) is a product over links.  Hence, integrating over the
gauge field amounts to doing a set of independent one--link integrals.
To write down and compute the one--link integrals, we need a good
notation. \smallskip

\begin{figure}
  \begin{center}
    \epsfig{file=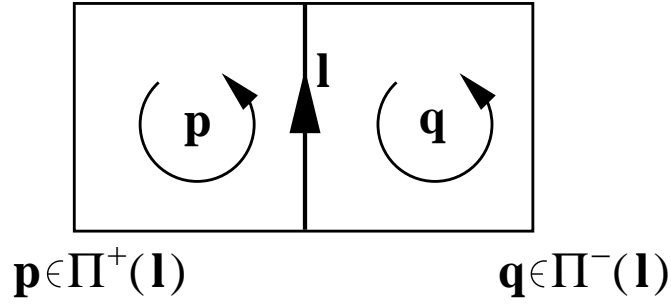,height=4cm}
  \end{center}
  \caption{The set of plaquettes adjacent to a link ${\bf l}$,
    $\Pi({\bf l})$, consists of two subsets defined by comparing
    orientations, which either match [$\Pi^+({\bf l})$] or don't match
    [$\Pi^-({\bf l})$].} \label{fig:twosets}
\end{figure}

Let $\Pi({\bf l})$ denote the set of oriented plaquettes ${\bf p}$
that contain the link ${\bf l}$ in their chain of boundary links.  In
formulas: ${\bf p} \in \Pi({\bf l})$ if $\partial {\bf p} = \pm{\bf l}
+ \ldots$; see Figure \ref{fig:twosets}.  The set $\Pi({\bf l})$
consists of two subsets: the plaquettes whose orientation agrees with
that of ${\bf l}$ ($\partial {\bf p} = + {\bf l} + \ldots$), and those
where the orientations disagree ($\partial{\bf p} = - {\bf l} +
\ldots$).  We write the decomposition into subsets as $\Pi({\bf l}) =
\Pi^+({\bf l}) \cup \Pi^-({\bf l})$.  Note that the cardinalities of
the two sets $\Pi^+$ and $\Pi^-$ are the same, since every plaquette
occurs twice, once with each orientation. \smallskip

Now we fix some link ${\bf l}$, write $U \equiv U({\bf l})$ and
$\Pi^\pm \equiv \Pi^\pm ({\bf l})$ for short, and denote the two
boundary sites of ${\bf l}$ by $n_1$ and $n_2$ (in other words, $n_1$
is the site where ${\bf l}$ begins, and $n_2$ is where ${\bf l}$
ends). Then the gauge field integral pertaining to link ${\bf l}$ is
\begin{equation}\label{onelink}
  I(\varphi,\bar\varphi) \equiv \int\limits_{{\rm U}(N_{\rm c})} \exp
  \left( \sum_{{\bf p} \in \Pi^+} \bar\varphi_{\bf p}(n_2) U
    \varphi_{\bf p}(n_1) \right) \exp \left( \sum_{{\bf q} \in \Pi^-}
    \bar\varphi_{\bf q}(n_1) U^{-1} \varphi_{\bf q}(n_2) \right) \, dU
  \;.
\end{equation}
Here we are using the same short--hand notation as in (\ref{bos_act}):
what the quadratic expressions really mean is
\begin{displaymath}
  \bar\varphi_{\bf p}(n_1) U \varphi_{\bf p}(n_2) = \sum_{i,j =
    1}^{N_{\rm c}} \sum_{a = 1}^{N_{\rm b}} \bar\varphi_{\bf
    p}^{i,a}(n_2) U^{ij} \varphi_{\bf p}^{j,a}(n_1) \;.
\end{displaymath}
In what follows we trade the integral over $U$ for a sum over dual
degrees of freedom.

\subsection{Quantization}

To proceed, we need some definitions.  The main idea is to carry out a
kind of ``quantization'' (not to be confused with quantization in the
usual sense of passing to a quantum Hamiltonian formulation of the
theory): for every color $i \in \{ 1, \ldots, N_{\rm c} \}$, every
flavor $a \in \{ 1, \ldots, N_{\rm b} \}$, and every oriented
plaquette ${\bf p} \in \Pi = \Pi^+ \cup \Pi^-$, we introduce a boson
annihilation operator $b_{\bf p}^{i,a}$ and the adjoint creation
operator ${b_{\bf p}^{i,a}}^\dagger$.  These operators obey the usual
boson commutation relations:
\begin{displaymath}
  [ b_{\bf p}^{i,a} , {b_{\bf q}^{j,b}}^\dagger ] = \delta_{\bf p,q}
  \delta^{i,j} \delta^{a,b} \;,
\end{displaymath}
(with the $b$'s and $b^\dagger$'s commuting amongst themselves), and
they act in a Fock space ${\cal V}$ with vacuum $| 0 \rangle$
characterized by
\begin{displaymath}
  b_{\bf p}^{i,a} | 0 \rangle = 0 \;.
\end{displaymath}
Elements of the gauge group, $U \in {\rm U}(N_{\rm c})$, act on ${\cal
  V}$ by unitary operators $T_U$ with the properties $T_U | 0 \rangle
= | 0 \rangle$ and
\begin{eqnarray}
  T_U b_{\bf p}^{i,a} T_U^\dagger &=& \sum_j b_{\bf p}^{j,a} {\bar
    U}^{ji} \label{UN_action_p} \qquad ({\bf p} \in \Pi^+) \;, \\ T_U
  b_{\bf q}^{i,a} T_U^\dagger &=& \sum_j b_{\bf q}^{j,a} U^{ji} \qquad
  ({\bf q} \in \Pi^-) \label{UN_action_q} \;.
\end{eqnarray}
Thus the $b_{\bf q}$ for ${\bf q} \in \Pi^-$ transform according to
the vector representation of ${\rm U}(N_{\rm c})$, and the $b_{\bf p}$
for ${\bf p} \in \Pi^+$ according to the covector representation
($\bar U = {U^{-1}}^{\rm T}$).  For the creation operators $b^\dagger$
the situation is reversed. \smallskip

Some mathematical background to this setup was given in Section
\ref{sec:howe}.  There, we explained how $T_U$ arises by
exponentiating an isomorphism of Lie algebras.  For present purposes,
just knowing the existence of $T_U$ and its specified properties will
be sufficient. \smallskip

Consider now any one of the factors in the integrand of the one--link
integral (\ref{onelink}), for ${\bf p} \in \Pi^+$.  Using the operator
$T_U$ (or rather the part, $T_U({\bf p})$, acting on the bosons
associated with ${\bf p}$) we can express it as a matrix element
between coherent states (index summations suppressed!):
\begin{displaymath}
  {\rm e}^{\bar\varphi_{\bf p}(n_2) \, U \varphi_{\bf p}(n_1)} =
  \left\langle 0 \big| {\rm e}^{ \bar\varphi_{\bf p}(n_2) \, b_{\bf
        p}} \, T_U({\bf p}) \, {\rm e}^{ b_{\bf p}^\dagger
      \varphi_{\bf p} (n_1)} \big| 0 \right\rangle \;.
\end{displaymath}
Verification is immediate by the ${\rm U}(N_{\rm c})$ invariance of
the vacuum ($T_U | 0 \rangle = | 0 \rangle$), relation
(\ref{UN_action_p}) and $\langle 0 | {\rm e}^{ \bar\varphi \, b } {\rm
  e}^{b^\dagger \varphi } | 0 \rangle = {\rm e}^{ \bar\varphi
  \varphi}$.  Similarly, for ${\bf q} \in \Pi^-$, we have
\begin{displaymath}
  {\rm e}^{\bar\varphi_{\bf q}(n_1) \, U^{-1} \varphi_{\bf q}(n_2)} =
  {\rm e}^{\varphi_{\bf q}(n_2) \, \bar U \bar\varphi_{\bf q}(n_1)} =
  \left\langle 0 \big| {\rm e}^{ \varphi_{\bf q}(n_2) \, b_{\bf q}} \,
    T_U({\bf q}) \, {\rm e}^{ b_{\bf q}^\dagger \bar\varphi_{\bf
        q}(n_1)} \big| 0 \right\rangle \;.
\end{displaymath}
Here we used $(U^{-1})^{ij} = \bar U^{ji}$ and the fact that switching
from ${\bf p} \in \Pi^+$ to ${\bf q} \in \Pi^-$ interchanges the
vector and covector representations of ${\rm U}(N_{\rm c})$.  Thus,
although the matrix $U$ was replaced by its complex conjugate $\bar
U$, the operator whose matrix element we take is still $T_U$.

\subsection{Projection on gauge singlets}

We return to the task of integrating over $U$.  The benefit of
``quantization'', i.e.~interpreting the integrand of the one--link
integral as the matrix element of an operator in Fock space ${\cal
  V}$, is that all dependence on the gauge field matrix $U$ now
resides in
\begin{displaymath}
  T_U = \prod_{{\bf p}\in \Pi^+} T_U ({\bf p}) \prod_{{\bf q}\in
    \Pi^-} T_U({\bf q}) \;,
\end{displaymath}which is just the operator satisfying the relations
(\ref{UN_action_p}, \ref{UN_action_q}).  Thus, doing the $U$--integral
has become very easy; given the last two formulas in the preceding
subsection, we immediately express the one--link integral as a
coherent state matrix element of $P_0 \equiv \int_{{\rm U} (N_{\rm
    c})} T_U \, dU$:
\begin{equation}\label{linkres}
\begin{split}
  \int\limits_{{\rm U}(N_{\rm c})} \exp \left( \sum_{{\bf p} \in
      \Pi^+} \bar\varphi_{\bf p}(n_2) U \varphi_{\bf p}(n_1) +
    \sum_{{\bf q} \in \Pi^-} \bar\varphi_{\bf q}(n_1) U^{-1}
    \varphi_{\bf q}(n_2) \right) \, dU &= \\
  \Big\langle 0 \Big| \exp \Big( \sum_{{\bf p} \in \Pi^+}
      \bar\varphi_{\bf p}(n_2) b_{\bf p} + \sum_{{\bf q} \in \Pi^-}
      \varphi_{\bf q}(n_2) b_{\bf q} \Big) \, P_0 \, \exp \Big(
      \sum_{{\bf p} \in \Pi^+} b_{\bf p}^\dagger \varphi_{\bf p}(n_1)
      &+ \sum_{{\bf q} \in \Pi^-} b_{\bf q}^\dagger \bar \varphi_{\bf
        q}(n_1) \Big) \Big| 0 \Big\rangle \;.
\end{split}
\end{equation}
What is the meaning of $P_0$?  According to (\ref{UN_action_p},
\ref{UN_action_q}), the operator $T_U$ acts on Fock space ${\cal V}$
by ${\rm U}(N_{\rm c})$ rotations.  If $|\psi\rangle$ is any state in
${\cal V}$, integrating the rotated state $T_U | \psi \rangle$ against
Haar measure $dU$ kills that part of the state which transforms
nontrivially under ${\rm U}(N_{\rm c})$, and leaves only the ${\rm
  U}(N_{\rm c})$ invariant part.  Hence $P_0$ is simply the operator
that projects on the ${\rm U}(N_{\rm c})$ invariant sector of ${\cal
  V}$! \smallskip

The present formalism would be useless if that sector --- let it be
denoted by ${\cal V}_0$ --- were some complicated and obscure subspace
of ${\cal V}$.  Fortunately, that's not the case and ${\cal V}_0$ has
a very transparent description, as follows. \smallskip

Consider the set of all boson pair creation operators of the form
\begin{equation}\label{pairs}
  E_{{\bf p q}}^{a b} = \sum_{i = 1}^{N_{\rm c}} {b_{\bf p}^{i ,
      a}}^\dagger {b_{\bf q}^{i,b}}^\dagger \;,
\end{equation}
where ${\bf p} \in \Pi^+$, ${\bf q} \in \Pi^-$, and $a, b \in \{ 1,
\ldots, N_{\rm b} \}$.  Because the ${b_{\bf p}^{i,a}}^\dagger$
transform as a ${\rm U}(N_{\rm c})$ vector and the ${b_{\bf
    q}^{j,b}}^\dagger$ as a ${\rm U}(N_{\rm c})$ covector, the
contraction formed by summing over equal colors $i = j = 1, \ldots,
N_{\rm c}$ is a ${\rm U}(N_{\rm c})$ scalar.  Thus the $E_{\bf
  pq}^{ab}$ are ${\rm U}(N_{\rm c})$ invariant, and so is every state
created by acting with an arbitrary polynomial in these operators on
the vacuum:
\begin{equation}\label{singlets}
  E_{{\bf p}_1 {\bf q}_1}^{a_1 b_1} E_{{\bf p}_2 {\bf q}_2}^{a_2 b_2}
  \cdots E_{{\bf p}_n {\bf q}_n}^{a_n b_n} | 0 \rangle \;.
\end{equation}
It is a beautiful and powerful theorem of classical invariant theory
\cite{Howe} that the states (\ref{singlets}) span ${\cal V}_0$;
i.e.~every ${\rm U}(N_{\rm c})$ invariant state in ${\cal V}$ can be
reached by repeatedly acting on the vacuum $|0\rangle$ with the pair
creation operators $E_{\bf pq}^{ab}$ and taking linear combinations.
\smallskip

An outline of the reasoning is as follows.  Let $r = N_{\rm b} \times
{\rm card} \, \Pi^+$, with ${\rm card} \, \Pi^+$ being the cardinality
of the set $\Pi^+$ (which is the same as the cardinality of $\Pi^-$).
For a $d$--dimensional hypercubic lattice the value of $r$ is $2(d -
1)$.  The antihermitian operators built from the $E_{\bf pq}^{ab}$ and
their adjoints, taken together with all their commutators, span a Lie
algebra that generates a unitary representation (on ${\cal V}$ with
the usual Fock space scalar product) of the noncompact group ${\rm
  U}(r,r)$.  Because the $E_{\bf pq}^{ab}$ are ${\rm U}(N_{\rm c})$
invariant, it is clear that the Fock space actions of the two Lie
groups ${\rm U}(N_{\rm c})$ and ${\rm U}(r,r)$ commute with each
other; they in fact centralize each other and constitute a dual pair
in the sense of R.~Howe.  \smallskip

Recall from Section \ref{sec:howe} that the representation of a dual
pair ${\rm U}(N_{\rm c}) \times {\rm U}(r,r)$ is irreducible on every
${\rm U}(N_{\rm c})$ isotypic component of ${\cal V}$. In particular,
this implies that ${\rm U}(r,r)$ acts irreducibly on the ${\rm
  U}(N_{\rm c})$ invariant sector ${\cal V}_0$ of ${\cal V}$.  This
irreducibility is just what we are claiming: every state in ${\cal
  V}_0$ can be reached by acting with the generators of ${\rm U}(r,r)$
on some ${\rm U}(N_{\rm c})$ invariant reference state.  (And, of
course, if we take that reference state to be the vacuum $| 0
\rangle$, it suffices to act with the pair creation operators $E_{\bf
  pq}^{ab} $.) \smallskip

Let us summarize where we were prior to the digression characterizing
the subspace ${\cal V}_0$: we had identified the integral $P_0 =
\int_{{\rm U} (N_{\rm c})} T_U \, dU$ as the projector onto ${\cal
  V}_0$, and we had expressed the one--link integral (\ref{onelink})
as a matrix element between coherent states; schematically,
(\ref{linkres}) is of the form
\begin{displaymath}
  I(\varphi,\bar\varphi) = \langle \varphi(n_2) | P_0 | \varphi(n_1)
  \rangle \;.
\end{displaymath}
In the following we will think about this matrix element as a trace:
\begin{displaymath}
  I(\varphi,\bar\varphi) = {\rm Tr}_{\cal V} (P_0 A_\varphi) = {\rm
    Tr}_{{\cal V}_0} \, A_\varphi \;, \quad \mbox{with } A_\varphi = |
  \varphi(n_1) \rangle \langle \varphi(n_2) | \;.
\end{displaymath}

We mention in passing that the color--flavor transformation in its
standard form, introduced first in a supersymmetric setting in
\cite{circular}, and applied to the present context in \cite{NS:02},
would result at the present stage of development if it was possible to
express the trace over ${\cal V}_0$ as an integral over ${\rm U}(r,r)$
coherent states.  However, this can only be done if $r \le N_{\rm c}
/2$, and fails for the range of $r$ values of physical interest. (The
problem is a problem of convergence: the ${\rm U}(r,r)$ representation
spaces ${\cal V}_0$ for $N_{\rm c} < 2r$ cannot be realized by
square--integrable holomorphic sections; cf.~the last part of Section
\ref{sec:discuss}.)  We must therefore proceed in a different manner.

\subsection{Integration over $\varphi,\bar\varphi$}
\label{sec:integrate}

We have written the integral (\ref{onelink}) over a single matrix $U
\equiv U({\bf l})$ as the trace of some operator $A_\varphi({\bf l})$
in the ${\rm U} (N_{\rm c})$ invariant sector ${\cal V}_0({\bf l})$ of
an auxiliary Fock space ${\cal V}({\bf l})$.  In the final step taken
now, we multiply all one--link integrals together and arrive at an
expression for the partition function (\ref{partfunc}) of the form
\begin{displaymath}
  Z = {\rm Tr}_{{\cal V}_0} \, A \;,
\end{displaymath}
where ${\cal V}_0 \equiv \bigotimes_{\bf l} {\cal V}_0({\bf l})$ is
the tensor product of the ${\rm U}(N_{\rm c})$ invariant spaces
associated with all the links, and the operator $A$ is the result of
integrating $\prod A_\varphi({\bf l})$ over the boson fields $\varphi,
\bar\varphi$.  It remains to describe $A$, which is what we do next.
\smallskip

Recall from Section \ref{sec:induced} that the boson $\varphi_{\bf p}$
hops from site to site of the chain of boundary links of the oriented
plaquette ${\bf p}$, and carries a color index $i \in \{ 1, \ldots,
N_{\rm c} \}$ and a flavor index $a \in \{ 1, \ldots, N_{\rm b} \}$.
$\varphi_{\bf p}$ enters through the mass term $m_{{\rm b},p}$ of
(\ref{bos_act}) and, by the manipulations of the previous subsection,
it appears as a parameter in the operators $A_\varphi = | \varphi(n_1)
\rangle \langle \varphi(n_2) |$ (schematic notation).  We now write
$m_{{\rm b},p} \equiv m_p$ for short.  \smallskip

While the boson fields were originally coupled by the link matrices
$U({\bf l})$, they are now completely uncoupled (the mass term does
not couple them, and the operators $A_\varphi$ under the trace don't
either.)  Thus we can do the boson field integration for each
component $\varphi_{\bf p}^{i,a}(n)$ separately.  So let us
concentrate on some $\varphi_{\bf p}^{i,a}(n)$ and calculate the
corresponding integral. \smallskip

We start the calculation by noting that $\varphi_{\bf p}^{i,a}(n)$ and
its complex conjugate $\bar\varphi_{\bf p}^{i,a}(n)$ both occur
exactly once as a coherent state parameter, multiplying some boson
operator $b$ or $b^\dagger$ in the exponents on the right--hand side
of (\ref{linkres}).  (The reason is that, since $n$ is a site visited
by the boundary chain of ${\bf p}$, there exist exactly two links in
$\partial {\bf p}$ that are attached to $n$.) To decide which they
multiply --- $b$, or $b^\dagger$ ---, let ${\bf l}_1$ and ${\bf l}_2$
be those two links in the boundary chain of ${\bf p}$ that have the
site $n$ as a boundary point.  Recalling that the lattice links ${\bf
  l}$ come with an a priori orientation determined by $\Lambda$, we
are led to distinguish between three cases.

\begin{figure}
  \begin{center}
    \epsfig{file=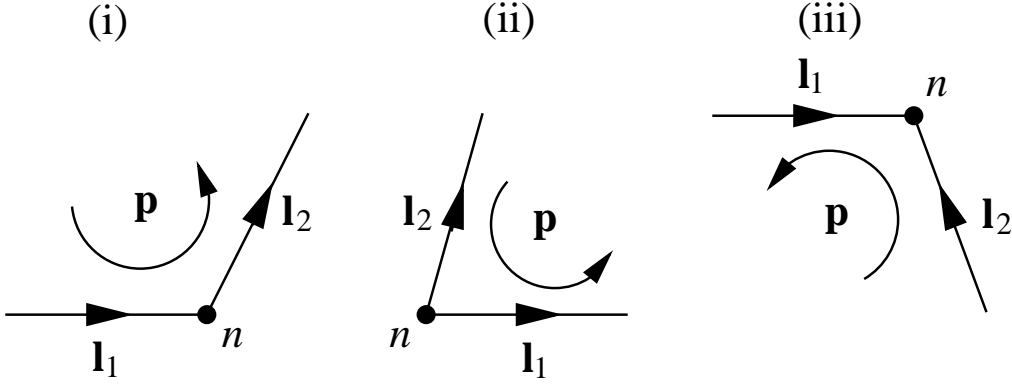,height=5cm}
  \end{center}
  \caption{For the purpose of doing the boson field integration,
    one must distinguish between the three types of corner shown
    here.}\label{fig:corners}
\end{figure}

\begin{enumerate}
\item Let the site $n$ be the end point of ${\bf l}_1$ and the
  starting point of ${\bf l}_2$, and let the orientations of ${\bf
    l}_1$ and ${\bf l}_2$ agree with that of $\partial{\bf p}$; see
  Figure \ref{fig:corners}(i).  Then, from (\ref{linkres}) we
  encounter the following integral:
  \begin{displaymath}
    T_{\bf p}(n) = \int\limits_{\mathbb C} d\varphi_{\bf p}(n)
    d\bar\varphi_{\bf p}(n) \, {\rm e}^{- m_{p} |\varphi_{\bf
        p}(n)|^2} {\rm e}^{b_{\bf p}^\dagger ({\bf l}_2) \varphi_{\bf
        p}(n)} | 0 \rangle \langle 0 | {\rm e}^{\bar \varphi_{\bf
        p}(n) b_{\bf p}({\bf l}_1)} \,.
  \end{displaymath}
  (The color and flavor indices play no important role here and have
  been omitted.)  Expanding the last two exponentials and doing the
  Gaussian integral we obtain
  \begin{displaymath}
    T_{\bf p}(n) = \sum_{k = 0}^\infty m_{p}^{-k} \frac{1}{k!} \big(
    b_{\bf p}^\dagger ({\bf l}_2) \big)^k | 0 \rangle \langle 0 |
    \big( b_{\bf p}({\bf l}_1) \big)^k \,.
  \end{displaymath}
  Since $(b^\dagger)^k | 0 \rangle / \sqrt{k!}$ is a normalized state,
  we see that what the operator $T_{\bf p}(n)$ does is to simply
  transfer bosons from link ${\bf l}_1$ to ${\bf l}_2$, while
  weighting each boson with the inverse of the mass $m_{p}$.  Note
  that this transfer happens for each color and flavor independently.
  A second subcase of the present type occurs when the orientations of
  ${\bf l}_1$ and ${\bf l}_2$ are opposite to that of $\partial{\bf
    p}$.  In that case the variables $\varphi_{\bf p}(n)$ and $\bar
  \varphi_{\bf p}(n)$ exchange their roles, but the operator that
  results on doing the integral is still the same.

\item Next, let the site $n$ be the {\it starting} point of both links
  ${\bf l}_1$ and ${\bf l}_2$, and let the orientation of $\partial
  {\bf p}$ agree (disagree) with that of ${\bf l}_1$ (resp.~${\bf
    l}_2$); see Figure \ref{fig:corners}(ii).  Then from
  (\ref{linkres}) we encounter the integral
  \begin{eqnarray*}
    C_{\bf p}(n) &=& \int\limits_{\mathbb C} d\varphi_{\bf p}(n)
    d\bar\varphi_{\bf p}(n) \, {\rm e}^{- m_{p} |\varphi_{\bf
        p}(n)|^2} \, {\rm e}^{b_{\bf p}^\dagger ({\bf l}_1)
      \varphi_{\bf p}(n) + b_{\bf p}^\dagger ({\bf l}_2)
      \bar\varphi_{\bf p}(n)} | 0 \rangle \langle 0 | \\ &=& \sum_{k =
      0}^\infty m_{p}^{-k} \frac{1}{k!} \big( b_{\bf p}^\dagger ({\bf
      l}_1) b_{\bf p}^\dagger ({\bf l}_2) \big)^k | 0 \rangle \langle
    0 | \,.
  \end{eqnarray*}
  Clearly, $C_{\bf p}(n)$ creates (an indefinite number of) boson
  pairs in normalized states, weighted by powers of the inverse mass.
  Again, this happens for each color and flavor independently.  If the
  orientation of ${\bf p}$ is reversed, $\varphi_{\bf p}(n)$ and $\bar
  \varphi_{\bf p}(n)$ again switch roles but the form of the operator
  $C_{\bf p}(n)$ remains unchanged.

\item Finally, let the site $n$ be the {\it end} point of both links
  ${\bf l}_1$ and ${\bf l}_2$ (and, although it makes no difference,
  let the orientation of $\partial{\bf p}$ agree with that of ${\bf
    l}_2$); see Figure \ref{fig:corners}(iii).  Then from
  (\ref{linkres}) we have
  \begin{eqnarray*}
    D_{\bf p}(n) &=& \int\limits_{\mathbb C} d\varphi_{\bf p}(n)
    d\bar\varphi_{\bf p}(n) \, {\rm e}^{- m_{p} |\varphi_{\bf p}
      (n)|^2} \, | 0 \rangle \langle 0 | {\rm e}^{\varphi_{\bf p} (n)
      b_{\bf p}({\bf l}_1) + \bar\varphi_{\bf p}(n) b_{\bf p}({\bf
        l}_2) } \\ &=& \sum_{k = 0}^\infty m_{p}^{-k} | 0 \rangle
    \langle 0 | \big( b_{\bf p}({\bf l}_1) b_{\bf p}({\bf l}_2)
    \big)^k / k!  \,.
  \end{eqnarray*}
  The operator $D_{\bf p}(n)$ now annihilates pairs of bosons (again
  in normalized states, and weighted by powers of the inverse mass;
  and if the orientation of ${\bf p}$ is reversed, the final result
  does not change).
\end{enumerate}

\subsection{Summary: dual theory}

In summary, doing the integral over the boson fields $\varphi, \bar
\varphi$ in the Fock space formalism produces operators that either
transfer bosons ($T$) from one link to a neighboring one, or
create/destroy boson pairs ($C/D$) on adjacent links.  In all these
processes, the plaquette label is conserved, and so are the color and
flavor quantum numbers.  (In the latter two processes, pairs of bosons
are always created/annihilated in identical color and flavor states.)
We are going to need names for the sets of transfer, creation and
destruction sites on $\partial {\bf p}$; let these be
$\mathfrak{t}_{\bf p}$, $\mathfrak{c}_{\bf p}$, and $\mathfrak{d}_{\bf
  p}$, respectively. \smallskip

\begin{figure}
  \begin{center}
  \epsfig{file=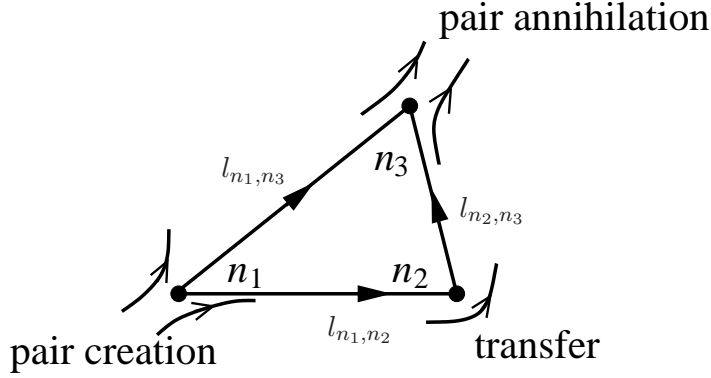,height=5cm}
  \begin{picture}(0,0)(300,0)
    \put(150,15){$l_{n_1,n_2}$}
    \put(110,75){$l_{n_1,n_3}$}
    \put(200,60){$l_{n_2,n_3}$}
  \end{picture}
  \end{center}
  \caption{Illustration of the operator $A_{\bf p}^{i,a}$ for the case of
    a triangular plaquette: at site $n_3$, bosons are annihilated in
    pairs; at site $n_2$, they are transferred according to the flow
    of the links; at site $n_1$, boson pairs are
    created.}\label{fig:dynamics}
\end{figure}

A simplification now comes from the fact that taking the trace picks
out the boson number conserving processes.  Using this to extract a
common boson mass weight factor, we conclude that the partition
function of the theory in the dual representation is
\begin{equation}\label{Zastrace}
  Z = {\rm Tr}_{{\cal V}_0} \left( \prod_{\pm \bf p} m_p^{- \hat
      N_{\bf p}} \prod_{a = 1}^{N_{\rm b}} \prod_{i = 1}^{N_{\rm c}}
    A_{\bf p}^{i,a} \right) \;,
\end{equation}
where $\hat N_{\bf p}$ is the operator counting the total number of
bosons associated with ${\bf p}$, and the operators $A_{\bf p}^{i,a}$
have the following structure:
\begin{displaymath}
  A_{\bf p}^{i,a} =
  \prod_{n \in \mathfrak{c}_{\bf p}} C_{\bf p}^{i,a}(n)
  \prod_{n \in \mathfrak{t}_{\bf p}} T_{\bf p}^{i,a}(n)
  \prod_{n \in \mathfrak{d}_{\bf p}} D_{\bf p}^{i,a}(n) \;.
\end{displaymath}
The left and right operators create resp.~destroy pairs of bosons in
identical states, while the middle ones simply transfer bosons.  (The
operator $A_{\bf p}^{i,a}$ is illustrated for the case of a triangular
plaquette in Figure \ref{fig:dynamics}.) The trace runs over the
linear space of states ${\cal V}_0$ spanned by all polynomials
\begin{displaymath}
  E_{{\bf p}_1 {\bf q}_1}^{a_1 b_1}({\bf l}_1) E_{{\bf p}_2 {\bf
      q}_2}^{a_2 b_2}({\bf l}_2) \cdots E_{{\bf p}_n {\bf q}_n}^{a_n
    b_n}({\bf l}_n) | 0 \rangle \quad \mbox{where} \quad E_{{\bf p
      q}}^{a b}({\bf l}) = \sum_{i = 1}^{N_{\rm c}} {b_{\bf p}^{i ,
      a}}^\dagger ({\bf l}) {b_{\bf q}^{i,b}}^\dagger ({\bf l}) \;,
\end{displaymath}
with ${\bf p}_j \in \Pi^+ ({\bf l}_j)$ and ${\bf q}_j \in \Pi^-({\bf
  l}_j)$. \smallskip

While the evaluation of this partition function remains a challenging
task in general, there are two conclusions we can draw immediately.
To arrive at the first one, we note that one may employ any complete
set of linearly independent states to compute the trace
(\ref{Zastrace}).  If the projection on the ${\rm U}(N_{\rm c})$
invariant sector ${\cal V}_0$ were absent, we could use a basis of
states labeled by occupation numbers $\{ n_{\bf p}^{i,a}({\bf l}) \}$,
but since the projection acts on the color degrees of freedom, this is
not possible unless $N_{\rm c} = 1$.  Nevertheless, the occupation
numbers {\it summed over colors},
\begin{displaymath}
  \sum_{i = 1}^{N_{\rm c}} n_{\bf p}^{i,a} ({\bf l}) \equiv n_{\bf
    p}^a ({\bf l}) \;,
\end{displaymath}
are still good quantum numbers.  Now reconsider the triangular
plaquette with corners $n_1, n_2, n_3$ in Figure \ref{fig:dynamics}.
Because the operator $T_{\bf p}^{i,a}(n_2)$ simply transfers bosons
without changing any of their quantum numbers, the contribution of a
state in ${\cal V}_0$ to the trace (\ref{Zastrace}) vanishes unless
\begin{displaymath}
  n_{\bf p}^{a} (l_{n_2,n_3}) = n_{\bf p}^{a} (l_{n_1,n_2}) \;.
\end{displaymath}
Similarly, because $C_{\bf p}^{i,a}(n_1)$ and $D_{\bf p}^{i,a}(n_3)$
create and destroy pairs of bosons in identical single--boson states,
the contribution of a state to (\ref{Zastrace}) vanishes unless
\begin{displaymath}
  n_{\bf p}^{a} (l_{n_1,n_2}) = n_{\bf p}^{a} (l_{n_1,n_3}) \;, \quad
  n_{\bf p}^{a} (l_{n_1,n_3}) = n_{\bf p}^{a} (l_{n_2,n_3}) \;.
\end{displaymath}
Altogether, this leads to the conclusion that the occupation numbers
(summed over colors) must be the same for each of the links in the
boundary chain of ${\bf p}$:
\begin{equation}\label{plaqonly}
  \sum_{i = 1}^{N_{\rm c}} n_{\bf p}^{i,a}({\bf l}) \equiv n_{\bf
    p}^{a} \quad (\mbox{independent of }{\bf l}) \;.
\end{equation}
By identical reasoning, this conclusion holds true not just for the
triangular plaquette ${\bf p}$ in Figure \ref{fig:dynamics}, but for
{\it any} plaquette on $\Lambda$. \smallskip

The second conclusion we can draw immediately results from the fact
that all states in ${\cal V}_0$ are created by the ${\rm U} (N_{\rm
  c})$ invariant pair operators $E_{\bf pq}^{ab}({\bf l})$, where
${\bf p} \in \Pi^+({\bf l})$ and ${\bf q} \in \Pi^-({\bf l})$.  If we
write ${\tilde n}_{\bf p} \equiv \sum_{a = 1}^{N_{\rm b}} n_{\bf p}^a$
for the occupation numbers summed over flavors (as well as colors),
this implies
\begin{equation}\label{fluxcons}
  \sum_{{\bf p} \in \Pi^+({\bf l})} {\tilde n}_{\bf p} - \sum_{{\bf q}
    \in \Pi^-({\bf l})} {\tilde n}_{\bf q} = 0
\end{equation}
for every link ${\bf l}$.  This equation is easy to interpret: it is
equivalent to the constraint on the field strength of the ${\mathbb
  Z}$--valued gauge field dual to the ${\rm U}(1)$ gauge field in
${\rm U}(N_{\rm c}) = {\rm SU} (N_{\rm c}) \times {\rm U}(1)$.  We
will come back to this in Section \ref{sec:U1_limit}.

\subsection{Wegner--Wilson loop}

Up to now we have concentrated on the partition function of the
induced gauge model.  To access its full physics, gauge--invariant
correlation functions such as the Wegner--Wilson loop must be
computed. \smallskip

A gauge--invariant correlation function that contains the
Wegner--Wilson loop and neatly fits into our formalism is constructed
as follows.  Let $C$ be any closed oriented contour of total length
$L$ on the $d$--dimensional cell complex $\Lambda$.  Viewing $C$ as a
one--chain, we can write it as a sum of oriented links: $C = {\bf l}_1
+ {\bf l}_2 + \ldots + {\bf l}_L$, where the end point of ${\bf l}_k$
is the starting point of ${\bf l}_{k + 1}$.  We then define $U(C)$ as
the ordered product
\begin{displaymath}
  U(C) = U({\bf l}_L) U({\bf l}_{L-1}) \cdots U({\bf l}_2) U({\bf
    l}_1) \;,
\end{displaymath}
and consider the expectation value $\langle \ldots \rangle$ of the
gauge--invariant function ${\rm Det} \big( 1 - \alpha_C \, U(C) \big)$
with respect to the statistical measure given by the partition
function of the lattice gauge theory.  This generates the
Wegner--Wilson loop $W(C)$ in the fundamental vector representation:
\begin{displaymath}
  \left\langle {\rm Det} \big( 1 - \alpha_C \, U(C) \big) \right\rangle
  = 1 - \alpha_C \, W(C) + {\cal O}(\alpha_C^2) \;, \qquad W(C) =
  \left\langle {\rm Tr} \, U(C) \right\rangle \;.
\end{displaymath}
We now set up the calculation of this correlation function, as
follows.  Let $\partial {\bf l}_k = n_{k+1} - n_k$, i.e.~$n_{k+1}$ is
the end point and $n_k$ the starting point of the link ${\bf l}_k$ in
the one--chain of $C$.  Introducing complex fermion fields $\psi_C,
\bar\psi_C$ associated with the sites visited by the contour $C$, we
modify the primordial action (\ref{bos_act}) by adding
\begin{displaymath}
  S_{C;\alpha_C} = \sum_{k=1}^L \Big( \bar\psi_C (n_k) \psi_C(n_k) -
  \alpha_C^{1/L} \bar\psi_C(n_{k+1}) U({\bf l}_k) \psi_C(n_k) \Big) \;.
\end{displaymath}
By the standard rules of fermionic integration we then have
\begin{displaymath}
  \left\langle {\rm Det} \big( 1 - \alpha_C \, U(C) \big)
  \right\rangle = \frac{\int {\rm e}^{- S_{\rm b} - S_{C;\alpha_C} }}
  {\int {\rm e}^{- S_{\rm b}}} \;.
\end{displaymath}
To pass to the dual description, we proceed in much the same manner as
before.  There are a few minor changes in order to take into account
the fermions, but these do not affect the general strategy: \smallskip

The one--link integral (\ref{onelink}) is modified for all links ${\bf
  l}$ in the contour $C$ by the additional presence of the fermion
fields.  To carry it out, we again ``quantize'', i.e.~we introduce
colorful fermion operators $f_C^i({\bf l})$ and ${f_C^i}^\dagger ({\bf
  l})$ which satisfy the canonical anticommutation relations and act
on Fock space with vacuum $| 0 \rangle$.  The ${\rm U}(N_{\rm c})$
transformation behavior is still determined by relative orientation,
now between ${\bf l}$ and $C$:
\begin{eqnarray*}
  T_{U({\bf l})} f_C^{i}({\bf l}) T_{U({\bf l})}^\dagger &=& \sum_j
  f_C^{j}({\bf l}) {\bar U}^{ji}({\bf l}) \quad \mbox{if} \quad C = +
  {\bf l} + \ldots \;, \\ T_{U({\bf l})} f_C^{i}({\bf l})
  T_{U({\bf l})}^\dagger &=& \sum_j f_C^{j}({\bf l}) U^{ji}({\bf
    l}) \quad \mbox{if} \quad C = - {\bf l} + \ldots \;.
\end{eqnarray*}
Coherent states work formally the same way for fermions and bosons; we
can still express ${\rm e}^{\bar\psi U \psi}$ as a matrix element of
$T_U$ between coherent states ${\rm e}^{f^\dagger \psi} | 0 \rangle$.
\smallskip

Doing the integral over $U({\bf l})$ with Haar measure, we still get a
Fock space projection operator $P_0({\bf l})$, although the ${\rm U}
(N_{\rm c})$ invariant subspace it projects onto, is enlarged for
every link ${\bf l}$ in $C$: there exist additional states, created by
boson--fermion pair operators
\begin{eqnarray}
  F_{C,{\bf q}}^{a}({\bf l}) &=& \sum_{i = 1}^{N_{\rm c}}
  {f_C^{i}}^\dagger ({\bf l}) {b_{\bf q}^{i,a}}^\dagger ({\bf l})
  \quad ({\bf q} \in \Pi^-({\bf l})) \quad \mbox{if} \quad C = +
  {\bf l} + \ldots, \quad \mbox{or} \label{bfpairs1} \\ F_{{\bf
      p},C}^{a}({\bf l}) &=& \sum_{i = 1}^{N_{\rm c}} {b_{\bf
      p}^{i,a}}^\dagger ({\bf l}) {f_C^{i}}^\dagger ({\bf l}) \quad
  ({\bf p} \in \Pi^+({\bf l})) \quad \mbox{if} \quad C = - {\bf l} +
  \ldots \;, \label{bfpairs2}
\end{eqnarray}
and by invariant pairs involving two fermions.  The latter, however,
will play no role ultimately, as we only want $W(C)$ in the {\it
  vector} representation (corresponding to a single--quark source), so
we refrain from writing them down.  The total set of ${\rm U}(N_{\rm
  c})$ invariant pair operators together with all their
(super--)commutators form a certain Lie superalgebra (of
$\mathfrak{gl}$ type), $\mathfrak{g}({\bf l})$. \smallskip

We denote the enlarged ${\rm U}(N_{\rm c})$ invariant subspace
associated with a link ${\bf l}$ on $C$ by ${\cal V}_0^\prime ({\bf
  l})$.  Howe's theorem \cite{Howe} on the multiplicity--free action
of dual pairs in Fock space [the dual pair now being ${\rm U}(N_{\rm
  c})$ with $\mathfrak{g}({\bf l})$] still holds in the present
generalized setting, so ${\cal V}_0^\prime ({\bf l})$ is still
obtained by acting with all ${\rm U}(N_{\rm c})$ invariant pair
operators on the vacuum $| 0 \rangle$.  \smallskip

Integration over the fermion sources $\psi, \bar\psi$ works the same
way as for bosons (Section \ref{sec:integrate}).  The final result of
doing the integration (for fixed color quantum number $i$) is an
operator which is denoted by $A_C^i$ and is defined as follows.
$A_C^i$ is a product over all sites $n$ visited by the contour $C$.
In the same way as was shown in detail for the bosons in Section
\ref{sec:integrate}, the definition of each factor $A_C^i(n)$ depends
on the a priori orientation of the links in $C$ that begin or end on
$n$.  For sites $n$ of type (i) (Figure \ref{fig:corners}), $A_C^i$
transfers fermions; for sites of type (ii) it creates pairs of
fermions; and for sites of type (iii) it annihilates pairs of
fermions. \smallskip

All this discussion is summarized in the following final formula:
\begin{equation}\label{genfunc}
  \left \langle {\rm Det} \big( 1 - \alpha_C \, U(C) \big) \right
  \rangle = Z^{-1} {\rm Tr}_{{\cal V}_{0;C}}^{\vphantom{\dagger}}
  \left( \alpha_C^{\hat N_{\rm f} / L} \prod_{i = 1}^{N_{\rm c}} A_C^i
    \prod_{\pm {\bf p}} \alpha_p^{\hat N_{\bf p} / L_p} \prod_{a =
      1}^{N_{\rm b}} A_{\bf p}^{i,a} \right) \;,
\end{equation}
where $\hat N_{\rm f}$ counts the total number of fermions, and ${\cal
  V}_{0;C}$ is the tensor product of the spaces ${\cal V}_0^ \prime
({\bf l})$ (resp.~${\cal V}_0({\bf l})$) for ${\bf l}$ contained
(resp.~not contained) in $C$.  The normalization $Z^{-1}$ is given by
(\ref{Zastrace}).

\subsection{Abelian limit: $G = {\rm U}(1)$}\label{sec:U1_limit}

In this subsection we briefly comment on the special case $N_{\rm c} =
1$, adopting the minimal model $N_{\rm b} = N_{\rm c} = 1$.  In that
model, both the color and flavor degrees of freedom are absent and
drastic simplifications occur. \smallskip

We begin with the expression for the partition function
(\ref{Zastrace}) as a trace over ${\cal V}_0$.  The evaluation of the
trace is simplified by the observation that the states of ${\cal V}_0$
for $N_{\rm c} = 1$ are in one--to--one correspondence with sets of
boson occupation numbers.  Thus for $N_{\rm b} = 1$ they are labeled
by $n_{\bf p} ({\bf l}) \in {\mathbb N} \cup \{ 0 \}$.  By
Eq.~(\ref{plaqonly}), these occupation numbers are ${\bf
  l}$--independent: $n_{\bf p}({\bf l}) \equiv {\tilde n}_{\bf p}$.
(We put the tilda here to reserve the name $n_{\bf p}$ for another set
of integer variables to be introduced presently.)  They also satisfy
the constraint (\ref{fluxcons}). \smallskip

The best way to deal with the constraint is to switch variables.
Recall that in the full set of occupation numbers, $\{ {\tilde n}_{\bf
  p} \}$, every $2$--cell ${\bf p}$ occurs twice: once with its proper
orientation $(+{\bf p})$, and once with its orientation reversed
$(-{\bf p})$.  We now switch variables from ${\tilde n}_{\pm{\bf p}}
\in {\mathbb N}\cup \{ 0 \}$ to
\begin{displaymath}
  n_{\bf p} = {\tilde n}_{\bf p} - {\tilde n}_{-{\bf p}} \in {\mathbb
    Z} \;, \quad \mbox{and} \quad l_{\bf p} = {\tilde n}_{\bf p} +
  {\tilde n}_{- {\bf p}} = |n_{\bf p}|, |n_{\bf p}| + 2, |n_{\bf p}| +
  4, \ldots \;.
\end{displaymath}
If we view ${\bf p} \mapsto n_{\bf p}$ as a $2$--chain $n$ on the
$d$--dimensional complex $\Lambda$, the constraint (\ref{fluxcons})
simply says that $n$ is closed: $\partial n = 0$.  Indeed, the
question whether ${\bf p}$ belongs to $\Pi^+({\bf l})$ or $\Pi^-({\bf
  l})$, and hence what sign is attributed to $\tilde n_{\bf p}$ in
(\ref{fluxcons}), is decided by comparing orientations.  This sign
information is concisely encoded in the boundary operator $\partial$ :
$2$--chains $\to$ $1$--chains. \smallskip

The variables $l_{\bf p}$, on the other hand, decouple from the
problem; the sum over each of them is a geometric series:
\begin{displaymath}
  \sum_{l_{\bf p} \in |n_{\bf p}| + 2({\mathbb N} \cup \{ 0 \})} m_p^{
    - L_p \times l_{\bf p} } = (1 - \alpha_p^2)^{-1} \alpha_p^{|n_{\bf
      p}|} \;,
\end{displaymath}
where $\alpha_p = 1 / m_p^{L_p}$ as before.  The power $L_p$ appears
because $\hat N_{\bf p}$ counts the total number of bosons associated
with ${\bf p}$ and the $L_p$ links about ${\bf p}$ all have the same
boson occupation number [Eq.~(\ref{plaqonly})].  Thus the partition
function (\ref{Zastrace}) takes the final form
\begin{displaymath}
  Z = \prod_p (1 - \alpha_p^2)^{-1} \sum_{n \, : \, \partial n = 0}
  \prod_{\bf p} \alpha_p^{|n_{\bf p}|} \;,
\end{displaymath}
which coincides with the expression that resulted from the standard
abelian duality transform; see Section \ref{sec:abelian}. \smallskip

Turning to $W(C)$, we observe that the same argument that gave
constant boson occupation numbers on plaquettes, leads to constant
fermion occupation numbers along the contour $C$.  To extract $W(C)$
from the generating function (\ref{genfunc}), we must place exactly
one fermion on each link ${\bf l}$ in $C$.  States in the ${\rm U}(1)$
invariant subspace ${\cal V}_0^\prime ({\bf l})$ are created by the
invariant pair operators (\ref{bfpairs1},\ref{bfpairs2}), which
implies that the creation of a fermion along $C$ is always accompanied
by the creation of exactly one boson.  Therefore, along $C$ the
relation (\ref{fluxcons}) is modified to
\begin{displaymath}
  \epsilon + \sum_{{\bf p} \in \Pi^+({\bf l})} {\tilde n}_{\bf p} -
  \sum_{{\bf q} \in \Pi^-({\bf l})} {\tilde n}_{\bf q} = 0 \;,
\end{displaymath}
where $\epsilon = +1$ if the orientation of $C$ agrees with that of
${\bf l}$, and $\epsilon = -1$ otherwise.  If we switch from the
variables ${\tilde n}_{\bf p}, {\tilde n}_{-{\bf p}}$ to $n_{\bf p},
l_{\bf p}$, this constraint becomes $\partial n = - C$.  Solving the
constraint by setting $n = \partial a - S$ with $\partial S = C$, we
obtain $W(C)$ in the form (\ref{dualWW}).

\section{Discussion}\label{sec:discuss}

The distinctive feature of the class of lattice models for ${\rm U}
(N_{\rm c})$ gluodynamics introduced here, is that they are induced
from a pre--theory with $N_{{\rm b}/{\rm f}}$ species of local bosons
and/or fermions.  The statistical measure of these lattice gauge
models is a product over elementary plaquettes, with each factor being
a ratio of determinants. \smallskip

The boson induced models have a critical point at unit mass, which
allows a continuum limit to be taken.  Our careful study in $d = 1 +
1$ dimensions showed that, if $N_{\rm b} \ge N_{\rm c} + 1$, this
continuum limit is ${\rm U}(N_{\rm c})$ Yang--Mills theory, with a
specific ratio of the ${\rm U}(1)$ and ${\rm SU}(N_{\rm c})$
couplings.  The ratio goes to unity for $N_{\rm b} \to \infty$.  In
contrast, the continuum limit for $N_{\rm b} = N_{\rm c}$ is {\it not}
Yang--Mills but an unusual theory, which exists as a
(super--)renormalizable quantum theory because the Cauchy distribution
on $\mathfrak{u}(N_{\rm c})$ converges under convolution to a stable
family of distributions distinct from the heat kernel family.
\smallskip

Going up in dimension enhances the collectivity of the gauge field (by
increasing the number of transverse gluons) and thus works in favor of
``universality''.  Therefore, whenever our model induces ${\rm U}
(N_{\rm c})$ Yang--Mills theory in two dimensions, we expect it to do
so in higher dimensions, at least generically. \smallskip

Although we concentrated on the special case of the gauge group being
${\rm U}(N_{\rm c})$, a very similar treatment is possible for all
classical compact Lie groups.  Proposition (\ref{thm1}), which is the
mathematical basis for the existence of a continuum limit, carries
over to the normalized distributions ${\rm Det}(1-\alpha U)^{- N_{\rm
    b}} dU$ on ${\rm Sp}(2N_{\rm c})$ and ${\rm SO}(2N_{\rm c})$, with
thresholds $N_{\rm b} \ge 2N_{\rm c} + 1$ and $N_{\rm b} \ge 2 N_{\rm
  c} - 1$, respectively. \smallskip

In the second part of the paper we subjected the boson induced gauge
model to a duality transformation (a variant of the color--flavor
transformation), which demonstrably reduces to standard abelian
duality for $G = {\rm U}(1)$.  The partition function of the theory in
the dual formulation is the trace of a color-- and flavor--diagonal
operator $\prod A_{\bf p}^{i,a}$ acting on a tensor product of modules
${\cal V}_0({\bf l})$ generated by quadratic ${\rm U}(N_{\rm c})$
invariants on lattice links ${\bf l}$.  Again, the dual formulation is
not restricted to ${\rm U}(N_{\rm c})$ but exists for other classical
compact gauge groups as well.  (This is because all of the compact Lie
groups ${\rm U}(N_{\rm c})$, ${\rm Sp}(2N_{\rm c})$ and ${\rm O}
(N_{\rm c})$ are placed in Howe duality with corresponding families of
noncompact Lie groups.  For ${\rm SU}(N_{\rm c})$ and ${\rm SO}
(N_{\rm c})$ the duality transform is more complicated
\cite{Budczies:01,boristilo1,bnsz} due to the existence of
nonquadratic invariants of ``baryon'' type .)  \smallskip

We do not know at present how far one can push the analysis of the
dual theory.  We believe it to be possible to develop a combinatorial
approach to handle the operator $\prod A_{\bf p}^{i,a}$, but the
details have not been worked out yet.  A more refined understanding of
the asymptotics of the modules ${\cal V}_0({\bf l})$ at large boson
number is also needed.  Although they are generated by quadratic
invariants, they are not {\it freely} generated in the range $N_{\rm
  b} > N_{\rm c}$, as is evident from the result $c_0(\alpha) \sim
(1-\alpha)^{-2N_{\rm b} N_{\rm c} + N_{\rm c}^2}$ [see (\ref{thm2c})]
for the ${\rm U}(N_{\rm b}, N_{\rm b})$ character $c_0 (\alpha) = {\rm
  Tr}_{{\cal V}_0} \alpha^{\hat N}$.  \smallskip

We do not understand at present exactly how the Howe duality transform
connects with other recent proposals, such as the quantum gravity
formulation of ${\rm SU}(2)_{d = 3}$ Yang--Mills theory in
\cite{diakonov}.  Also, to bring our induced gauge models closer to
continuum gauge theories of current interest, one may ask whether they
can be extended to models with robust supersymmetry on the lattice
\cite{kaplan}.  Another natural question to ask is whether one can
construct a coherent--state resolution of the identity operator on the
modules ${\cal V}_0$, and by adding an extra time direction to
four--dimensional space--time pass to a five--dimensional continuum
theory (of topological gravity?) via the coherent--state path integral
method. \smallskip

Let us finish by commenting on just the issue of existence of a
coherent--state resolution.  Consider the Howe pair ${\rm U}(N_{\rm
  c}) \times {\rm U}(N_{\rm b}, N_{\rm b})$ acting in a Fock space
${\cal V}$ with $2 N_{\rm b} N_{\rm c}$ species of bosons, as
described in Section \ref{sec:howe}.  We are interested in the ${\rm
  U}(N_{\rm c})$ invariant sector ${\cal V}_0$.  In particular, we
would like a coherent--state integral representation for the invariant
Hermitian scalar product on ${\cal V}_0$.  For $N_{\rm c} \ge 2 N_{\rm
  b}$ this is a standard problem with the following standard solution.
Let $M$ be the Hermitian symmetric space formed by dividing ${\rm U}
(N_{\rm b},N_{\rm b})$ by its maximal compact subgroup.  $M$ is
modeled by complex $N_{\rm b} \times N_{\rm b}$ matrices $Z$ with
noncompact domain $Z^\dagger Z < 1$; and $g \equiv \begin{pmatrix} A
  &B \\ C &D \end{pmatrix} \in {\rm U} (N_{\rm b},N_{\rm b})$ acts on
the points $Z$ of $M$ by
\begin{displaymath}
  g \cdot Z = (DZ + C) (BZ + A)^{-1} \;.
\end{displaymath}
The coherent--state expression for any $\varphi \in {\cal V}_0$ is a
holomorphic section $\varphi(Z)$, transforming under $g \in {\rm
  U}(N_{\rm b},N_{\rm b})$ as
\begin{displaymath}
  \left( {\cal D}^{N_{\rm c}}(g) \varphi \right)(Z) = {\rm Det}^{-
    N_{\rm c}} (D - Z B) \, \varphi \left( g^{-1} \cdot Z \right) \;.
\end{displaymath}
The invariant Hermitian scalar product of two sections $\varphi_1$
and $\varphi_2$ is
\begin{displaymath}
  \left\langle \varphi_1 | \varphi_2 \right\rangle = \int\limits_{Z
    Z^\dagger < 1} \overline{\varphi_1(Z)} \varphi_2(Z) \, {\rm Det}
  ^{N_{\rm c} - 2 N_{\rm b}}(1 - Z^\dagger Z) \, dZ d\bar Z \;,
\end{displaymath}
where $dZ d\bar Z$ is a flat density (suitably normalized), and
invariance means
\begin{displaymath}
  \langle \varphi_1 | \varphi_2 \rangle = \langle {\cal D}^{N_{\rm c}}
  (g) \varphi_1 | {\cal D}^{N_{\rm c}}(g) \varphi_2 \rangle \quad
  \mbox{for} \quad g \in {\rm U}(N_{\rm b},N_{\rm b}) \;.
\end{displaymath}

All this makes perfect sense as long as $N_{\rm c}$ is big enough (or
$N_{\rm b}$ small enough).  However, when $N_{\rm c}$ is decreased
below $2 N_{\rm b}$, the density ${\rm Det}^{N_{\rm c} - 2 N_{\rm
    b}}(1 - Z^\dagger Z) \, dZ d\bar Z$ becomes singular at the
boundary of the symmetric domain $Z^\dagger Z < 1$, and the
coherent--state expression for the Hermitian scalar product ceases to
exist in the form given.  This does not mean that it ceases to exist
altogether.  Indeed, for the case $N_{\rm b} = N_{\rm c}$ we can
easily see how to fix the problem.  The singularity at the boundary
indicates that the proper measure to use {\it is concentrated in the
  boundary}.  Now, the boundary of the $Z$--model for $M$ always
contains the unitary group ${\rm U} (N_{\rm b})$ (the set of solutions
of $Z^\dagger Z = 1$) as a ${\rm U}(N_{\rm b}, N_{\rm b})$ orbit,
i.e.~if $U$ is in ${\rm U}(N_{\rm b})$, then so is its image $g^{-1}
\cdot U = (D - UB)^{ -1} (UA - C)$.  A straightforward computation
shows that the Haar measure $dU$ on ${\rm U}(N_{\rm b})$ transforms
under the action of $g^{-1}$ as ${g^{-1}}^\ast (dU) = | {\rm Det}(D -
UB) |^{-2 N_{\rm b}} dU$.  For the case $N_{\rm c} = N_{\rm b} \equiv
N$ (and only in that case) the factor multiplying $dU$ is canceled by
the multiplier in the transformation law for sections $\varphi(Z)$, so
that the boundary integral
\begin{displaymath}
  \left\langle \varphi_1 | \varphi_2 \right\rangle = \int_{
    {\rm U}(N)} \overline{\varphi_1(U)} \varphi_2(U) \, dU
\end{displaymath}
is invariant, and (by uniqueness) is the coherent--state integral
representation of the Hermitian scalar product on ${\cal V}_0$.  This
implies that, while the sections $\varphi (z)$ are sections on the
$2N^2$--dimensional symmetric domain $M$, the complete information
about them in the special case $N_{\rm b} = N_{\rm c} \equiv N$ at
hand is already encoded in the values they take on approaching the
$N^2$--dimensional part ${\rm U}(N) \subset \partial M$ of the
boundary $\partial M$. \smallskip

We do not understand the details of the analogous construction of an
invariant boundary measure for $N_{\rm b} \not= N_{\rm c} < 2N_{\rm
  b}$, although we know on general grounds that such a construction
must exist.  Having a detailed understanding of that construction
would enable us to extend the color--flavor transformation to the
whole domain $N_{\rm c} < 2 N_{\rm b}$.

\appendix
\section{Appendix: Calculations for the One--Plaquette Model}
\label{appendixA}

\subsection{Fermion induced model}\label{sec:appferms}

We compute the expectation value of the Wilson loop, $W(\alpha_{\rm
  f},0)$, for the fermion induced model on a lattice consisting of a
single plaquette (Section \ref{sec:0d}). \smallskip

Doing the same steps as in the proof of statement (\ref{thm1}),
Section \ref{sec:proof}, we obtain
\begin{displaymath}
  W(\alpha_{\rm f},0) = N_{\rm c} \langle {\rm e}^{{\rm i} \theta_1}
  \rangle_{\alpha_{\rm f}} / \langle 1 \rangle_{ \alpha_{\rm f}} \;,
\end{displaymath}
where
\begin{displaymath}
  \langle F \rangle_{\alpha} = \int\limits_{[0,2\pi]^{N_{\rm c}}}
  F({\rm e}^{{\rm i}\theta_1}, \ldots ) \prod_{j=1}^{N_{\rm c}} | 1 -
  \alpha {\rm e}^{{\rm i} \theta_j} |^{2N_{\rm f}} \, \prod_{k < l} |
  {\rm e}^{{\rm i}\theta_k} - {\rm e}^{{\rm i}\theta_l} |^2 \,
  d\theta_1 \cdots d\theta_{N_{\rm c}} \;.
\end{displaymath}
Putting $z_k = {\rm e}^{{\rm i}\theta_k}$, we rewrite this as a
multiple complex contour integral over the unit circle ${\rm U}(1)
\subset {\mathbb C}$ :
\begin{equation}\label{contour}
  \langle F \rangle_\alpha = (-{\rm i})^{N_{\rm c}^2}
  \oint\limits_{{\rm U}(1)^{N_{\rm c}}} F( z_1, \ldots) \prod_{j =
    1}^{N_{\rm c}} \frac{ (z_j - \alpha)^{N_{\rm f}} (1 - \alpha
    z_j)^{N_{\rm f}} } { z_j^{N_{\rm c} + N_{\rm f}}} \prod_{k < l}
  (z_k - z_l)^2 dz_1 \cdots dz_{N_{\rm c}} \;.
\end{equation}
The integrand is holomorphic on ${\mathbb C} \setminus \{ 0 \}$ for
each of the variables $z_k$, which suggests evaluating the integral by
contracting all integration contours to zero.  There is a pole of
order $N_{\rm c} + N_{\rm f}$ at zero in each variable of the
normalization integral $(F \equiv 1)$.  In the numerator $\langle z_1
\rangle_\alpha$ of the expectation value, the order of the pole in the
distinguished variable $z_1$ is reduced by one.  Thus, by contracting
the contours and evaluating the residues at zero, we obtain
\begin{displaymath}
  W(\alpha_{\rm f},0) = N_{\rm c} (N_{\rm c} + N_{\rm f} - 1)
  \times \frac{ \partial_{z_1}^{N_{\rm c} + N_{\rm f}-2}
    \partial_{z_2}^{N_{\rm c} + N_{\rm f}-1} \cdots
    \partial_{z_{N_{\rm c}}}^{N_{\rm c} + N_{\rm f}-1} f(z_1, \ldots,
    z_{N_{\rm c}}; \alpha_{\rm f})}{ \partial_{z_1}^{N_{\rm c} + N_{\rm
        f}-1} \cdots \partial_{z_{N_{\rm c}}}^{N_{\rm c} + N_{\rm
        f}-1} f(z_1, \ldots ,z_{N_{\rm c}}; \alpha_{\rm f})}
  \Bigg|_{z_1 = \ldots = z_{N_{\rm c}} = 0} \;,
\end{displaymath}
where $f$ is the function
\begin{displaymath}
  f(z_1, \ldots, z_{N_{\rm c}}; \alpha) = \prod_{j = 1}^{N_{\rm c}}
  (z_j - \alpha)^{N_{\rm f}} (z_j - \alpha^{-1} )^{N_{\rm f}} \prod_{k <
    l} (z_k - z_l)^2 \;.
\end{displaymath}
For small values of $N_{\rm c}$ and $N_{\rm f}$ the derivatives are
easy to evaluate using computer algebra.  The results obtained in this
way for $N_{\rm c} = 3$ and $N_{\rm f} = 1, \ldots, 6$ were shown in
Figure \ref{fig:fermions}.

\subsection{Boson induced model}\label{sec:appbos}

Passing from the fermion induced model to its bosonic analog amounts
to doing an analytic continuation, from positive integers $N_{\rm f}$
to negative integers $-N_{\rm b}$.  Given Eq.~(\ref{contour}) of the
previous subsection, this continuation yields $W(0,\alpha_{\rm b}) =
N_{\rm c} \langle z_1 \rangle_{\alpha_{\rm b}} / \langle 1
\rangle_{\alpha_{\rm b}}$ with
\begin{equation}\label{contourb}
  \langle F \rangle_\alpha = (-{\rm i})^{N_{\rm c}^2}
  \oint\limits_{{\rm U}(1)^{N_{\rm c}}} F( z_1, \ldots) \frac{
    \prod_{k < l} (z_k - z_l)^2 } {\prod_{j = 1}^{N_{\rm c}}
    z_j^{N_{\rm c} - N_{\rm b}} (z_j - \alpha)^{N_{\rm b}} (1 - \alpha
    z_j)^{N_{\rm b}} } \, dz_1 \cdots dz_{N_{\rm c}} \;.
\end{equation}
The strategy again is to contract all integration contours to zero.
As compared with the fermionic case, we now encounter $N_{\rm
  b}$--fold poles at $z_j = \alpha_{\rm b}$ in addition to the poles
at $z_j = 0$ that occur when $N_{\rm b}$ is less than $N_{\rm c}$.  We
separately consider the three cases $N_{\rm b} > N_{\rm c}$, $N_{\rm
  b} = N_{\rm c}$, and $N_{\rm b} < N_{\rm c}$.

\paragraph{$N_{\rm b} > N_{\rm c}$ :}
In this case the integrand has an $N_{\rm b}$--fold pole at $z_j =
\alpha_{\rm b}$ for $j = 1, \ldots, N_{\rm c}$ and no poles at $z_j =
0$.  Application of the residue theorem yields
\begin{displaymath}
  W(0,\alpha_{\rm b}) = N_{\rm c} \frac{\partial^{N_{\rm b} -
      1}_{z_1} \cdots \partial^{N_{\rm b} - 1}_{z_{N_{\rm c}}} z_1
    f(z_1, \ldots , z_{N_{\rm c}}; \alpha_{\rm b})} {\partial^{N_{\rm
        b} - 1}_{z_1} \cdots \partial^{N_{\rm b} - 1}_{z_{N_{\rm c}}}
    f(z_1, \ldots , z_{N_{\rm c}}; \alpha_{\rm b})} \Bigg|_{ z_1 =
    \ldots = z_{N_{\rm c}} = \alpha_{\rm b}} \;,
\end{displaymath}
with $f$ being the function
\begin{displaymath}
  f(z_1, \ldots , z_{N_{\rm c}}; \alpha) = \prod_{j = 1}^{N_{\rm c}}
  \frac{z_j^{N_{\rm b} - N_{\rm c}}} {(z_j - \alpha^{-1})^{N_{\rm b}}}
  \prod_{k < l} (z_k - z_l)^2 \;.
\end{displaymath}
Again, for small values of $N_{\rm c}$ and $N_{\rm b}$ we have
employed computer algebra to evaluate the derivatives.  The results
for $N_{\rm c} = 3$ and $N_{\rm b} = 1, \ldots, 6$ were shown in
Figure \ref{fig:bosons}.

\paragraph{$N_{\rm b} = N_{\rm c}$ :}
As before, the integrand has apparent poles (now of order $N_{\rm b} =
N_{\rm c}$) at $z_j = \alpha_{\rm b}$ for $j = 1, \ldots, N_{\rm c}$
and no poles at $z_j = 0$.  The previous formula for $W(0,\alpha_{\rm
  b})$ is still valid, but now it is possible to go further and get an
answer in closed form.  For that purpose we note that the polynomial
$\prod_{k < l} (z_k - z_l)^2$ has degree $N_{\rm c}(N_{\rm c} - 1)$
and can be viewed as the square of a Vandermonde determinant, which in
turn is a sum over permutations:
\begin{eqnarray*}
  \prod_{k < l} (z_k - z_l)^2 &=& \left( \sum_{\pi \in S_{N_{\rm c}}}
    {\rm sgn}(\pi) \, (z_1 - \alpha_{\rm b})^{\pi(1) - 1} \cdots
    (z_{N_{\rm c}} - \alpha_{\rm b})^{\pi(N_{\rm c}) - 1} \right)^2 \\
  &=& \sum_{\pi, \pi^\prime \in S_{N_{\rm c}}} {\rm sgn}(\pi
  \pi^\prime) \, (z_1 - \alpha_{\rm b})^{\pi(1) + \pi^\prime(1) - 2}
  \cdots (z_{N_{\rm c}} - \alpha_{\rm b})^{\pi(N_{\rm c}) +
    \pi^\prime(N_{\rm c}) - 2} \;.
\end{eqnarray*}
Inserting this double sum into the integral we see that almost all of
its terms give rise to an integrand which is actually holomorphic at
$z = \alpha_{\rm b}$ (and hence holomorphic everywhere in the unit
disc) for at least one of the integration variables.  A true
singularity in every one of the variables occurs only for the terms
with equal exponents,
\begin{displaymath}
  \pm (z_1 - \alpha_{\rm b})^{N_{\rm c} - 1} \cdots (z_{N_{\rm c}} -
  \alpha_{\rm b})^{N_{\rm c} - 1} \;,
\end{displaymath}
which arise from the permutations with $\pi(j) + \pi^\prime(j) =
N_{\rm c} + 1$ (for all $j$).  There exist $N_{\rm c} !$ of such terms
in the double sum, and these are the only ones that give a nonzero
contribution to the integral.  Looking at (\ref{contourb}) we see that
the resulting poles are simple in each variable, so
\begin{displaymath}
  \langle 1 \rangle_{\alpha_{\rm b}} = (2\pi)^{N_{\rm c}} N_{\rm c}!
  \left( 1 - \alpha_{\rm b}^2 \right)^{- N_{\rm c}^2} \;,
\end{displaymath}
and
\begin{displaymath}
  W(0,\alpha_{\rm b}) = N_{\rm c} \langle z_1 \rangle_{\alpha_{\rm
      b}} / \langle 1 \rangle_{\alpha_{\rm b}} = N_{\rm c} \alpha_{\rm
    b} \;,
\end{displaymath}
which is the exact result quoted in the main text.

\paragraph{$N_{\rm b} < N_{\rm c}$ :}
In this case there exist poles inside the unit circle both at $z_j =
\alpha_{\rm b}$ \emph{and} at $z_j = 0$.  Hence we must investigate the
analytic structure of the integrand when $q$ variables are sent to $z
= \alpha_{\rm b}$ and $N_{\rm c} - q$ variables to $z = 0$.  The poles
at these locations compete with the zeroes from the numerator
$\prod_{k < l} (z_k - z_l)^2$.  By power counting one easily sees that
the integrand at the nominal singularities is actually holomorphic in
at least one of the variables unless $q = N_{\rm b}$.  In other words,
nonzero contributions to the integral come only from sending $N_{\rm
  b}$ variables to $z = \alpha_{\rm b}$, and the remaining $N_{\rm c} -
N_{\rm b}$ variables to $z = 0$.  Adapting the Vandermonde determinant
expansion of $\prod_{k < l} (z_k - z_l)^2$ to such a location, say
$z_1 = \ldots = z_{N_{ \rm b}} = \alpha_{\rm b}$ and $z_{N_{\rm b}+1} =
\ldots = z_{N_{\rm c}} = 0$, we see that the corresponding Laurent
series starts as follows:
\begin{displaymath}
  \frac{ \prod_{k < l} (z_k - z_l)^2 } {\prod_{j = 1}^{N_{\rm c}} (z_j
    - \alpha_{\rm b})^{N_{\rm b}} z_j^{N_{\rm c} - N_{\rm b}} } =
  N_{\rm b} ! (N_{\rm c} - N_{\rm b}) ! \prod_{j = 1}^{N_{\rm b}} (z_j
  - \alpha_{\rm b})^{- 1} \prod_{j = N_{\rm b} + 1}^{N_{\rm c}} z_j^{-
    1} + \ldots \;,
\end{displaymath}
where the terms omitted are regular in at least one variable.
\smallskip

Since there exist $\binom{N_{\rm c}}{N_{\rm b}}$ ways to divide
$N_{\rm c}$ objects into two sets of $N_{\rm b}$ resp.~$N_{\rm c} -
N_{\rm b}$ objects, and $\binom{N_{\rm c}}{N_{\rm b}} \times N_{\rm
  b}!  (N_{\rm c} - N_{\rm b})! = N_{\rm c}!$, the normalization
integral is
\begin{displaymath}
  \langle 1 \rangle_{\alpha_{\rm b}} = (2\pi)^{N_{\rm c}} N_{\rm c}!
  \left( 1 - \alpha_{\rm b}^2 \right)^{- N_{\rm b}^2} \;.
\end{displaymath}
In the numerator $\langle z_1 \rangle_{\alpha_{\rm b}}$ a nonzero
contribution results only when the distinguished variable $z_1$ is
sent to $z = \alpha_{\rm b}$ (as opposed to $z = 0$).  This happens in
a fraction $N_{\rm b} / N_{\rm c}$ of all cases, and so the Wilson
loop expectation value is reduced by this very factor:
\begin{displaymath}
  W(0,\alpha_{\rm b}) = N_{\rm c} \frac{ \langle z_1
    \rangle_{\alpha_{\rm b}} } { \langle 1 \rangle_{\alpha_{\rm b}} }
  = N_{\rm c} \frac{N_{\rm b}}{N_{\rm c}} \alpha_{\rm b} = N_{\rm b}
  \alpha_{\rm b} \;.
\end{displaymath}

\bibliographystyle{../my2}

\begin{thebibliography}{10}
\expandafter\ifx\csname url\endcsname\relax
  \def\url#1{\texttt{#1}}\fi
\expandafter\ifx\csname urlprefix\endcsname\relax\def\urlprefix{URL }\fi
\providecommand{\eprint}[2][]{\url{#2}}

\bibitem{hamber} H.W.~Hamber, {\it Lattice gauge theories at large
    $N_{\rm f}$}, Phys. Lett.~\textbf{126B} (1983) 471--474

\bibitem{bander} M.~Bander, {\it Equivalence of lattice gauge theories
    and spin theories}, Phys. Lett.~\textbf{126B} (1983) 463--466

\bibitem{hasenfratz} A.~Hasenfratz and P.~Hasenfratz, {\it The
    equivalence of the ${\rm SU}(N)$ Yang--Mills theory with a purely
    fermionic model}, Phys. Lett.~\textbf{297B} (1992) 166--170

\bibitem{indQCD} V.A.~Kazakov and A.A.~Migdal, {\it Induced QCD at
    large $N$}, Nucl. Phys. \textbf{B 397} (1993) 214--238;
  arXiv:hep-th/9206015

\bibitem{boristilo2} B.~Schlittgen and T.~Wettig, \emph{The
    color--flavor transformation and lattice QCD}, Nucl. Phys. {\bf B}
  (Proc. Suppl.) {\bf 119} (2003) 956--961; arXiv:hep-lat/0208044

\bibitem{NS:02} S.~Nonnenmacher and Y.~Shnir, \emph{The color-flavor
    transformation of induced QCD}, \\ arXiv:hep-lat/0210002

\bibitem{wegner} F.~Wegner, {\it Duality in generalized Ising models
    and phase transitions without local order parameter},
  J.~Math.~Phys.~{\bf 12} (1971) 2259--2272

\bibitem{polyakov} A.M.~Polyakov, {\it Compact gauge fields and the
    infrared catastrophe}, Phys.~Lett.~\textbf{59B} (1975) 82--84

\bibitem{banks} T.~Banks, J.~Kogut, and R.~Myerson, {\it Phase
    transitions in abelian lattice gauge theories}, Nucl. Phys.
  \textbf{B 129} (1977) 493--510

\bibitem{guth} A.~Guth, {\it Existence proof of a nonconfining phase
    in four--dimensional ${\rm U}(1)$ lattice gauge theory},
  Phys.~Rev.~\textbf{D 21} (1980) 2291--2307

\bibitem{juergtom} J.~Fr\"ohlich and T.~Spencer, {\it Massless phases
    and symmetry restoration in abelian gauge theories and spin
    systems}, Commun.~Math.~Phys.~{\bf 83} (1982) 411--454

\bibitem{kogut83} J.B.~Kogut, {\it The lattice gauge theory approach
    to quantum chromodynamics}, Rev. Mod. Phys.~{\bf 55} (1983)
  775--836

\bibitem{diakonov} D.~Diakonov and V.~Petrov, {\it Yang--Mills theory
    in three dimensions as quantum gravity theory},
  arXiv:hep-th/9912268

\bibitem{Hw_pop} R. Howe, {\it Dual pairs in physics: harmonic
    oscillators, photons, electrons, and singletons}, Lect.~in
  Appl.~Math.~{\bf 21} (1985) 179--207

\bibitem{Howe} R.~Howe, {\it Remarks on classical invariant theory},
  Trans.~Amer.~Math.~Soc.~{\bf 313} (1989) 539--570

\bibitem{circular} M.R.~Zirnbauer, {\it Supersymmetry for systems with
    unitary disorder: circular ensembles}, J.~Phys.~\textbf{A 29}
  (1996) 7113--7136; arXiv:chao-dyn/9609007

\bibitem{icmp97} M.R.~Zirnbauer, {\it The color--flavor transformation
    and a new approach to quantum chaotic maps}, in: Proceedings of
  the XIIth International Congress of Mathematical Physics, Brisbane,
  1997; arXiv:chao-dyn/9810016

\bibitem{Budczies:01} J.~Budczies and Y.~Shnir, {\it Color--flavor
    transformation for the special unitary group and application to
    low--energy QCD}, in: Proceedings of the 15th International
  Workshop on High Energy Physics and Quantum Field Theory, Tver,
  Russia, 2000; arXiv:hep-lat/0101016

\bibitem{boristilo1} B.~Schlittgen and T.~Wettig, {\it Color--flavor
    transformation for the special unitary group},
  Nucl.~Phys.~\textbf{B 632} (2002) 155--172; arXiv:hep-lat/0111039

\bibitem{bnsz} J.~Budczies, S.~Nonnenmacher, Y.~Shnir and
  M.R.~Zirnbauer, {\it $1+1$--dimensional bary\-ons from the ${\rm
      SU}(N)$ color--flavor transformation}, Nucl. Phys. \textbf{B
    635} (2002) 309--356; arXiv:hep-lat/0112018

\bibitem{thesis} J.~Budczies, \emph{The Color Flavor Transformation
    and its Applications to Quantum Chromodynamics}, Ph.D. thesis,
  Universit\"at zu K\"oln (2002); \\
  \verb+http://kups.ub.uni-koeln.de/volltexte/2003/355/+

\bibitem{detdet} S.~Nonnenmacher and M.R.~Zirnbauer, {\it Det--Det
    correlations for quantum maps: dual pair and saddle--point
    analyses}, J. ~Math.~Phys.~\textbf{43} (2002) 2214--2240;
  arXiv:math-ph/0109025

\bibitem{wilson} K.~Wilson, {\it Confinement of quarks}, Phys. Rev.
  \textbf{D 10} (1974) 2445--2459

\bibitem{kogut79} J.B.~Kogut, {\it Lattice gauge theory and spin
    systems}, Rev.~Mod.~Phys.~{\bf 51} (1979) 659--713

\bibitem{creutz} M.~Creutz, \emph{Quarks, Gluons and Lattices}
  (Cambridge University Press, 1983)

\bibitem{balaban} The proof is given in a long series of papers by
  T.~Balaban, the last of which is: {\it Large field renormalization.
    II. Localization, exponentiation and bounds for the R operation},
  Commun.~Math.~Phys.~{\bf 122} (1989) 355--392

\bibitem{knapp} A.~Knapp, \emph{Lie Groups Beyond an Introduction}
  (Princeton University Press, 1996)

\bibitem{witten} E.~Witten, {\it On quantum gauge theories in two
    dimensions}, Commun.~Math.~Phys.~{\bf 141} (1991) 153--209

\bibitem{weyl} H.~Weyl, \emph{The Classical Groups} (Princeton
  University Press, 1939)

\bibitem{sternberg} P.~Bamberg and S.~Sternberg, {\it A Course in
    Mathematics for Students of Physics}, vol.~2 (Cambridge University
  Press, 1990)

\bibitem{kaplan} A.G.~Cohen, D.B.~Kaplan, E.~Katz, and M.~Unsal, {\it
    Supersymmetry on a Euclidean spacetime lattice (I): a theory with
    four supercharges}, arXiv:hep-lat/0302017

%
%
%
%
\end{thebibliography}

\end{document}